\documentclass[prx,aps,twocolumn,showpacs,10pt]{revtex4-1}
\usepackage{amsmath,amssymb,graphicx}
\usepackage{epsfig}
\usepackage{textcomp}
\usepackage{color}
\usepackage{epstopdf}
\usepackage{array}

\usepackage{amssymb}
\usepackage{amsmath}
\usepackage{graphicx}
\usepackage{mathrsfs}
\usepackage[utf8]{inputenc}
\usepackage[english]{babel}
\usepackage{amsthm}
\usepackage{mathtools}

\usepackage{ifthen}
\newcommand{\sifref}[1]
{
\ifthenelse{\equal{#1}{fig:SRGRateEquations}}{S1}{
\ifthenelse{\equal{#1}{fig:SRGSwitchingTimes}}{S2}{
\ifthenelse{\equal{#1}{fig:SRGTimeDependentPDF}}{S3}{
\ifthenelse{\equal{#1}{fig:SRGPrefactorOptimization2D-h35}}{S4}{
\ifthenelse{\equal{#1}{fig:SRGPrefactorOptimization2D-h365}}{S5}{
\ifthenelse{\equal{#1}{fig:SRGPrefactorOptimization2D-h375}}{S6}{
\ifthenelse{\equal{#1}{fig:SRGPrefactorOptimization2D-h40}}{S7}{
\ifthenelse{\equal{#1}{fig:SRGLikelihood-h35}}{S8}{
\ifthenelse{\equal{#1}{fig:SRGLikelihood-h365}}{S9}{
\ifthenelse{\equal{#1}{fig:SRGLikelihood-h375}}{S10}{
\ifthenelse{\equal{#1}{fig:SRGLikelihood-h40}}{S11}{
\ifthenelse{\equal{#1}{fig:SRGSimTheoryComp}}{S12}{
\ifthenelse{\equal{#1}{fig:SRGFitPDF}}{S13}{
\ifthenelse{\equal{#1}{fig:MRNARateEquations}}{S14}{
\ifthenelse{\equal{#1}{fig:MRNASwitchingTimes}}{S15}{
\ifthenelse{\equal{#1}{fig:MRNAPDFs}}{S16}{
\ifthenelse{\equal{#1}{fig:MRNAPrefactorOptimization2D-b5}}{S17}{
\ifthenelse{\equal{#1}{fig:MRNAPrefactorOptimization2D-b4}}{S18}{
\ifthenelse{\equal{#1}{fig:MRNAPrefactorOptimization2D-b3}}{S19}{
\ifthenelse{\equal{#1}{fig:MRNAPrefactorOptimization2D-b2}}{S20}{
\ifthenelse{\equal{#1}{fig:MRNAPrefactorOptimization2D-b1}}{S21}{
\ifthenelse{\equal{#1}{fig:MRNALikelihood-b5}}{S22}{
\ifthenelse{\equal{#1}{fig:MRNALikelihood-b4}}{S23}{
\ifthenelse{\equal{#1}{fig:MRNALikelihood-b3}}{S24}{
\ifthenelse{\equal{#1}{fig:MRNALikelihood-b2}}{S25}{
\ifthenelse{\equal{#1}{fig:MRNALikelihood-b1}}{S26}{
\ifthenelse{\equal{#1}{fig:MRNASimTheoryComp}}{S27}{
\ifthenelse{\equal{#1}{fig:MRNAFitPDF}}{S28}{
\ifthenelse{\equal{#1}{fig:SRGBifurcation}}{S29}{
Unknown siref #1??}}}}}}}}}}}}}}}}}}}}}}}}}}}}}
}

\begin{document}
\title{Reconstructing an epigenetic
landscape using a genetic `pulling' approach}
\author{Michael Assaf$^{1*}$, Shay Be'er$^{1}$ and Elijah Roberts$^{2*}$}
\affiliation{$\;^1$ Racah Institute of Physics, Hebrew University of Jerusalem, Jerusalem 91904, Israel\\
$\;^2$ Department of Biophysics, Johns Hopkins University, Baltimore, MD 21218, USA\\
$\;^*$ Correspondence to michael.assaf@mail.huji.ac.il or eroberts@jhu.edu}
\begin{abstract}
Cells use genetic switches to shift between alternate stable gene expression states, {\it e.g.}, to adapt to new environments or to follow a developmental pathway. Conceptually, these stable phenotypes can be considered as attractive states on an epigenetic landscape with phenotypic changes being transitions between states. Measuring these transitions is challenging because they are both very rare in the absence of appropriate signals and very fast. As such, it has proven difficult to experimentally map the epigenetic landscapes that are widely believed to underly developmental networks. Here, we introduce a new nonequilibrium perturbation method to help reconstruct a regulatory network's epigenetic landscape. We derive the mathematical theory needed and then use the method on simulated data to reconstruct the landscapes. Our results show that with a relatively small number of perturbation experiments it is possible to recover an accurate representation of the true epigenetic landscape. We propose that our theory provides a general method by which epigenetic landscapes can be studied. Finally, our theory suggests that the total perturbation impulse required to induce a switch between metastable states is a fundamental quantity in developmental dynamics.
\end{abstract}
\maketitle

\subsection{Introduction}

\noindent{}The presence of many overlapping feedback-based circuits within a cell's regulatory network has been theorized to give rise to a cellular epigenetic landscape (also called a phenotype landscape) with many metastable states~\cite{Chang2006mmd,Smits2006pvb}. Fluctuations in the cell's state due to molecular noise \cite{McAdams1997smg,Elowitz2000son,Paulsson2000rsf,Thattai2001ing,Ozbudak2002rne,Elowitz2002sge,Blake2003neg,Sasai2003sge,Rosenfeld2005grs,Golding2005rkg,Newman2006spa,Yu2006pge,Shahrezaei2008ads} randomly drive the cell along this epigenetic landscape bounded by the so-called quasi-potential barriers separating the metastable states \cite{Hasty2000nsa,Kepler2001str,Aurell2002eaf}. Most of the time the system dwells in the vicinity of one of these metastable states undergoing small random excursions about it. Occasionally, however, a rare, large fluctuation can move the system from one basin of attraction (of a metastable state) to another \cite{Roma2005ope,Schultz2007mls,Choi2008sse,Morelli2008efr,Mehta2008esn,Leisner2009kgs,Zong2010lsi,Wang2010kpt,Assaf2011dsg,Assaf2013end,Dixit2015itr,Ge2015spt,Roberts2015dsg,Ge2018rss}.

The stability of these phenotypic states, quantified by the mean first passage time (MFPT) or mean switching time (MST) to transition from one state to another solely via fluctuations, is typically very long to ensure stable phenotypes \cite{Walczak2005art}, and yet cells must transition quickly and deterministically once the proper signal is received \cite{Eldar2010frn,Golding2011dml,Balazsi2011cdm,Ghusinga2017fta}. Such noise-driven switches, using positive and negative feedback loops, regulate diverse decision-making processes including persistence \cite{Balaban2004bpa,Rocco2013spf}, bet-hedging \cite{Veening2008beb,Nozoe2017ifl}, gradient decoding \cite{Barkai2009rgd,Sharma2016gsb}, differentiation \cite{Raj2010vge}, phage infection \cite{Zeng2010dms}, and resource sensing \cite{Roberts2011nci,levy2011competitive,Earnest2013dli}.

In developmental processes, the regulatory network guides a developing cell through a series of transitions by moving from one metastable state to another along the quasi-potential landscape~\cite{Huang2009nhc,Garcia-Ojalvo2012tsm,Corson2017snd}. Stochastic fluctuations have been observed to be involved in several developmental processes \cite{Chubb2017sbd} and developmental transitions may involve quick passage through a number of intermediate states \cite{Antolovic2019tsd}. Indeed, cellular reprogramming under strong perturbations follows a barrier crossing process along a one-dimensional order parameter \cite{Pusuluri2017crd}.

Using signals to guide a cell's state artificially along an epigenetic landscape could open new avenues to treating disease using induced pluripotent stem cells and must also underlie natural differentiation processes \cite{Bargaje2017cps,Kaity2018rot}. A theory to describe the work required to transition a cell between metastable states would be valuable in developing detailed models of differentiation networks, and designing differentiation protocols. Yet, reconstructing the cellular epigenetic landscape of a real biological phenotype from steady-state experimental data is usually impossible due to the extreme rareness of the transitions.

Here, we describe a new approach for studying cellular decision landscapes using perturbations. The idea is similar in principle to single-molecule force spectroscopy studies of protein-folding landscapes, allowing one to extract transition information from force-spectroscopy pulling experiments~\cite{hummer2003kinetics,Dudko2008tai}. By pulling a macromolecule or molecular complex at a sufficient force, rare transitions in single molecules such as ligand-receptor dissociation~\cite{florin1994adhesion}, unfolding of a protein~\cite{kellermayer1997folding}, or unzipping of nucleic acids~\cite{liphardt2001reversible} can be experimentally observed. The authors in Refs.~\cite{hummer2003kinetics,Dudko2008tai} have devised a theoretical method, in the framework of the Kramers theory, that allows translating the distribution of rupture forces that can be measured experimentally, into the force-dependent lifetime of the system. In our case, starting with a cellular regulatory network, we apply an external ``force" or ``pulling" to perturb the network in the direction of the desired change. Here, pulling can represent, \textit{e.g.}, adding a time-dependent force to the protein's expression rate, such that the system is pushed closer to the switching barrier and can switch with an increased probability. At this point, the statistics of the response of the system, \textit{i.e.} the statistics of switching events in the presence of such pulling force, are then used to infer the topology of the landscape, which allows evaluating the lifetime of the various metastable states.

To compute the response of the system to external pulling, we employ a semi-classical approach in the spirit of the Wentzel–Kramers–Brillouin (WKB) theory~\cite{Dykman1994lfo,Kessler2007erf,Meerson2008nup,Escudero2009srm,Assaf2010ems,Assaf2011dsg,Roberts2015dsg} in order to treat the underlying chemical master equation describing the stochastic dynamics of the regulatory network. This formalism allows us to transform the master equation into a set of Hamilton equations which can be dealt with analytically or numerically. We then solve these equations under a prescribed external perturbation with given magnitude and duration and compute the change in the switching probability due to the external pulling. Finally, we use our semiclassical solution in a maximum likelihood framework to infer the model's parameters, which allows reconstructing the epigenetic landscape of the network.

We present our method on two prototypical model systems: a one-dimensional (1D) system of a self-regulating gene, and a two-dimensional (2D) system of mRNA-protein positive feedback loop. We then discuss how our model can be generalized to higher-dimensional systems.

\subsection{Switching in the absence of an external perturbation}
Our starting point is an effective 1D model for the dynamics of the protein of interest. It is assumed that the protein is expressed and degraded according to the following set of birth-death reactions
\begin{equation}\label{reactions}
n\xrightarrow[]{\Lambda_n}n+1,\;\;\;\;\;n\xrightarrow[]{{\cal M}_n}n-1,
\end{equation}
where $\Lambda_n$ and ${\cal M}_n$ are the expression and degradation rates, respectively, and $n$ is the protein copy number.

Neglecting intrinsic noise, the mean number of proteins $\bar{n}$ satisfies the following \textit{deterministic} rate equation
\begin{equation}\label{RE}
\dot{\bar{n}}=\Lambda_{\bar{n}}-{\cal M}_{\bar{n}}.
\end{equation}
We are interested in a scenario where this rate equation has (at least) three fixed points: $n_1<n_2<n_3$, where $n_1$ and $n_3$ are stable fixed points corresponding to the \textit{low} and \textit{high} phenotypes, while $n_2$ is an intermediate unstable fixed point. One model system that exhibits this property is a protein that positively regulates itself -- a self-regulating gene (SRG). While our analysis below is done for generic $\Lambda_n$ and ${\cal M}_n$, in all our simulations we have chosen the birth and death rates to satisfy
\begin{equation}\label{Hillrates}
\lambda(q) = \alpha_0 + (1-\alpha_0)\frac{q^h }{q^h+\beta^h},\;\;\;\;\mu(q)=q.
\end{equation}
Here $\lambda(q)=\Lambda_n/N$ and $\mu(q)={\cal M}_n/N$ are rescaled expression and degradation rates, $q=n/N$ is the protein density, while $N$ is the typical system size, assumed to be large, which represents the typical protein copy number in the \textit{high} state. Furthermore, $\alpha_0$ is the rescaled baseline expression rate, $h$ is the Hill exponent, and $\beta$ is the midpoint of the Hill function. Fig.~S1 shows an example of rate equation~(\ref{RE}) using rates~(\ref{Hillrates}) when the system has three fixed points.

Once intrinsic noise is accounted for, these stable fixed points become metastable, and noise-induced switching between $n_1$ and $n_3$ or \textit{vice versa}, occurs. To account for intrinsic noise, we write down the so-called chemical master equation describing the dynamics of $P_n(t)$ -- the probability to find $n$ proteins at time $t$:
\begin{eqnarray}\label{ME}
\dot{P}_n=\Lambda_{n-1}P_{n-1}+{\cal M}_{n+1}P_{n+1}-(\Lambda_n+{\cal M}_n)P_n.
\end{eqnarray}

Let us first consider the case of switching in the absence of external perturbations. Here, one can find an exact expression for the mean switching time (MST), by computing the mean time it takes the system to cross the unstable boundary starting from a state with $n$ proteins~\cite{Gardiner2004hsm}. Yet, since the resulting expression is highly cumbersome, throughout the text we instead use the WKB method~\cite{Dykman1994lfo} to compute the MST, or switching probability.

To set the stage for the WKB method, let us assume without loss of generality that the system starts in the vicinity of the  \textit{low} stable fixed point $n_1$. Assuming the typical system's size is large, $N\gg 1$, the resulting MST is expected to be exponentially long, see below. In this case,  prior to switching the system enters a long-lived metastable state which is centered about $n_1$. Indeed, starting from any initial condition $n_0<n_2$, after a short ${\cal O}(1)$ relaxation time, the dynamics of the probability distribution function can be shown to satisfy the metastability ansatz: $P(n\leq n_2,t)\simeq \pi(n)e^{-t/\tau}$, while $\sum_{n>n_2}P(n)=1-e^{-t/\tau}$~\cite{Dykman1994lfo,Assaf2006stm,Kessler2007erf,Meerson2008nup,Escudero2009srm,Assaf2010ems,assaf2017wkb}. Here, $\tau$ is the MST, $\pi(n)$ is called the quasi-stationary distribution (QSD), which determines the \textit{shape} of the metastable state, and it is evident that the probability to be at $n>n_2$ is negligibly small at times $t\ll \tau$.

We now plug this ansatz into master equation~(\ref{ME}), and neglect the exponentially small term proportional to $\tau^{-1}$ (see below). Employing the WKB ansatz $\pi(n)\equiv \pi(q)\sim\exp[-NS(q)]$ on the resulting quasistationary master equation, where $S(q)$ is the action function, yields  a stationary Hamilton-Jacobi equation $H(q,\partial_q S)=0$, with the Hamiltonian being
\begin{equation}\label{hamil0}
H_0(q,p)=(e^p-1)\left[\lambda(q)-e^{-p}\mu(q)\right].
\end{equation}
Here $p=\partial_q S$ is called the momentum in analogy to classical mechanics, while the subscript $0$ stands for the unperturbed case. To find the optimal path to switch -- the path the system takes with an overwhelmingly large probability during a switching event~\cite{dykman1979theory,freidlin1998random} -- we need to find a nontrivial heteroclinic trajectory, $p_0(q)$, connecting the saddles $(q,p)=(q_1,0)$ and $(q_2,0)$, where  $q_1=n_1/N$ and $q_2=n_2/N$~\cite{Dykman1994lfo,Kessler2007erf,Meerson2008nup,Escudero2009srm,Assaf2010ems}. Equating $H_0=0$ yields
\begin{equation}\label{p0}
p_0(q)=\ln\left[\mu(q)/\lambda(q)\right].
\end{equation}
Thus, the action function is found by integrating: $S(q)=\int p(q')dq'$. The MST between the \textit{low} and \textit{high} states can be shown to satisfy in the leading order~\cite{Dykman1994lfo,Meerson2008nup,Escudero2009srm,assaf2017wkb}
\begin{equation}
\tau_{\mbox{\tiny{low}}\to \mbox{\tiny{high}}}\sim e^{N{\cal S}_0^{lh}}
\end{equation}
where ${\cal S}_0^{lh}=S(q_2)-S(q_1)=\int_{q_1}^{q_2}\ln[\mu(q)/\lambda(q)]dq,$ is the switching barrier between the \textit{low} and \textit{high} states, in the absence of an external force. Similarly, $\tau_{\mbox{\tiny{high}}\to \mbox{\tiny{low}}}\sim e^{N{\cal S}_0^{hl}}$, where
${\cal S}_0^{hl}=S(q_2)-S(q_3)=\int_{q_3}^{q_2}\ln[\mu(q)/\lambda(q)]dq,$ is the switching barrier between the \textit{high} and \textit{low} states. For $N\gg 1$, these MSTs are indeed exponentially large thus validating our a-priori metastability assumption (see Fig.~S2). Note that, in the unperturbed case, the pre-factor of $\tau$ can be accurately found as well~\cite{Escudero2009srm,Newby2015bsa}.

In the following, rather than the MST, we will be interested in computing ${\cal P}^{lh}$ and ${\cal P}^{hl}$ -- the switching \textit{probabilities} over some time $t\ll \tau$ starting from the \textit{low} to \textit{high} and \textit{high} to \textit{low} states, respectively. For example, starting from the vicinity of $n_1$, ${\cal P}^{lh}$ is determined by the fraction of stochastic realizations of process~(\ref{reactions}) that cross $n_2$ in a given time $t$ out of the total number of realizations. Using the metastability ansatz, and demanding that the total probability be unity, we have ${\cal P}^{lh}=\sum_{n>n_2}P(n,t)\simeq 1-e^{-t/\tau}\simeq t/\tau$, where the last approximation holds for $t\ll\tau$; that is, ${\cal P}^{lh}$ is exponentially small at $t\ll\tau$. As a result, in the absence of external force, and using a similar argument for the calculation of ${\cal P}^{hl}$, the switching probabilities up to some arbitrary time $t\ll \tau$ satisfy in the leading order
\begin{eqnarray}\label{prob}
\hspace{-4mm}{\cal P}^{lh}\sim \tau_{\mbox{\tiny{low}}\to \mbox{\tiny{high}}}^{-1}\sim e^{-N{\cal S}_0^{lh}},\;\;{\cal P}^{hl}\sim \tau_{\mbox{\tiny{high}}\to \mbox{\tiny{low}}}^{-1}\sim e^{-N{\cal S}_0^{hl}}\!,
\end{eqnarray}
where logarithmic corrections depending on the arbitrary time $t$ and the pre-factor entering $\tau$ have been omitted.

\subsection{Switching in the presence of an external perturbation}

\textit{Low-to-high switch.\;}
Let us begin by studying the case of \textit{low} to \textit{high} switch in the presence of an external perturbation. To do so, we add an external time-dependent force to the protein's expression rate, $\Lambda_n\to\Lambda_n+\phi(t)$, where $\phi(t)$ is applied for a finite duration $T$
such that
\begin{equation}\label{force}
\phi(t)=\begin{cases}
0 & t<0\;\; \mbox{or} \;\;t>T, \\
F & 0<t<T.
\end{cases}
\end{equation}
As shown in Fig.~S3, the result of this perturbation is that the system is pushed nearer to the switching barrier and with some increased probability can then switch to the \textit{high} state. The switching probability depends on both the force $F$ and duration $T$ of the perturbation. Note that, in general, the system does not need to relax to a new quasi-stationary distribution during the perturbation. We now compute the dependence of the change in the \textit{low} to \textit{high} switching barrier ${\cal S}^{lh}$ on $F$ and $T$.

\begin{figure}[t!]
\begin{center}
\includegraphics{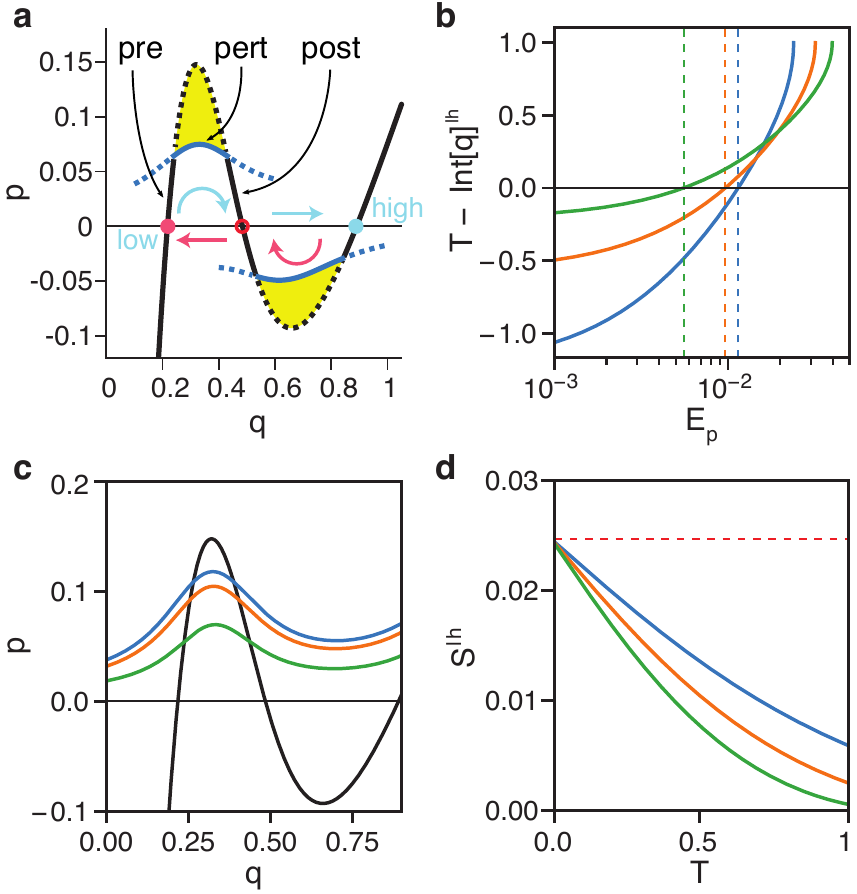}
\end{center}
\vspace{-0.5cm}
\caption{{\bf Calculation of the optimal switching path for the self-regulating gene.} (a) Illustration of the optimal paths to switching without (black) and with (blue) perturbation. The path segments are labeled as pre-perturbation (pre), perturbation (pert), and post-perturbation (post), see text. The top path shows \textit{low} to \textit{high} switching and the bottom path shows \textit{high} to \textit{low}. The yellow areas give the decrease in the switching barriers due to the perturbation. In the \textit{low} to \textit{high} switch we have taken $F=0.15$ and $E_p=0.01$ such that $T=1.1$, whereas for the \textit{high} to \textit{low} we have taken $F=0.2$ and $E_p=0.005$ such that $T=2.06$. The other parameters are $\alpha_0=0.2$, $\beta=0.562$ and $h=4$. (b) Plot of $T-\int 1/\dot{q}\,dq$ [see equation (12)] vs $E_p$ for the \textit{low} to \textit{high} switch under three different values of the perturbation strength $F$: 0.15 (blue), 0.20 (orange), 0.25 (green). Also shown are (c) momentum $p$ vs coordinate $q$ and (d) the action ${\cal S}_0^{lh}$ vs perturbation time $T$ for the same three $F$ values.}
\label{fig:TheorySummary}
\end{figure}

Given the time-dependent protocol $\phi(t)$ [Eq.~(\ref{force})] for the change in the protein's expression rate, one can perform a similar WKB analysis as done above in the unperturbed case. This yields two distinct Hamiltonians: the unperturbed Hamiltonian~(\ref{hamil0}) before and after the external perturbation has been applied, and the Hamiltonian during the perturbation with an elevated expression rate:
\begin{equation} \label{hamilpertlh}
H_p(q,p) = (e^p-1)\left[\lambda(q)+F-e^{-p}\mu(q)\right],
\end{equation}
where the subscript $p$ stands for the perturbed case. Each of the two Hamiltonians is an integral of motion on the corresponding time interval. Here, the optimal switching path $[q_{op}(t),q_{op}(t)]$ starts at the saddle point $(q,p)=(q_1,0)$ well before the perturbation has been applied, and ends at the saddle point $(q,p)=(q_2,0)$, well after the perturbation has been applied. It can be found by matching three separate trajectory segments: the pre-perturbation, perturbation, and post-perturbation segments (see Fig.~\ref{fig:TheorySummary}a)~\cite{assaf2009population,vilk2018population,israeli2020population}.

The matching conditions at times $t=0$ and $t=T$ are provided by the continuity of the functions $q(t)$ and $p(t)$~\cite{continuity}. The pre- and post-perturbation segments must have a zero energy, $E=0$, so they are parts of the original zero-energy trajectory, $p_0(q)$, see Eq.~(\ref{p0}). Yet, for the perturbation segment, the energy $E=E_p$ is nonzero and a-priori unknown.
It parameterizes the intersection points $q_1^{p}$ and $q_2^{p}$ between the unperturbed zero-energy line $p_0(q)$ [Eq.~(\ref{p0})] and the perturbed path, $p_{p}(q)$~\cite{assaf2009population,vilk2018population,israeli2020population}. The latter is the solution of $H_p(q,p)=E_p$, yielding:
\begin{equation} \label{pint}
p_{p}(q)=\ln\left\{\left[B+\sqrt{B^2 - 4AC}\right]/(2A)\right\},
\end{equation}
where $A$, $B$ and $C$ are functions of $q$ and satisfy $A=\lambda(q)+F$, $B=\lambda(q)+F+\mu(q)+E_p$, and $C=\mu(q)$.

To determine the energy $E_p$, we demand that the duration of the perturbation be $T$~\cite{assaf2009population,vilk2018population,israeli2020population}. Thus, we have:
\begin{equation} \label{Ep}
T = \int_0^T dt = \int_{q_1^{p}(E_p)}^{q_2^{p}(E_p)}\frac{dq}{\dot{q}[q,p_{p}(q,E_p)]},
\end{equation}
where $q_{1,2}^{p}(E_p)$ are the intersection points between the unperturbed $p_0(q)$ and perturbed $p_{p}(q)$ trajectories, and $\dot{q}(q,p)=dq/dt$ is given by Hamilton's equation $\dot{q} = \partial H_p/\partial p = [\lambda(q)+F]e^p - \mu(q)e^{-p}$. Therefore, plugging $p_p(q)$ from Eq.~(\ref{pint}) into $\dot{q}$, Eq.~(\ref{Ep}) becomes:
\begin{equation} \label{EpT}
T=\int_{q_1^{p}(E_p)}^{q_2^{p}(E_p)}\left(B^2-4AC\right)^{-1/2}dq,
\end{equation}
where $A$, $B$ and $C$ are given below Eq.~(\ref{pint}). Finally, using the fact that the action satisfies ${\cal S}=\int_{-\infty}^{\infty}\{p_{op}(t)\dot{q}_{op}(t)-H[q_{op}(t),p_{op}(t),t]\}dt$~\cite{escudero2008persistence,assaf2008population},
and recalling that $dS = (\partial S/\partial t)dt+(\partial S/\partial q)dq$,
we  arrive at the perturbed switching barrier from the \textit{low} to \textit{high} states:
\begin{equation} \label{actiongeneral}
{\cal S}^{lh} = {\cal S}_0^{lh} - \int_{q_1^{p}(E_p)}^{q_2^{p}(E_p)} \left[p_0(q) - p_p(q,E_p) \right]dq - E_p T,
\end{equation}
where $E_p = E_p(F,T)$ can be found from Eq.~(\ref{EpT}), ${\cal S}_0^{lh}=\int_{q_1}^{q_2}p_0(q)dq$ is the unperturbed switching barrier from the \textit{low} to \textit{high} states, and we have used the fact that $\int_0^T H_p dt=E_p T$.

Note that, for the birth and death rates of the SRG model~(\ref{Hillrates}), Eq.~(\ref{EpT}) has no closed form solution. To study the switching behavior under such perturbation, we first numerically evaluate the integral equation to find the matching $E_p$ (see Fig.~\ref{fig:TheorySummary}b). Given a numerical value for $E_p$, we then use Eq.~(\ref{actiongeneral}) to calculate the perturbed action (Fig.~\ref{fig:TheorySummary}c-d). Finally, we use Eq.~(\ref{prob}) to calculate the perturbed switching probability ${\cal P}^{lh}$.

\begin{figure}[t!]
\begin{center}
\includegraphics{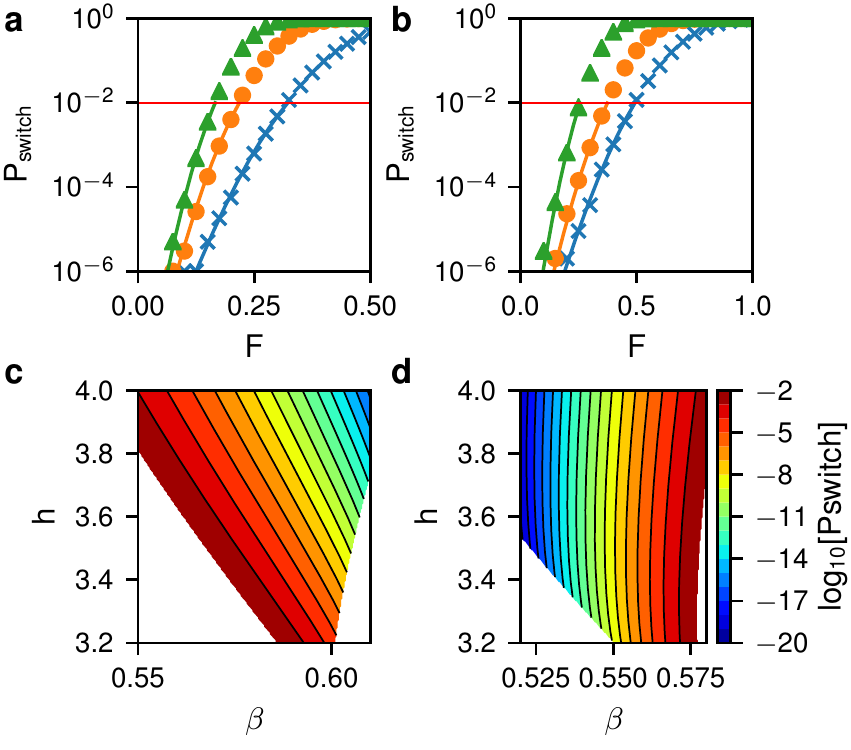}
\end{center}
\vspace{-0.5cm}
\caption{{\bf Change in the switching probability with perturbation.} (a) Change in switching probability vs perturbation strength $F$ for the \textit{low} to \textit{high} switch. Symbols and lines give numerical and theoretical values, respectively. A constant prefactor of 0.15 was used with the theory. Data are shown for three values of the perturbation time $T$: 0.5 (blue $\times$), 0.75 (orange $\circ$), 1.0 (green $\bigtriangleup$).  (b) Change in switching probability vs perturbation strength for the \textit{high} to \textit{low} switch with perturbation times 0.75 (blue $\times$), 1.0 (orange $\circ$), and 1.5 (green $\bigtriangleup$), and a prefactor of 0.2. (c+d) Switching probabilities in the $\beta$ vs $h$ plane for the \textit{low} to \textit{high} switch (c) and the \textit{high} to \textit{low} switch (d).}
\label{fig:SRGSimTheoryComp}
\end{figure}

To check our theory, we compared the theoretical dependence of ${\cal P}^{lh}$ on $F$ and $T$ against Monte Carlo simulations \cite{Roberts2013lmh}. We calculated ${\cal P}^{lh}$ for a range of $F$ values for three different perturbation times $T$. Because we are trying to develop a theory that is directly relatable to biological experiments, the $F$ values were limited such that ${\cal P}^{lh} > 1 \times 10^{-6}$. Detecting a cell phenotype with a frequency of one per million cells is at the limit of feasibility using flow cytometry techniques. We also limited the comparison to ${\cal P}^{lh} < 1 \times 10^{-2}$, above which the switching barrier ${\cal S}^{lh}$ starts to become low enough such that the WKB approximation is invalidated~\cite{Escudero2009srm}. Figure~\ref{fig:SRGSimTheoryComp}a, shows excellent agreement between theory and numerics.

Finally, our result for the switching probability in the aftermath of an external perturbation [Eq.~(\ref{actiongeneral})] can be simplified in three particular limits: (i) close to the bifurcation limit, where the \textit{low} and intermediate fixed points merge and the switching barrier vanishes, (ii) for weak external force, $F\ll 1$, and (iii) in the case of $h\to\infty$, \textit{i.e}, a very steep regulatory function. Close to the bifurcation limit, we find that ${\cal S}^{lh}$  depends only on the \textit{impulse} of the perturbation, $FT$, see Discussion and Appendix A; in the case of weak force, we show that the increase in the switching probability is exponential in $F$, see Appendix B, while in the limit of $h\to\infty$, where the expression rate is given by a heaviside step function, we find an explicit expression for  ${\cal S}^{lh}$ as function of $F$ and $T$, see Appendix C.
~\\

\textit{High-to-low switch.\;}
We now turn to the case of switching from the \textit{high} to \textit{low} states in the presence of an external perturbation. Here, switching can be driven, {\it e.g.}, by increasing the protein's degradation rate, $\mu(n)\to\mu(n)[1+\phi(t)]$, where $\phi(t)$ is given by Eq.~(\ref{force}), which yields the perturbed Hamiltonian
\begin{equation} \label{hamilperthl}
H_p(q,p) = (e^p-1)\left[\lambda(q)-e^{-p}\mu(q)(1+F)\right].
\end{equation}
As a result, the intersection points $q_2^{p}$ and $q_3^{p}$ are now determined by equating $p_0(q)$ from Eq.~(\ref{p0}) with the perturbed path $p_{p}(q)$, given by Eq.~(\ref{pint}), with $A=\lambda(q)$, $B=\lambda(q)+\mu(q)(1+F)+E_p$, and $C=\mu(q)(1+F)$.

To find $E_p$, we use Eq.~(\ref{hamilperthl}) to write Hamilton's equation $\dot{q} = \partial H_p/\partial p = \lambda(q)e^p - \mu(q)(1+F)e^{-p}$. Replacing the lower integration limit of Eq.~(\ref{Ep}) by $q_3^{p}(E_p)$, we find
\begin{equation} \label{EpThl}
T=\int_{q_2^{p}(E_p)}^{q_3^{p}(E_p)}\left(B^2-4AC\right)^{-1/2}dq,
\end{equation}
where $A$, $B$ and $C$ are given below Eq.~(\ref{hamilperthl}), and we have swapped the integration limits such that the integrand is positive.
Finally, the switching barrier from the \textit{high} to \textit{low} states is given by Eq.~(\ref{actiongeneral}) upon replacing the lower integration limit by $q_3^{p}(E_p)$, and ${\cal S}_0^{lh}$ by ${\cal S}_0^{hl}$.

Fig.~\ref{fig:SRGSimTheoryComp}b shows a comparison of the perturbed \textit{high} to \textit{low} switching probabilities from Monte Carlo simulations with ${\cal P}^{hl}$ calculated using the above action along with Eq.~(\ref{prob}). They are again in excellent agreement.

Notably, in all of our calculations we have assumed a square pulse. Yet, such a pulse is practically impossible to realize experimentally. Instead, one expects experimental signals to be noisy, and to gradually rise and drop slowly rather than instantaneously. Nevertheless, we claim that the exact form of the pulse will not change the above results qualitatively. For example if instead of an instantaneous rise and drop, we have a linear rise in the pulse up to some maximal value, and a linear drop back to the original value, one can use the same mathematical formalism as above. Indeed, in this case the optimal path will be comprised of five segments rather than three: a pre- and post-perturbation segment, a perturbed segment, and two segments with intermediate values of force corresponding to the \textit{average} force value in the regimes of linear increase and decrease of the pulse.

\subsection{Inference of the epigenetic landscape}

We now proceed to our main idea which is to use our theoretical formalism to infer the epigenetic landscape of a regulatory network, given experimental data of the network's response to external perturbations. The goal is to find the set of parameters for the regulatory network that best recapitulate the observed responses. We first set out to determine the feasibility of inferring the parameters from the perturbation data.

For the SRG, two key parameters that control the shape of the landscape are $\beta$, which influences the barrier position, and $h$, which influences the landscape steepness. We desired to know to what extent these two parameters could be independently distinguished using only the switching probability. To this end, we used our theory to calculate the dependence of ${\cal P}^{lh}$ and ${\cal P}^{hl}$ on $\beta$ and $h$. As can be seen in Fig.~\ref{fig:SRGSimTheoryComp}c+d, when switching either from \textit{low} to \textit{high} or from \textit{high} to \textit{low}, $\beta$ and $h$ can be changed simultaneously to maintain the same switching probability. This corresponds, \textit{e.g.}, to moving the barrier position closer to the starting state while increasing the height of the barrier. Yet, by considering switching in \textit{both} directions simultaneously, both $\beta$ and $h$ are uniquely constrained, and there is only one pair ($\beta$,$h$) consistent with both the \textit{low} to \textit{high} and \textit{high} to \textit{low} pulling.

To perform parameter inference we adopted a  \textit{maximum likelihood} approach. To generate synthetic experimental data we performed Monte Carlo simulations of the stochastic process~(\ref{reactions}), and measured for various values of $F$ and $T$ the number of realizations $k$, out of $m$ total realizations, that switched phenotypes after some designated time. For all of our simulations we used $m= 1 \times 10^{6}$. Given the switching probability ${\cal P}$ for each realization, and assuming that the sequence of ``experiments" or numerical realizations is independent and identically distributed, the probability $P(k)$ that exactly $k$ realizations out of $m$ switch is given by a binomial distribution
\begin{equation}
P(k)=\binom{m}{k} {\cal P}^k(1-{\cal P})^{m-k}.
\end{equation}

The likelihood of parameters $\theta$ producing the observed data $k$ given all of the various experimental $F$ and $T$ conditions is then given by the product of all $P(k)$ values
\begin{equation}
\hspace{-0.5mm}{\cal L}(\theta|k)=\!\!\!\!\!\prod_{\{T_i,F_j\}}\!\!\!\!\!P_{\theta}(k_{i,j})=\!\!\prod_{i,j}\!\binom{m}{k_{i,j}} \!{\cal P}_{\theta}^{k_{i,j}}(1\!-\!{\cal P}_{\theta})^{m-k_{i,j}},
\label{likelihood}
\end{equation}
where $i$ and $j$ denote the indices of the current values of $T$ and $F$, and $k_{i,j}$ denotes the number of realizations that switched given that $T=T_i$ and $F=F_j$. Note that, the likelihood function in Eq.~(\ref{likelihood}) includes both \textit{low} to \textit{high} and \textit{high} to \textit{low} pulling experiments.

Importantly, the probability of success ${\cal P}$ is given by Eq.~(\ref{prob}); as we have shown, it depends, in addition to $T$ and $F$, on the parameters $\theta$ defining the birth and death rates. By maximizing the likelihood function ${\cal L}$, we find the most probable parameter set for the birth and death rates $\Lambda_n$ and ${\cal M}_n$, given the perturbation data.

\begin{figure}[t!]
\begin{center}
\includegraphics{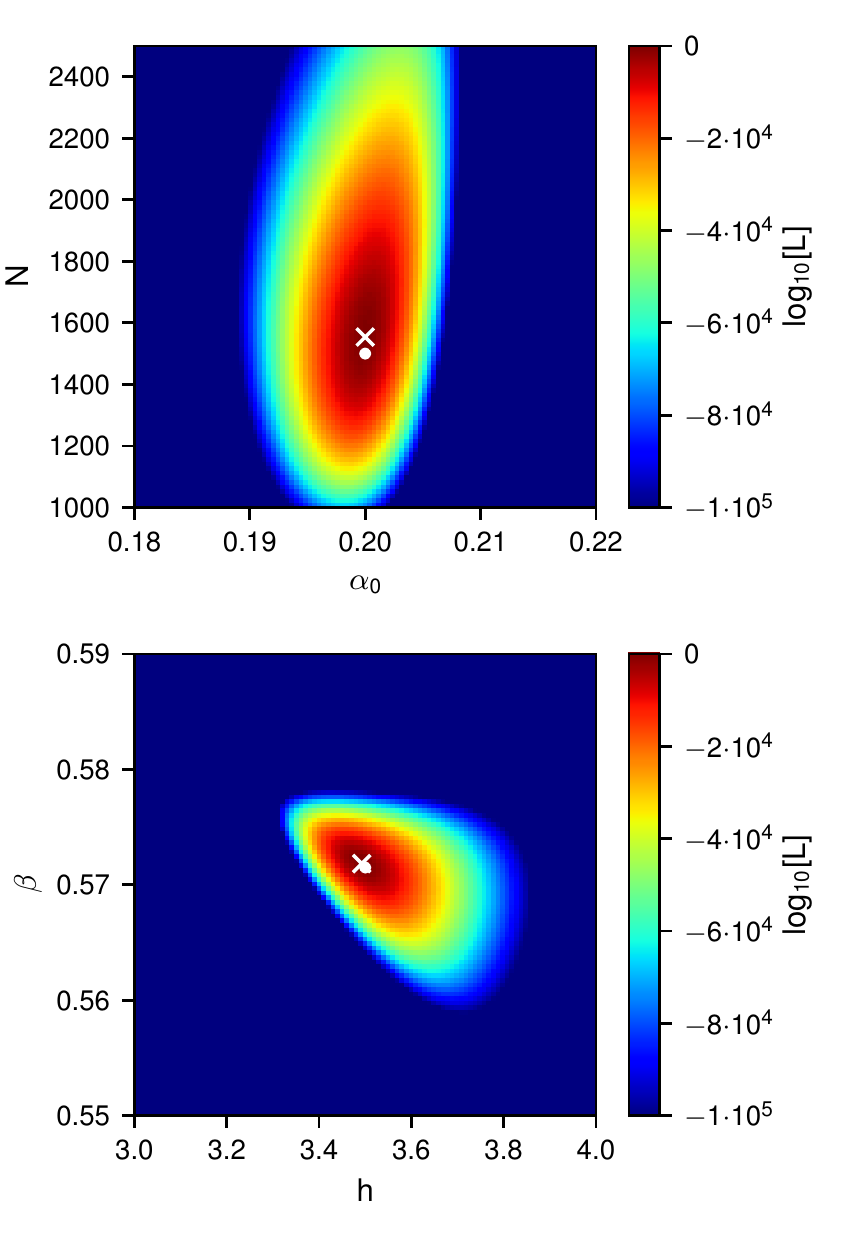}
\end{center}
\vspace{-1.0cm}
\caption{{\bf Log-likelihood function for inference of model parameters.} (top) The likelihood in the $N$ vs $\alpha_0$ plane where other parameters are fixed to their MLE. (bottom) The same for the $\beta$ vs $h$ plane. The white $\times$ symbols show the MLE and the white $\circ$ symbols show the true parameter values. The likelihood function is computed based on the data presented in Fig.~\ref{fig:SRGSimTheoryComp}.}
\label{fig:SRGLikelihood}
\end{figure}

We used the synthetic data set shown in Fig.~\ref{fig:SRGSimTheoryComp}a+b along with Eq.~(\ref{likelihood}) to infer the maximum likelihood estimate (MLE) for the three model parameters $N$, $\beta$, and $h$. We assume that $\alpha_0$ can be obtained directly from experimental measurement of the ratio of the stable fixed points. Again, we used only $F$ and $T$ values with $1 \times 10^{-6} < {\cal P} < 1 \times 10^{-2}$, which amounted to $\sim$35 experimental conditions combined from both switching directions. Fig.~\ref{fig:SRGLikelihood} and Fig.~S8 show the likelihood distribution resulting from the inference. The MLE was $N=1554$, $\beta=0.5718$, and $h=3.492$, which was in excellent agreement with the true parameter values of $N=1500$, $\beta=0.5715$, and $h=3.5$.

Because the WKB is a logarithmic theory, there is a preexponent in Eq.~(\ref{prob}) that must be estimated in order to compute the absolute value of ${\cal P}$. Typically, in a WKB theory this prefactor is obtained from fitting the functional dependence of the theory to the data. Here, we took the approach of obtaining the prefactor for both ${\cal P}^{lh}$ and ${\cal P}^{hl}$ directly from the likelihood estimation. We maximized likelihood over a range of prefactors and then used the set with the highest likelihood for all remaining calculations (see Fig.~S4). Comparison of the theory with optimized prefactors and parameters to the true parameters shows that the optimization leads to a moderate increase in likelihood while maintaining the excellent fits vs $T$ and $F$ (see Fig.~S12a+b).

\begin{figure}[t!]
\begin{center}
\includegraphics{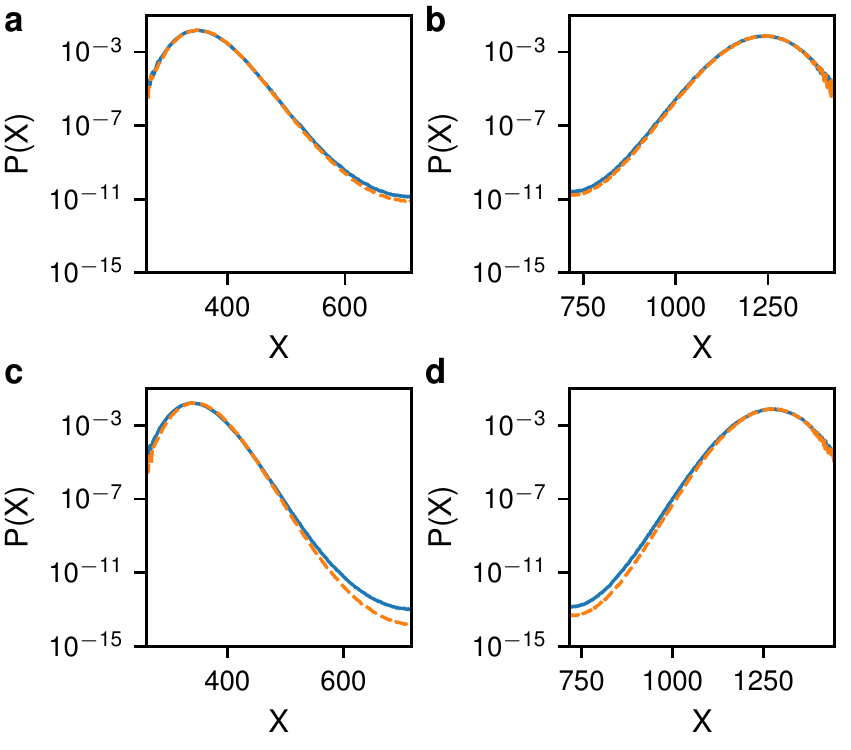}
\end{center}
\vspace{-0.5cm}
\caption{{\bf Comparison of actual and inferred probability distributions for the self-regulating gene.} (a) The actual (solid blue) and inferred (dashed orange) PDFs for the \textit{low} state, where $X$ denotes the protein copy number. (b) The same for the \textit{high} state. The model parameters were $N=1500$, $\alpha_0=0.2$, $\beta=0.5715$ and $h=3.5$ (a+b) and $h=3.65$ (c+d).}
\label{fig:SRGFitPDF}
\end{figure}

Finally, we used the MLE parameters to reconstruct the stationary probability density function (PDF) that corresponds to the epigenetic landscape. Fig.~\ref{fig:SRGFitPDF}a+b shows a comparison of the inferred and true PDFs, which we calculated for a given set of parameters using an enhanced sampling technique \cite{Klein2019aec}. The agreement is again excellent and shows that by using only $\sim$35 perturbation data points we are able to successfully reconstruct the epigenetic landscape of the model.

To further test our ability to use the theory to infer the network's PDF, we tested several other SRG parameters sets with increasing switching barrier heights (see Figs.~S4--S12). Fig.~\ref{fig:SRGFitPDF}c+d shows that for $h=3.65$ a greater discrepancy appears between the inferred and actual landscapes, with $\sim$5\% error in the height of the switching barrier. As the switching barrier continues to increase so does the estimated error (Fig.~S13). For $h=4.0$ the error is $\sim$10\% of the barrier height. However, at this value of $h$ the MST is $\sim{}10^{17}$. At these very long switching times, emanating from the large landscape steepness, additional perturbation points with ${\cal P} < 1 \times 10^{-6}$ may be necessary to accurately infer landscapes with a smaller error margin.

\subsection{One-state mRNA-protein model}

\textit{The unperturbed case.\;}
Above, we used the SRG as a basis to infer the landscape of a 1D switch. To see how our method can be generalized to higher-dimensional systems, we now repeat the calculations done above for a 2D system: the one-state mRNA-protein model with positive feedback that displays bistability. We explicitly account for mRNA noise which has been shown to greatly affect the switching properties in genetic circuits \cite{Assaf2011dsg}.

We consider a one-state gene-expression model where transcription depends on the protein copy number via positive feedback. The deterministic rate equation describing the dynamics of the average numbers of mRNA and proteins, respectively denoted by $\bar{m}$ and $\bar{n}$, satisfies:
\begin{equation}\label{REmrna}
\dot{\bar{m}}=\Lambda_{\bar{n}}/b-\gamma\bar{m}\;;\;\;\;\dot{\bar{n}}=\gamma b\bar{m}-\bar{n}.
\end{equation}
Here, $\gamma\gg 1$ is the mRNA degradation rate (relevant \textit{e.g.} for bacterial systems~\cite{Shahrezaei2008ads}), $\gamma b$ is the protein translation rate, such that $b$ is the burst size (the number of proteins created from a single instance of mRNA) and all rates are rescaled by the protein's degradation rate or cell division rate. Furthermore, $\Lambda_{\bar{n}}$ is a sigmoid-like function that ensures bistability (see Fig.~S14). By choosing the mRNA transcription rate to be $\Lambda_{\bar{n}}/b$, we made sure that the fixed points of the protein satisfy $\Lambda_{\bar{n}}=\bar{n}$, which coincide with those of Eq.~(\ref{RE}) for the SRG, upon choosing ${\cal M}_{\bar{n}}=\bar{n}$.

To find the switching probability we write down the master equation describing the dynamics of $P_{m,n}$ -- the probability to find $m$ mRNA molecules and $n$ proteins:
\begin{eqnarray}\label{MEmrna}
\hspace{-5mm}&&\dot{P}_{m,n}\!=\![\Lambda_n/b](P_{m-1,n}\!-\!P_{m,n})\!+\!\gamma b m (P_{m,n-1}\!-\!P_{m,n})\nonumber\\
\hspace{-5mm}&&+\gamma[(m\!+\!1) P_{m\!+\!1,n}\!-\!mP_{m,n}]+(n+1)P_{m,n\!+\!1}\!-\!nP_{m,n}.
\end{eqnarray}
Following the SRG calculations above, we use the metastable ansatz $P_{m,n}=\pi_{m,n}e^{-t/\tau}$ in Eq.~(\ref{MEmrna}), and employ the WKB approximation, $\pi_{m,n}=\pi(x,y)=e^{-NS(x,y)}$. Here $S(x,y)$ is the action, $N\gg 1$ is the typical protein copy number at the \textit{high} state, and $x=m/N$ and $y=n/N$ are the mRNA and protein concentrations, respectively. This yields a stationary Hamilton-Jacobi equation $H(x,y,\partial_x S,\partial_y S)=0$ with Hamiltonian~\cite{Vardi2013bye,assaf2017wkb}
\begin{equation}\label{Hammrna}
H\!=\!y(e^{-p_y}-1)+\gamma b x(e^{p_y}-1)+\gamma x(e^{-p_x}-1)+\frac{\lambda(y)}{b}(e^{p_x}-1)\!,
\end{equation}
where $p_x=\partial_x S$ and $p_y=\partial_y S$ are the mRNA and protein associated momenta, respectively, and $\lambda(y)=\Lambda(y)/N$.

The switching path from the \textit{low} to \textit{high} states (or vice versa) corresponds to a heteroclinic trajectory of Hamiltonian~(\ref{Hammrna}) connecting the saddle points $(x,y,p_x,p_y)=(y_{low}/(\gamma b),y_{low},0,0)$ and $(y_{high}/(\gamma b),y_{high},0,0)$ in the 4D phase space; it can be found by solving the Hamilton equations $\dot{x}=\partial_{p_x} H$, $\dot{y}=\partial_{p_y} H$, $\dot{p}_x=-\partial_{x} H$, and $\dot{p}_y=-\partial_{y} H$, which read
\begin{eqnarray}
&&\dot{x}=[\lambda(y)/b]e^{p_x}\!-\!\gamma xe^{-\!p_x},\;\;\dot{p}_x=\gamma b(1\!-\!e^{p_y})\!+\!\gamma(1\!-\!e^{-\!p_x}),\nonumber\\
&&\dot{y}=\gamma b xe^{p_y}\!-\!ye^{-\!p_y},\;\;\dot{p}_y=1\!-\!e^{-\!p_y}\!+\![\lambda'(y)/b](1\!-\!e^{p_x})\!.\label{Hameqs}
\end{eqnarray}
While a numerical solution can be found for any set of parameters, in order to make analytical progress we consider the limit where the mRNA lifetime is short compared to that of the protein, $\gamma\gg 1$, which holds \textit{e.g.}, in bacteria. In this limit, the mRNA concentration and momentum, $x(t)$ and $p_x(t)$, instantaneously equilibrate to some  (slowly varying) functions of $y$ and $p_y$~\cite{assaf2008noise}. Putting $\dot{x}=\dot{p}_x=0$ in the first two of Eqs.~(\ref{Hameqs}), we obtain $e^{-p_x}=b(1-e^{p_y})+1$ and $x=[\lambda(y)/(\gamma b)]/[b(1-e^{p_y})+1]^{2}$.
Using these relations in Hamiltonian~(\ref{Hammrna}) we arrive at a \textit{reduced} Hamiltonian for $y$ and $p_y$ only. Denoting $y\equiv q$ and $p_y\equiv p$, the effective 1D Hamiltonian reads~\cite{Vardi2013bye,assaf2017wkb}
\begin{equation}\label{hamilmran0}
H_0=q(e^{-p}-1)-\lambda(q)\frac{1-e^{p}}{b(1-e^{p})+1},
\end{equation}
where the subscript $0$ denotes the unperturbed case. This Hamiltonian effectively accounts for the fact that the proteins are produced in geometrically distributed bursts with mean $b$, which in turn asymptotically accounts for the mRNA noise when $\gamma\gg 1$. This Hamiltonian is our starting point for treating this system under external perturbation, and serves as the unperturbed Hamiltonian, similarly as Hamiltonian~(\ref{hamil0}). Note that, as done for the SRG model above, using this unperturbed Hamiltonian [Eq.~(\ref{hamilmran0})], one can find the unperturbed action which yields the PDF, $P_{m,n}$, and MST, in the absence of external perturbation, see Fig.~S15+S16.

~\\
\textit{The perturbed case.\;}
We now repeat the calculations done for the SRG in the perturbed case. Note that, here instead of perturbing the protein's expression and degradation rates, we perturb those of the mRNA. While both cases can be studied theoretically, we desired to study the impact of transcriptional perturbations as being more closely aligned with existing experimental techniques.

We start by perturbing the mRNA's transcription rate $\Lambda_n\to\Lambda_n+\phi(t)$, where $\phi(t)$ is given by Eq.~(\ref{force}). As before, the optimal path is made of three segments: an unperturbed segment before the onset of perturbation, a perturbed segment while the perturbation is applied, and an unperturbed segment after the perturbation has terminated. The unperturbed segment is found by equating Hamiltonian~(\ref{hamilmran0}) to zero
\begin{equation}\label{p0mrna}
p_0(q)=\ln \{[(b+1)q]/[b q+\lambda(q)]\}.
\end{equation}
The perturbed segment can be found by using Hamiltonian~(\ref{hamilmran0}) with the perturbed transcription rate
\begin{equation} \label{hamilpertlhmrna}
H_p(q,p) = q(e^{-p}-1)-[\lambda(q)+F]\frac{1-e^{p}}{b(1-e^{p})+1},
\end{equation}
and equating it to $E_p$; here the subscript $p$ stands for perturbation. The resulting perturbed segment reads
\begin{equation}\label{pintmrna}
p_{p}(q)=\ln\left\{\left[B+\sqrt{B^2 - 4AC}\right]/(2A)\right\},
\end{equation}
where $A=\lambda(q)+F+b(E_p+q)$, $B=E_p(b+1)+\lambda(q)+F+q(1+2b)$, and $C=(1+b)q$.
To determine the energy $E_p$, we use Eq.~(\ref{Ep}) with $\dot{q}$ found from Hamiltonian~(\ref{hamilpertlhmrna}). By doing so, condition~(\ref{Ep}) becomes:
\begin{equation}\label{EpTmrna}
T=\int_{q_1^{p}(E_p)}^{q_2^{p}(E_p)}\frac{2A+b\left(2A-B-\sqrt{B^2-4AC}\right)}{2A\sqrt{B^2-4AC}}dq.
\end{equation}

Finally, the action is given by Eq.~(\ref{actiongeneral}) with $E_p$ from Eq.~(\ref{EpTmrna}), while
${\cal S}_0^{lh}=\int_{q_1}^{q_2}p_0(q)dq$ is the unperturbed action from the \textit{low} to \textit{high} states.

Next, we perturb the degradation rate of the mRNA such that $\gamma m$  becomes $\gamma(1+F)m$. As a result, after some algebra the perturbed Hamiltonian becomes
\begin{equation} \label{hamilperthlmrna}
H_p(q,p) = q(e^{-p}-1)-\lambda(q)\frac{1-e^{p}}{b(1-e^{p})+1+F}.
\end{equation}
Equating $H_p(q,p)=E_p$, the perturbed segment yields
\begin{equation}\label{pintmrnahl}
p_{p}(q)=\ln\left\{\left[B-\sqrt{B^2 - 4AC}\right]/(2A)\right\},
\end{equation}
where $A=\lambda(q)+b(E_p+q)$, $B=E_p(1+b+F)+\lambda(q)+q(1+2b+F)$, and $C=q(1+b+F)$.
To determine the energy $E_p$, we use Eq.~(\ref{Ep}) with $\dot{q}$ found from Hamiltonian~(\ref{hamilperthlmrna}). By doing so, condition~(\ref{Ep}) becomes:
\begin{equation}\label{EpTmrnahl}
T=\int_{q_2^{p}(E_p)}^{q_3^{p}(E_p)}\frac{2A(1+F)+b\left(2A-B+\sqrt{B^2-4AC}\right)}{2A\sqrt{B^2-4AC}}dq,
\end{equation}
where we have swapped the integration limits such that the integrand is positive.
Finally, the action is given by Eq.~(\ref{actiongeneral}) with $E_p$ from Eq.~(\ref{EpTmrnahl}), while
${\cal S}_0^{hl}=\int_{q_3}^{q_2}p_0(q)dq$ is the unperturbed action from the \textit{high} to \textit{low} states.

\begin{figure}[t!]
\begin{center}
\includegraphics{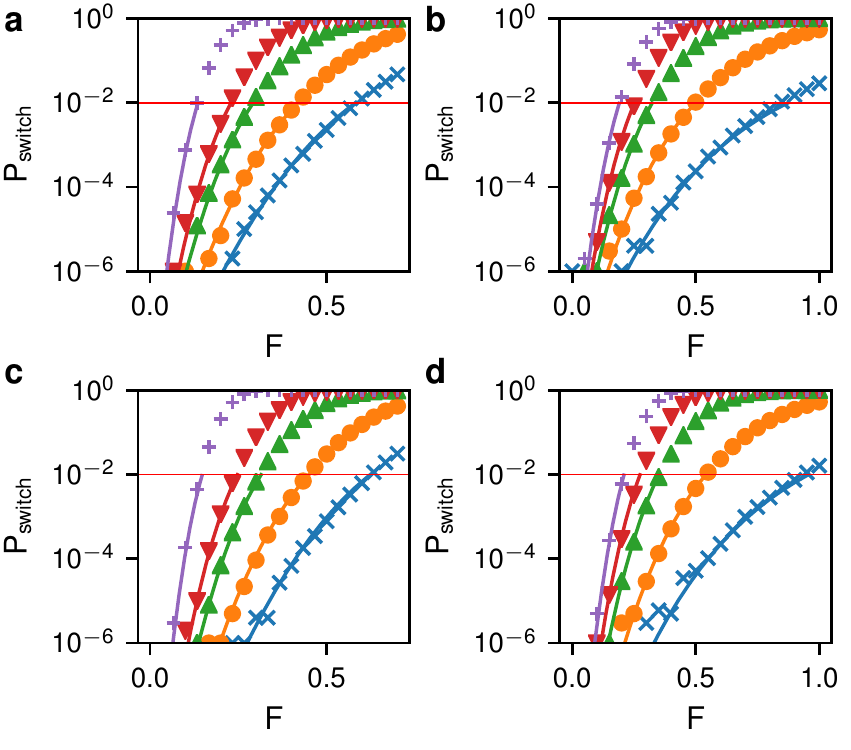}
\end{center}
\vspace{-0.5cm}
\caption{{\bf Perturbation effect on the mRNA-protein model.} (a) Change in switching probability vs perturbation strength $F$ for the \textit{low} to \textit{high} switch with $b=4$. Symbols and lines give numerical and theoretical values, respectively. Theoretical values were calculated using the maximum likelihood parameter estimates. Data are shown for five values of the perturbation time $T$: 0.35 (blue $\times$), 0.5 (orange $\circ$), 0.75 (green $\bigtriangleup$), 1.0 (red $\bigtriangledown$), 2.0 (purple $+$). (b) Change in switching probability vs $F$ for the \textit{high} to \textit{low} switch with $T= $ 1.0 (blue $\times$), 1.5 (orange $\circ$), 2.25 (green $\bigtriangleup$), 3.0 (red $\bigtriangledown$), 4.5 (purple $+$). (c+d) As above except for $b=3$.}
\label{fig:MRNASimTheoryComp}
\end{figure}

\begin{figure}[t!]
\begin{center}
\includegraphics{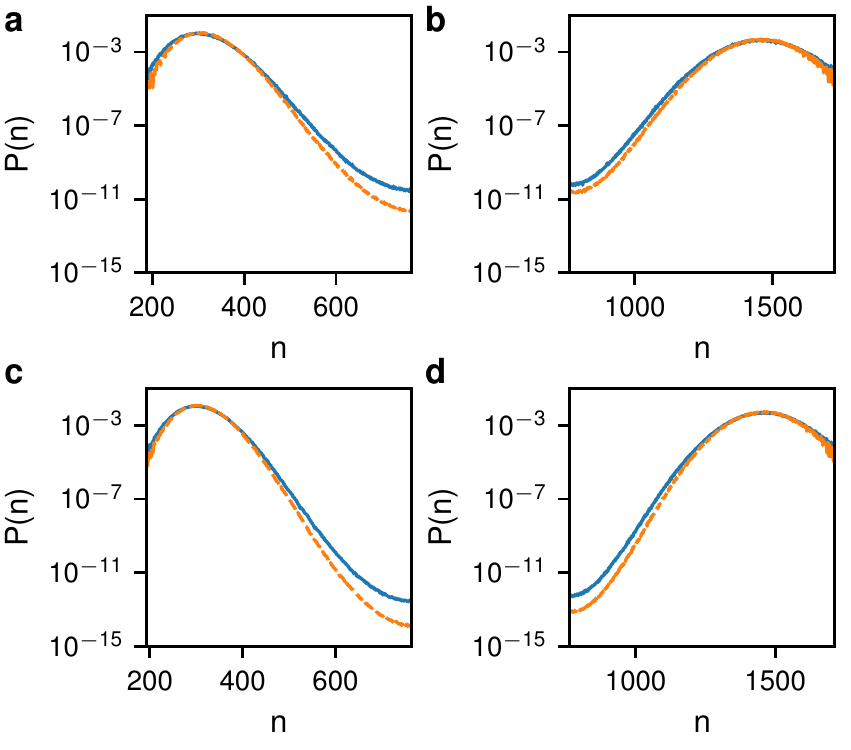}
\end{center}
\vspace{-0.5cm}
\caption{{\bf Comparison of actual and inferred probability distributions for the mRNA-protein model.} (a) The actual (solid blue) and inferred (dashed orange) PDFs for the \textit{low} state with $b=4$. (b) The same for the \textit{high} state. (c+d) The same as (a+b) except for $b=3$.}
\label{fig:MRNAFitPDF}
\end{figure}

To test the mRNA-protein model, we again ran sets of Monte Carlo simulations to calculate the dependence of ${\cal P}^{lh}$ and ${\cal P}^{hl}$ on $F$ for five values of $T$ with $b=4$. We then performed MLE estimation from these data, see Fig.~S18+S23. Fig.~\ref{fig:MRNASimTheoryComp}a+b shows an excellent agreement between the simulations and theory with the MLE parameters. Likewise, the reconstructed PDFs shown in Fig.~\ref{fig:MRNAFitPDF}a+b are in good agreement with the actual PDFs.

Finally, we wanted to study the impact of increasing the barrier height while maintaining the position of the fixed points. To this end, we varied $b$ in a range of $1$--$5$ (Fig.~S17-S28). For parameter sets with longer switching times the low $F$ region is not well sampled (Fig.~\ref{fig:MRNASimTheoryComp}c+d), which leads to an increased error in the predicted switching barrier (Fig.~\ref{fig:MRNAFitPDF}c+d). With $b=1$ and a switching time of $1 \times 10^{20}$ the relative error in the barrier height is $\sim$20\%. As with the SRG, these errors could be reduced by including lower probability events in the MLE.

\subsection{Discussion}

\subsubsection{Generic models}
We would now like to apply our methodology to arbitrarily complex networks, not only the simple models discussed above. A good example in this realm is the genetic toggle switch where two proteins negatively regulate each other using additional transcription factors~\cite{Allen2005srs,lipshtat2006genetic,Biancalani2015gts}. While a generalization to higher-dimensional systems is highly nontrivial, in what follows we will attempt to lay the theoretical grounds for such a generalization.

Let us consider a gene-regulatory network with $M$ species, $\mathbf{n}=(n_A,n_B,\dots,n_M)$, describing \textit{e.g.}, $M$ different proteins that regulate each other, where $n_A,n_B,\dots,n_M$ denote the copy numbers of the various proteins $A$, $B$, $\dots$, $M$. It is our aim to find an effective landscape for the protein of interest, say $A$. In general, the production of $A$ is regulated by all other proteins including $A$, while the degradation takes the usual form:
\begin{equation}
n_A\xrightarrow{\phi(n_A,n_B,\dots,n_M)}n_A+1,\quad\quad n_A\xrightarrow{n_A}n_A-1,
\end{equation}
where $\phi(n_A,n_B,\dots,n_M)$ is a function of all proteins or transcription factors in the circuit including $A$. We seek to infer an effective birth or production rate $\tilde{\phi}(n_A)$, which is a 1D projection of the M-dimensional production rate $\phi(n_A,n_B,\dots,n_M)$. This will allow to effectively describe the dynamics of protein $A$, and to infer the same marginal PDF of $A$ obtained in the M-dimensional case. Moreover, this will allow one to find the quasi-potential landscape of $A$ and the relative stability of its phenotypes.

Previously, when the regulatory network was known, we have used the master equation to account for demographic noise. Here, since we have an effective birth-death process for $A$, and the relation between the drift and diffusion is a-priori unknown, we will instead describe the stochastic dynamics of $A$ by a Langevin equation:
\begin{equation} \label{LangevinEqA}
\dot{q}=f(q)+\sqrt{D(q)/N}\eta(t).
\end{equation}
Here, $q=n_A/N$ is the density of $A$, and $N$ is its typical copy number in the \textit{high} state. Moreover, $f(q)=\tilde{\phi}(n_A)/N-q\,$ and $D(q)$ are the effective drift and diffusion functions, see below, while $\eta(t)$ is a delta-correlated normal
random variable with mean $0$ and variance $1/dt$.

In Eq.~(\ref{LangevinEqA}) both $f(q)$ and $D(q)$ are unknown and need to be found from perturbation experiments as done above. As we are interested in bistable systems, $f(q)$ has to be (at least) a cubic polynomial, to give rise to three fixed points at the deterministic level. We will assume for concreteness that the production rate is given by a Hill function such that: $f(q)=\alpha_0+(1-\alpha_0)q^h/(q^h+\beta^h)-q$. Naturally, other functional forms for $f(q)$ are possible as long as they yield three fixed points.

Choosing a diffusion function is more intricate. In general, when a master equation is approximated by the van-Kampen system size expansion, one finds that the diffusion function $D(q)$ entering the resulting Fokker-Planck (or Langevin) equation, satisfies $D(q)=\lambda(q)+\mu(q)$~\cite{Gardiner2004hsm,assaf2017wkb}, where $\lambda(q)$ and $\mu(q)$ are the birth and death rates, respectively. Since a bistable system is obtained when $f(q)=\lambda(q)-\mu(q)$ is (at least) a cubic polynomial, we argue that taking $D(q)$ as a cubic polynomial $D(q)=D_0+D_1q+D_2q^2+D_3q^3$ should suffice to describe the effective noise in generic systems. In simpler cases, see below, a lower-order polynomial may also suffice.

\begin{figure}[t!]
\begin{center}
\includegraphics[width=9.108cm]{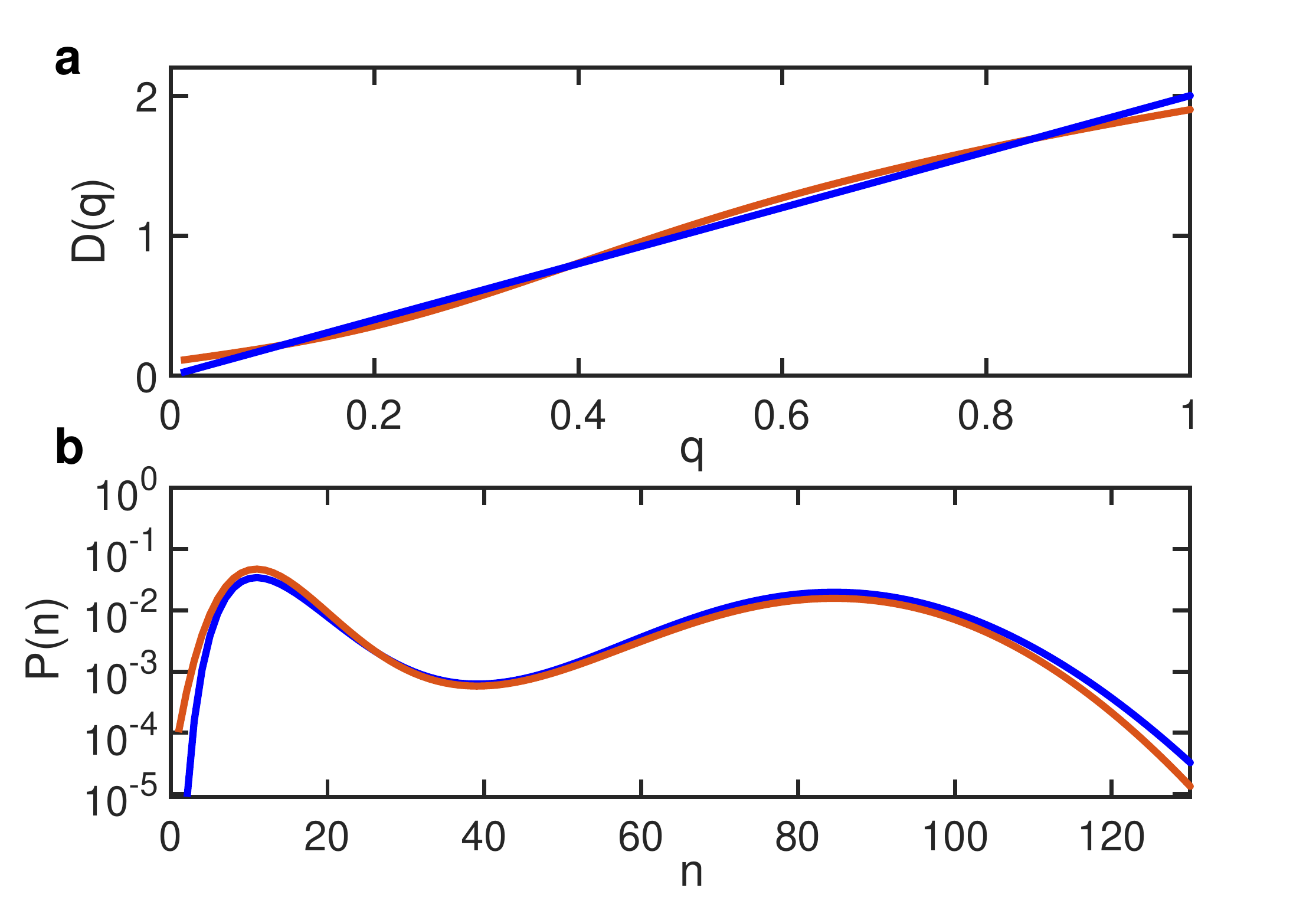}
\end{center}
\vspace{-0.5cm}
\caption{{\bf Comparison between the exact and approximated diffusion functions and probability distributions for the SRG model.} (a) The diffusion function obtained from a van-Kampen system size expansion (red line) as a function of $q=n/N$, compared with its linear fit (blue line). (b) The PDF given by Eq.~(\ref{PDFFP}) (red line) as a function of $n$ compared with the approximated PDF (blue line) computed with the linear diffusion function. Here, parameters were $\alpha_0=0.1$, $\beta=0.5$, $h=3$ and $N=100$, such that the fixed points satisfy $q_1\simeq 0.11$, $q_2\simeq 0.39$ and $q_3\simeq 0.85$.}
\label{fig:diffusioncoeff}
\end{figure}

For example, applying the van-Kampen system size expansion on the SRG model, master equation~(\ref{ME}) becomes
\begin{eqnarray}\label{FPE}
\hspace{-5.0mm}\frac{\partial P(q,t)}{\partial t}=-\frac{\partial}{\partial q}[f(q)P(q,t)]+\frac{1}{2N}\frac{\partial^2}{\partial q^2}[D(q)P(q,t)],
\end{eqnarray}
where $f(q)$ and $D(q)$ are defined above, and $\lambda(q)$ and $\mu(q)$ are defined in Eq.~(\ref{Hillrates}).
The zero-current, stationary solution of this Fokker-Planck equation reads~\cite{Gardiner2004hsm,assaf2017wkb}
\begin{equation}\label{PDFFP}
P(n)\simeq {\cal A}\exp\left[N\int_{q}\frac{2f(q)}{D(q)}dq\right],
\end{equation}
where to remind the reader, $n=Nq$ is the protein copy number, and ${\cal A}$ is a normalization constant, such that $\int_{0}^{\infty}P(n)dn=1$~\cite{footNote}.
Now, let us observe what happens when we replace $D(q)$ by a simple polynomial. In Fig.~\ref{fig:diffusioncoeff}a shown is a comparison between the diffusion function, $D(q)=\lambda(q)+\mu(q)$, and a linear fit, $D(q)=2q$. The fact that the curves agree well is not surprising, as in the vicinity of the fixed points, $\lambda(q)=\mu(q)=q$, and thus $D(q)\simeq 2q$. In Fig.~\ref{fig:diffusioncoeff}b we compare the PDF, given by Eq.~(\ref{PDFFP}), with the approximated PDF, obtained by substituting $D(q)=2q$ in Eq.~(\ref{PDFFP}). As one can see, even a linear approximation of $D(q)$ yields a decent agreement between the PDFs. As explained above, for generic systems we argue that a cubic polynomial should suffice in order to accurately capture the switching landscape.

Now, as done above, we propose to use perturbation experiments to infer the effective 1D drift and diffusion functions, using the MLE. To do so, we can either add a temporary perturbation of magnitude $F$ and duration $T$ to the production rate of $A$, or increase its degradation rate by a factor of $(1+F)$ as before. The problem is that if we apply a force in an experiment, involving all proteins $A$, $B$, $\dots$, $M$, it is not at all clear how this force is projected onto the 1D space we are interested in. To continue, we can denote by ${\cal F}$ the 1D projection of the force $F$ applied in the M-dimensional space, such that ${\cal F}=\psi(F)$, where $\psi$ is some unknown function. In simple cases, for which the projection of the switching trajectory from a M-dimensional to a 1D space does not include multiple crossings of the switching barrier, we expect the function $\psi$ to be monotone increasing with $F$ and unique. However, for generic systems this is not the case, and finding $\psi$ is expected to be more involved, requiring a significant theoretical and numerical effort.

Assuming for a moment that the effective 1D force is known, we can continue as in the simple cases discussed above. Using the WKB ansatz $\pi(q)\sim e^{-NS(q)}$ in the (stationary version of the) Fokker-Planck equation~(\ref{FPE}), to leading order in $N\gg 1$ one arrives at a stationary Hamilton-Jacobi equation with an unperturbed Hamiltonian:
\begin{equation}
H_0(q,p)=p\left[f(q)+pD(q)/2\right],
\end{equation}
where $p=dS/dq$ is the conjugate momentum. As a result, the unperturbed optimal path, $p_0(q)$, satisfies
$p_0(x)=-2f(q)/D(q)$, which allows finding the action barrier as before, both from the \textit{low} to \textit{high}, and from the \textit{high} to \textit{low} states, see text below Eq.~(\ref{p0}).

At this point, we can repeat the calculations done above in the perturbed case, by taking $\lambda(q)\to\lambda(q)+{\cal F}$, and $\mu(q)\to\mu(q)(1+{\cal F})$. This allows finding the perturbed action barrier, both from the \textit{low} to \textit{high}, and \textit{high} to \textit{low} states, which then allows one, using multiple switching experiments and the MLE, to extract the parameters defining $f(q)$ and $D(q)$. However, we have not yet determined how the effective 1D force ${\cal F}$ depends on the original pulling force $F$, and thus, applying this theory to realistic experiments remains far from being trivial. A possible way to study this functional dependence is to look at a 1D projection of deterministic trajectories of the full M-dimensional system upon applying a constant force for a finite duration. Here, understanding, \textit{e.g.}, the influence of the perturbation on the relaxation dynamics near a fixed point, or other dynamical properties, may allow one to get insight on how ${\cal F}$ depends on $F$. Yet, we leave this task to a future publication.

\subsubsection{Dependence of switching probability on impulse}
In physical terms $FT$ represents the total \textit{impulse} we apply to the system, which is equal to the force exerted on a particle multiplied by the duration of the force. In a mechanical system, when a constant force $F$ is applied on a particle for a duration $T$ in the direction of the particle's momentum, in the absence of dissipation or heat production, the particle's momentum is increased by $FT$. This increase is independent on $F$ or $T$ separately; that is, applying a small force for a long duration is equivalent to applying a large force for a short duration.

This relationship is exactly what we observe for small impulses in our system. Fig.~\ref{fig:SRGBifurcation} shows the change in the switching barrier \textit{e.g.} between the \textit{low} and \textit{high} states, $\Delta{\cal S}\equiv {\cal S}^{lh}-{\cal S}^{lh}_0$ as a function of total impulse $FT$. One can see that for low $FT$ the change in the switching barrier depends linearly on the product $FT$ and not on $F$ or $T$ separately. However, as the impulse increases the change in switching barrier is no longer a unique function of $FT$. As the impulse duration $T$ is increased (high $FT$ with low $F$) not all of the impulse results in a reduction in the switching barrier. This discrepancy indicates that there is some sort of dissipation or uncontrolled heat/entropy production in the system.

In contrast, when the system is near bifurcation we do expect the change in action to be a unique function of $FT$. In Appendix A we derive a simple analytical expression for the dependence of the switching barrier on $FT$ close to the bifurcation limit. As can be seen in Fig.~S29, the effect of the perturbation depends only on the product $FT$ in this case.

\begin{figure}[t!]
\begin{center}
\includegraphics{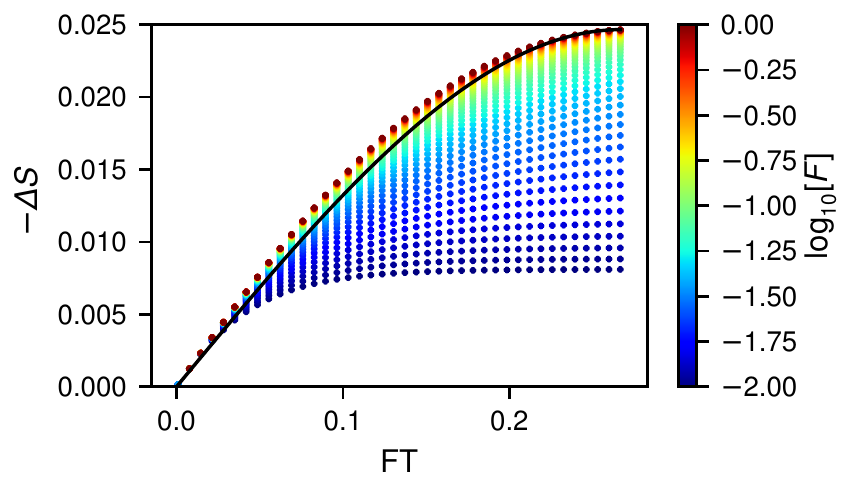}
\end{center}
\vspace{-0.5cm}
\caption{{\bf Dependence of change in the switching barrier on impulse.} Distribution of the perturbed switching barrier $\Delta{}S = {\cal S}^{lh}-{\cal S}^{lh}_{0}$ versus the total applied impulse $F\,T$ for the SRG model. Each symbol represents a perturbation with a different $F$, which is given by the color. The solid line shows the bifurcation theory given in Eq.~(\ref{actionbifur}).}
\label{fig:SRGBifurcation}
\end{figure}

Even though the switching barrier is a unique function of $FT$ close to bifurcation, see Eq.~(\ref{actionbifur}), we can see that the change in switching barrier is linear with $FT$ only at low impulse. Defining the \textit{efficiency} of inducing a switch by the change in the switching barrier divided by the impulse, $\Delta{\cal S}/(FT)$, from Eq.~(\ref{actionbifur}) we see that the efficiency decreases with $FT$. That is, the process of inducing a switch becomes less efficient as $FT$ is increased, while efficiency is maximized for weak impulses with vanishingly-small dissipation.

\subsubsection{Role of perturbation energy $E_p$}
What is the physical meaning of the perturbation energy $E_p$ which appears throughout our derivation? Mathematically, it is determined by a complicated function of the force $F$ and its duration $T$. However, looking at the result close to bifurcation (see Appendix A), the energy $E_p$ can be written as $E_p=E_0[1-(FT)^2/4]$, where $E_0$ is the maximal value of $E_p$ which is obtained as $F$ and/or $T$ vanish. Plugging this result into Eq.~(\ref{actionbifur}), and substituting $FT=2\sqrt{1-E_p/E_0}$, we find that for small impulses $\Delta{\cal S}/{\cal S}_0\sim \sqrt{E_0-E_p}$. This indicates that, in analogy to quantum mechanics, given a quasi-potential landscape $S(q)$, $E_0-E_p$ can be viewed as the ``energy excess" the particle receives to cross the switching barrier of height $E_0$. As a result, for $E_p=E_0$, the switching barrier remains unchanged (corresponding to $F=0$ and/or $T=0$), while for $E_p=0$, the energy excess is maximal corresponding to the absence of a switching barrier, leading to instantaneous switching.

\subsection{Conclusion}
Here, we have introduced a new theory for describing the effect of nonequilibrium perturbations on biological regulatory networks with metastable states, {\it i.e.}, epigenetic networks. Our theory can be used to infer the epigenetic landscape of a regulatory network by fitting a model using a series of perturbations of varying strength. The shape of the landscape is mapped out and reconstructed from the perturbation responses. The data needed for the fitting are purposely chosen to be reasonable biological observables.

The principle of such an experiment would be to apply a genetic or biochemical perturbation to the network, such as by introducing an inducible gene using transfection and/or silencing expression using siRNA. A response is measured as a function of the strength and duration of the perturbation. Unlike other theories that relate switching dynamics to fluctuations along the epigenetic landscape, our theory does not require detailed time-lapse imaging to collect data. One simply needs to record the fraction of cells that switch phenotypes at some time in the future after the perturbation. Such data can be quickly collected for millions of cells using flow cytometry. Using the response data the parameters of a regulatory model can be inferred and the epigenetic landscape numerically reconstructed. Our method, therefore, has great potential to be used to help decipher complex biological developmental trajectories.

Our theory also provides insights into the fundamental physics of how various signals induce state transitions in cells. As more complex cellular reprogramming is undertaken, it will become increasingly important to model how cells can be induced to make transitions between states. The impulse that we identified in our theory is one way to measure the work required to change phenotypic states. Further theoretical advances will be required to extend our understanding of important developmental techniques such as creation of induced pluripotent stem cells and cell reprogramming and differentiation.

\subsection*{Acknowledgments}
The authors would like to thank Chris Bohrer, Tom Israeli and Yen Ting Lin for useful discussions. ER acknowledges support from the National Science Foundation under grant PHY-1707961. MA and SB acknowledge support from the Israel Science Foundation grant No. 300/14 and the United States-Israel Binational Science Foundation grant No. 2016-655.

\subsection*{Appendix A: Bifurcation Limit}
In this section we show how the results derived for the SRG can be drastically simplified close to the bifurcation limit, where the stable and unstable fixed points merge. Without loss of generality, our analysis below will focus on the \textit{low} to \textit{high} switch, namely from $q_1$ to $q_3$, where the analysis of the \textit{high} to \textit{low} switch is identical.

Let us denote by $\epsilon\equiv (q_2-q_1)/2\ll 1$, such that $2\epsilon$ is the distance between the two fixed points. Let us also denote by $q_m=(q_1+q_2)/2$ the mid point between the stable and unstable fixed points. As a result, we can write $q_1=q_m-\epsilon$ and $q_2=q_m+\epsilon$.  It has been shown in previous works that in problems of switching between metastable states, close to bifurcation the momentum $p$ scales as $\epsilon^2$~\cite{Escudero2009srm,Assaf2010ems}. As a result, it is convenient to rescale the coordinate and momentum as follows:
\begin{equation}
\tilde{q} = (q-q_m)/\epsilon,\;\;\;\;\;\tilde{p} = p/\epsilon^2,
\end{equation}
where $\tilde{q},\tilde{p}$ are ${\cal O}(1)$. Note, that the fixed points in the rescaled coordinate become $\tilde{q}_1=-1$ and $\tilde{q}_2=1$.

We further denote $R(q)=\lambda(q)-\mu(q)$, such that in the absence of external forcing the mean-field rate equation becomes $\dot{q}=R(q)$. Since $R(q)$ can be approximated by a parabola in the regime $q_1<q<q_2$, we have: $R(q)\simeq ({\cal R}/2)(q-q_1)(q-q_2) \simeq ({\cal R}/2)\epsilon^2(\tilde{q}^2-1)$, where ${\cal R}\equiv R''(q_m)$ is a positive constant. Therefore, at the midpoint $R(q=q_m)=-({\cal R}/2)\epsilon^2$ is negative, and $R'(q_m)=0$, since the parabola has a minimum at $q=q_m$. Using these results, and denoting by ${\cal D}\equiv\lambda(q_m)+\mu(q_m)$, we now expand the time-dependent Hamiltonian [Eq.~(\ref{hamilpertlh}) with $F$ replaced by $\phi(t)$] up to ${\cal O}(\epsilon^4)$ in the vicinity of $q=q_m$ and $p=0$. This results in
\begin{equation}
H(p,q,t)\simeq \tilde{p}\phi(t)\epsilon^2
+\tilde{p}\left\{\frac{\tilde{p}}{2}\left[{\cal D}+\phi(t)\right] +\frac{{\cal R}}{2}\left(\tilde{q}^2-1\right)\right\}\epsilon^4,
\end{equation}
where we have expanded $e^{\pm p} \simeq 1 \pm \epsilon^2\tilde{p} + (\epsilon^4/2)\tilde{p}^2$, and have also expanded $\lambda(q)$ and $\mu(q)$ around $q=q_m$, up to second order in $\varepsilon$.

In the absence of external force, $\phi(t)=0$, the unperturbed optimal path satisfies $H=0$, which yields the unperturbed trajectory
\begin{equation} \label{p0bifur}
\tilde{p}_0(q) = {\cal R}(1-\tilde{q}^2)/{\cal D}.
\end{equation}
In the presence of an external force, $\phi(t)=F$, the perturbed Hamiltonian becomes in the leading order
\begin{equation}\label{Hpbifur}
H_p=\tilde{p}F\epsilon^2,
\end{equation}
independent of $\tilde{q}$. Thus, equating $H_p=E_p$ yields the perturbed optimal path, which becomes constant here
\begin{equation}\label{ppbifur}
\tilde{p}(q)=E_p/(F\epsilon^2)=\tilde{E}_p/F,
\end{equation}
where we have defined the rescaled energy $\tilde{E}_p=E_p/\epsilon^2$.
Equating the unperturbed and perturbed optimal paths, Eqs.~(\ref{p0bifur}) and (\ref{ppbifur}), we find the intersection points $\tilde{q}^p_{1,2}$ to be $\tilde{q}^p_1=-\tilde{q}^p_2=-[1-\tilde{E}_p {\cal D}/(F{\cal R})]^{1/2}$.

Let us now find the rescaled energy given the duration of the external perturbation $T$. Using Eqs.~(\ref{Ep}) and (\ref{Hpbifur}), and the fact that $\dot{q}=\partial H_p/\partial p=F$, we have
\begin{equation}
T = \frac{1}{F}\int_{q_1^{p}(E_p)}^{q_2^{p}(E_p)}dq = \frac{2\epsilon}{F}\tilde{q}^p_2,
\end{equation}
from which we can extract $E_p$ as a function of $T$:
\begin{equation} \label{Epbifur}
\tilde{E}_p=(F{\cal R})/{\cal D}\left[1-\left(F\tilde{T}/2\right)^2\right],
\end{equation}
where $\tilde{T}=T/\epsilon$. Note, that the result is valid as long as $\tilde{T}\leq 2/F$, which means that $T\leq 2\epsilon/F$. The is because when the system is close to bifurcation, a very small force, $F\sim \epsilon$ is sufficient to cause a deterministic switch. Therefore, if $F\sim \epsilon$, we have $T={\cal O}(1)$. Also note that, by using Eq.~(\ref{Epbifur}), the intersection points become $\tilde{q}^p_{1,2}=\mp F\tilde{T}/2$; here, at the maximal value of $\tilde{T}=2/F$ we obtain $E_p=0$, since $\tilde{q}^p_{1,2}=q_{1,2}=\mp 1$ (\textit{i.e.}, the unperturbed and perturbed fixed points coincide).

Having found the perturbation energy, the correction to the switching barrier is given by Eq.~(\ref{actiongeneral}).
Transforming to the rescaled coordinate and momentum, using Eqs.~(\ref{p0bifur}), (\ref{ppbifur}) and (\ref{Epbifur}), and using the definition of $\tilde{T}$, we finally have
\begin{equation}\label{actionbifur}
{\cal S}^{lh}={\cal S}^{lh}_0\left\{1-\frac{3F\tilde{T}}{4}\left[1-\frac{(F\tilde{T})^2}{12}\right]\right\},
\end{equation}
where ${\cal S}^{lh}_0 = 4{\cal R}\epsilon^3/(3{\cal D})$, and the result is valid as long as $F\tilde{T}<2$. Fig.~S29 shows a comparison of Eq. (\ref{actionbifur}) and the full theory when the system is near bifurcation. Note, that as $\tilde{T}$ approaches $2/F$, action~(\ref{actionbifur}) approaches zero, which invalidates the WKB theory. The latter is valid as long as $N{\cal S}\gg 1$, which limits the duration and/or magnitude of the external force.

\subsection*{Appendix B: Weak noise limit}
In this section we derive the switching barrier under a weak external perturbation. Here, one must have a long perturbation duration; otherwise the effect is negligible. We will henceforth assume for simplicity that $E_p=0$, corresponding to a long perturbation duration, see below.

When $E_p=0$, the perturbed and unperturbed optimal paths for switching intersect at $q_1$ and $q_2$, such that $q^p_{1,2}=q_{1,2}$.
Therefore, putting $E_p=0$ and expanding in $F\ll 1$, the perturbed optimal path~(\ref{pint}) becomes
\begin{equation}
p_p(q)\simeq \ln\left[\mu(q)/\lambda(q)\right]-F/\lambda(q)=p_0(q)-F/\lambda(q).
\end{equation}
As a result, using Eq.~(\ref{actiongeneral}), and the fact that $E_p\simeq 0$, the correction to the switching barrier in the case of weak force, drastically simplifies and becomes
\begin{equation}
{\cal S}^{lh}={\cal S}^{lh}_0-F\int_{q_1}^{q_2}\frac{dq}{\lambda(q)}.
\end{equation}
Note, that in this case, the duration of the external perturbation is simply given by $T=\int_{q_1}^{q_2}dq/\dot{q}$, where $\dot{q}=\lambda(q)-\mu(q)$ is the unperturbed rate equation.

\subsection*{Appendix C: Heaviside Step Function}
In this section we consider the case of a translation rate with a very large Hill exponent in Eq.~(\ref{Hillrates}). In the limit $h\to\infty$ the translation rate becomes a step function
\begin{equation}
\lambda(q)=\alpha_0 + (1-\alpha_0)\Theta(q-\beta)
\end{equation}
where $\Theta(z)$ is a heaviside step function. In this case, the mean-field rate equation has three fixed points: $q_1=\alpha_0$, $q_2=\beta$ and $q_3=1$, where $q_1$ and $q_3$ are stable, while $q_2$ is unstable.

Let us begin by computing the correction to the switching barrier from the \textit{low} to \textit{high} states. Here the unperturbed switching barrier satisfies ${\cal S}^{lh}_0=\alpha_0-\beta+\beta\ln(\beta/\alpha_0)$~\cite{Roberts2015dsg}. Going along the same lines as above, we can compute the unperturbed and perturbed optimal paths for switching using Eqs.~(\ref{p0}) and (\ref{pint}), as well as the intersection points between these paths, which satisfy $q_1^p=(E_p+F)\alpha_0/F$ and $q_2^p=\beta$. In this case, one can explicitly find $E_p$ using the expression for $\dot{q}$ and Eq.~(\ref{EpT}). The result is
\begin{equation}
E_p(F,T)=\frac{F\beta}{e^T (F+\alpha_0)-F}-e^{-T} F.
\end{equation}
Using this result and Eq.~(\ref{actiongeneral}), the correction to the switching barrier from the \textit{low} to \textit{high} states is:
\begin{equation}
{\cal S}^{lh}\! =\!{\cal S}^{lh}_0+F(1-e^{-T})+\beta\left\{T\!-\!\ln\left[e^T(1\!+\!F/\alpha_0)\!-\!F/\alpha_0\right]\right\}\!.
\end{equation}

We now move to compute the correction to the switching barrier from the \textit{high} to \textit{low} states. Here the unperturbed switching barrier satisfies ${\cal S}^{hl}_0=1-\beta+\beta\ln\beta$. Going along the same lines as above, we can compute the unperturbed and perturbed optimal paths for switching using Eqs.~(\ref{p0}) and (\ref{pint}) as well as the intersection points between these paths, which satisfy $q_3^p=1-E_p/F$ and $q_2^p=\beta$. In this case, one can also explicitly find $E_p$ using the expression for $\dot{q}$ and Eq.~(\ref{Ep}) with the lower integration limit replaced by $q_3^p(E_p)$. The result is
\begin{equation}
E_p(F,T)=\frac{F\left[e^{-(F+1)T}F+1-(F+1)\beta\right]}{e^{(F+1)T}+F}.
\end{equation}
Using this result  and Eq. (\ref{actiongeneral}), the correction to the switching barrier from the \textit{high} to \textit{low} states is:
\begin{eqnarray}
\hspace{-3mm}{\cal S}^{hl} &=& {\cal S}^{lh}_0+(F+1)T\beta+\nonumber\\
&+&\frac{F}{F+1}[e^{-(F+1)T}-1]+\beta\ln\left[\frac{F+1}{e^{(F+1)T}+F}\right].
\end{eqnarray}

\onecolumngrid

\newpage

\section{\LARGE{Supplementary Material}}
\setcounter{figure}{0} \renewcommand{\thefigure}{S\arabic{figure}}

\begin{figure}[h!]
    \begin{center}
    \includegraphics{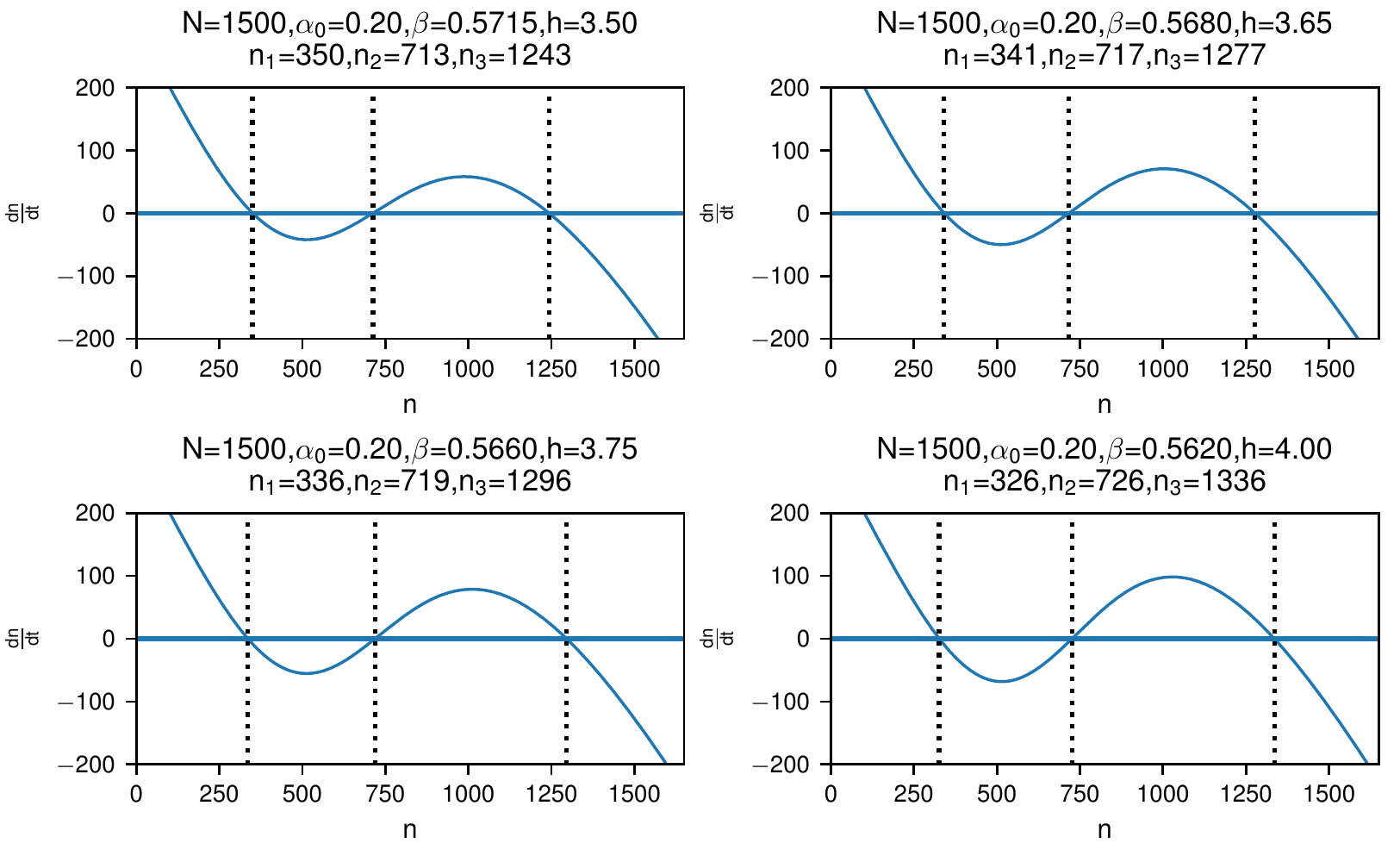}
    \end{center}
    \caption{{\bf Deterministic rate equations for the self-regulated gene model.} Value of $\frac{dn}{dt}$ as a function of $n$ for the deterministic model of the SRG. The positions of the three fixed points are given by the dotted lines. Parameters for each panel are as indicated.}
    \label{fig:SRGRateEquations}
\end{figure}

\newpage

\begin{figure}[h!]
    \begin{center}
    \includegraphics{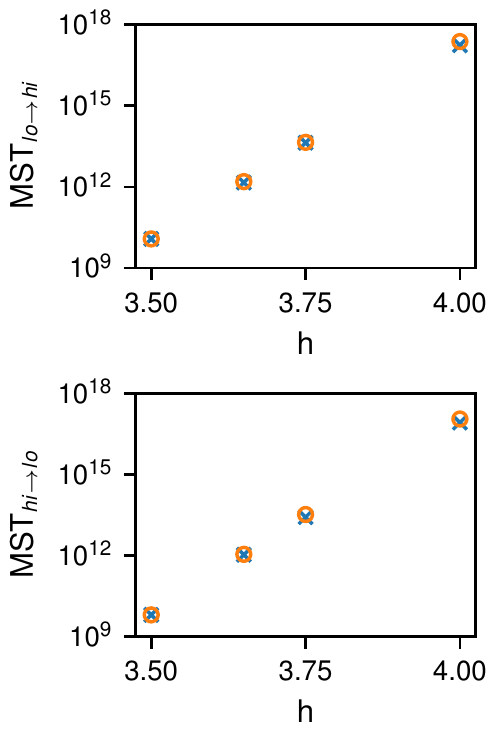}
    \end{center}
    \caption{{\bf Mean switching times for the self-regulated gene model.} (top) The MST to go from the $low$ state to the $high$ state vs $h$ calculated from numerical simulations (blue $\times$) and WKB theory (orange $\circ$) as given by Eq.~(7) in the main text. The WKB points are multiplied by a constant preexponential factor of 20.4. (bottom) The same for the $high$ to $low$ state with a preexponent of 11.94. All other parameters are as in Figure \ref{fig:SRGRateEquations}.}
    \label{fig:SRGSwitchingTimes}
\end{figure}
\newpage

\begin{figure}[h!]
    \begin{center}
    \includegraphics{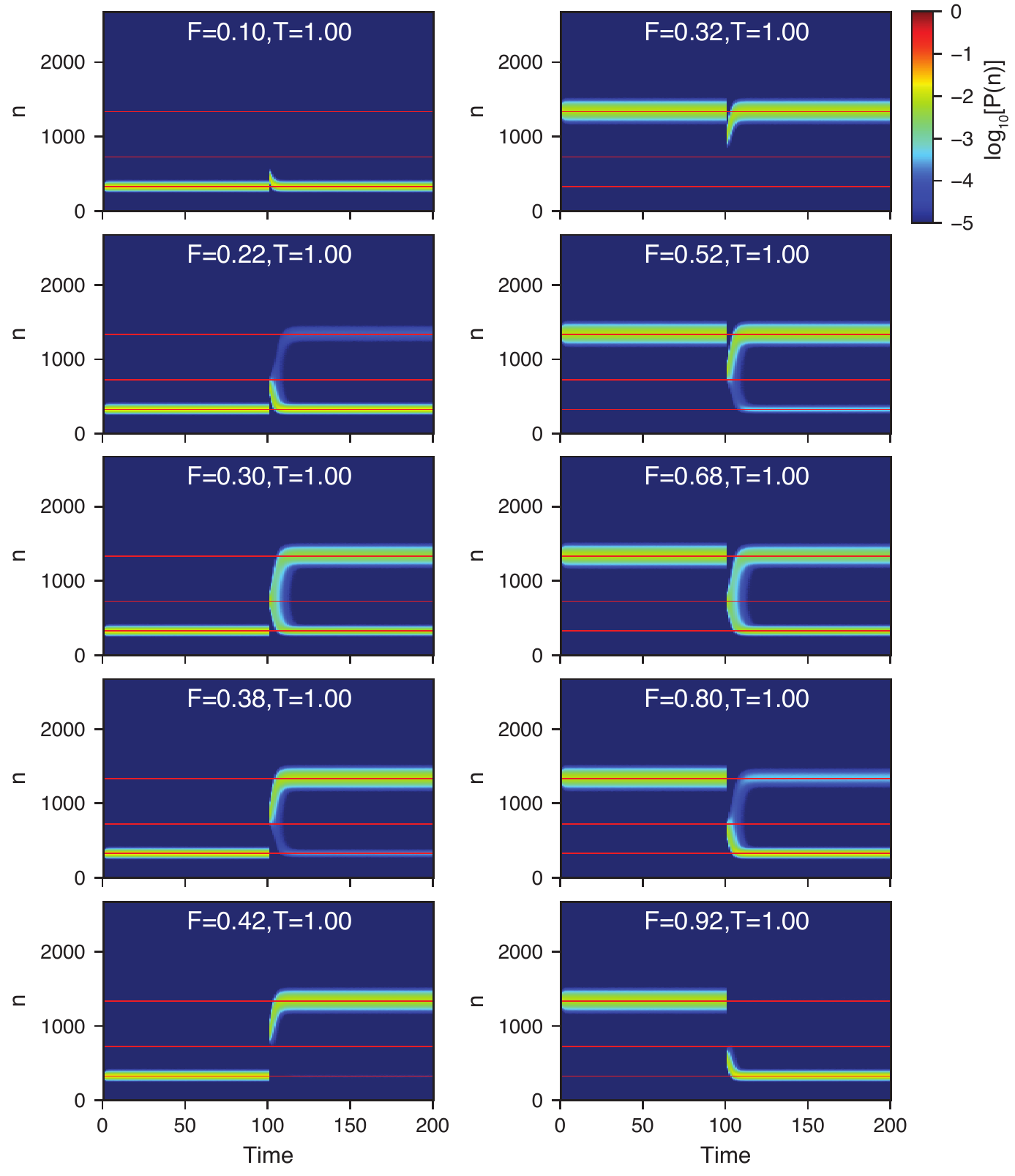}
    \end{center}
    \caption{{\bf Time dependent probability distribution for pulling on the self-regulated gene model.} The plots show the probability distribution versus time for pulling from $low \to high$ (left column) and $high \to low$ (right column). The simulations were initialized at the appropriate fixed point and allowed to equilibrate until t=100.0 before pulling was initiated. A pulling force of strength F was applied for the specified time T and then removed. The simulations then continued running until t=200.0 to relax and then the switching statistics were calculated. Red lines show the locations of the deterministic fixed points. Parameters for the SRG model were $N=1500$, $\alpha_0=0.2$, $\beta=0.562$, and $h=4.0$.}
    \label{fig:SRGTimeDependentPDF}
\end{figure}
\newpage

\begin{figure}[h!]
    \begin{center}
    \includegraphics{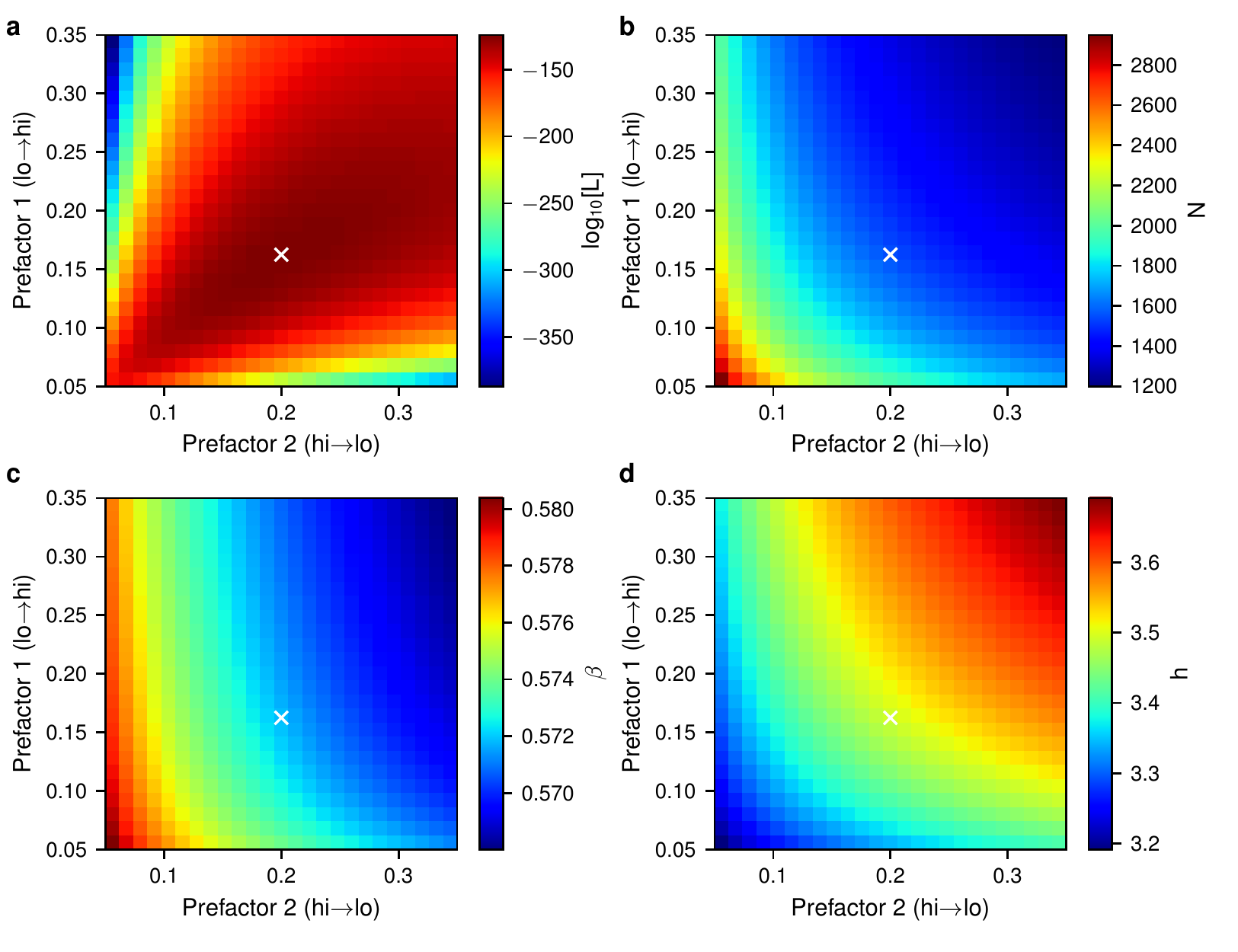}
    \end{center}
    \vspace{-0.2in}
    \caption{{\bf Prefactor dependence during maximum likelihood fitting of the SRG model with $h=3.5$.} (a) The maximum likelihood score obtained for each prefactor pair during fitting. Fitting was performed with $\alpha$ fixed to its true value, as described in the main text, using simulation data obtained from the SRG model with parameters $N=1500$, $\alpha_0=0.2$, $\beta=0.5715$, $h=3.5$. The prefactor pair with the highest likelihood score is marked with a white $\times$ in each panel. (b-d) The dependence of the maximum likelihood estimate for the parameters $N$, $\beta$, and $h$, respectively, on the prefactors.}
    \label{fig:SRGPrefactorOptimization2D-h35}
\end{figure}
\newpage

\begin{figure}[h!]
    \begin{center}
    \includegraphics{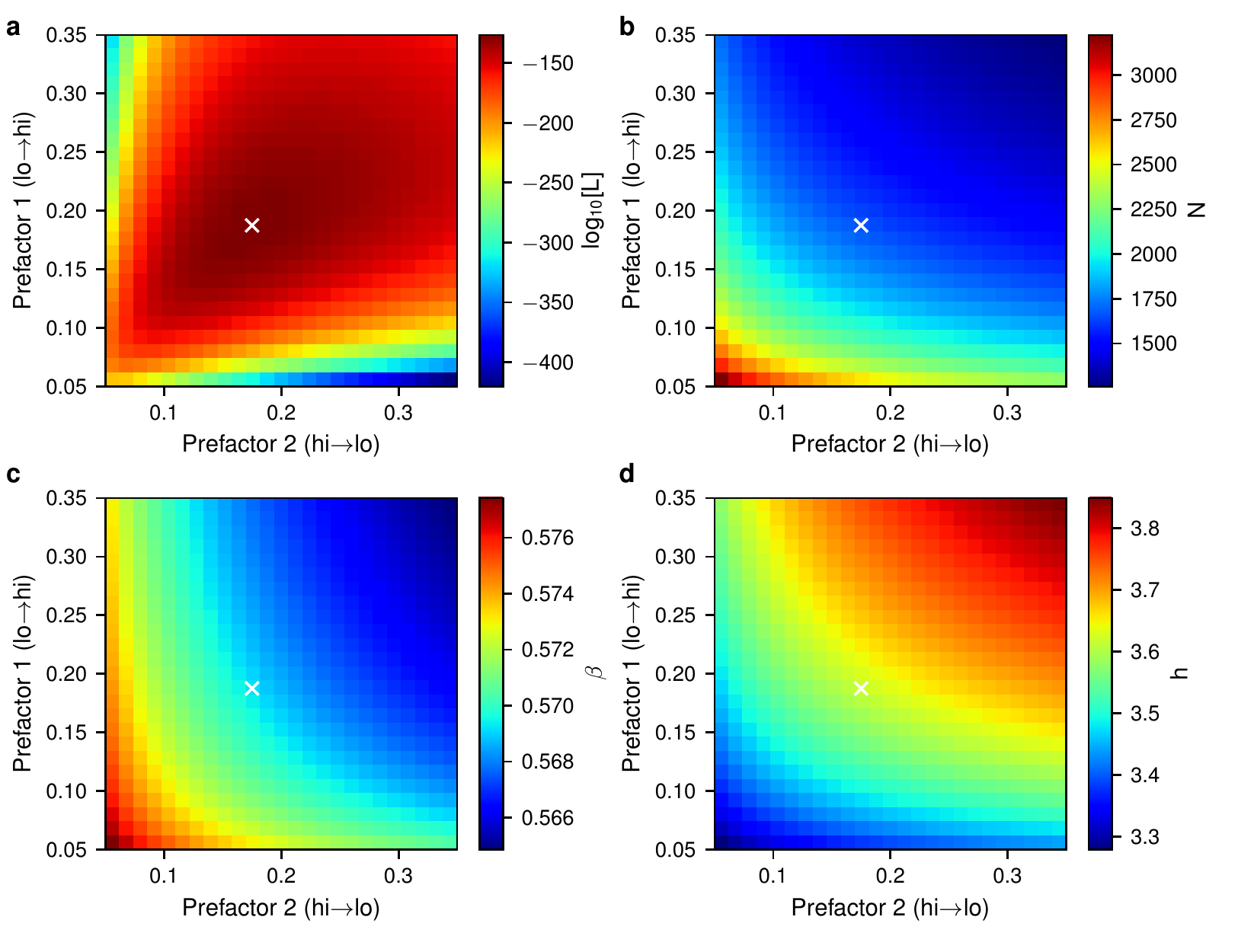}
    \end{center}
    \vspace{-0.2in}
    \caption{{\bf Prefactor dependence during maximum likelihood fitting of the SRG model with $h=3.65$.} (a) The maximum likelihood score obtained for each prefactor pair during fitting. Fitting was performed with $\alpha$ fixed to its true value, as described in the main text, using simulation data obtained from the SRG model with parameters $N=1500$, $\alpha_0=0.2$, $\beta=0.568$, $h=3.65$. The prefactor pair with the highest likelihood score is marked with a white $\times$ in each panel. (b-d) The dependence of the maximum likelihood estimate for the parameters $N$, $\beta$, and $h$, respectively, on the prefactors.}
    \label{fig:SRGPrefactorOptimization2D-h365}
\end{figure}
\newpage

\begin{figure}[h!]
    \begin{center}
    \includegraphics{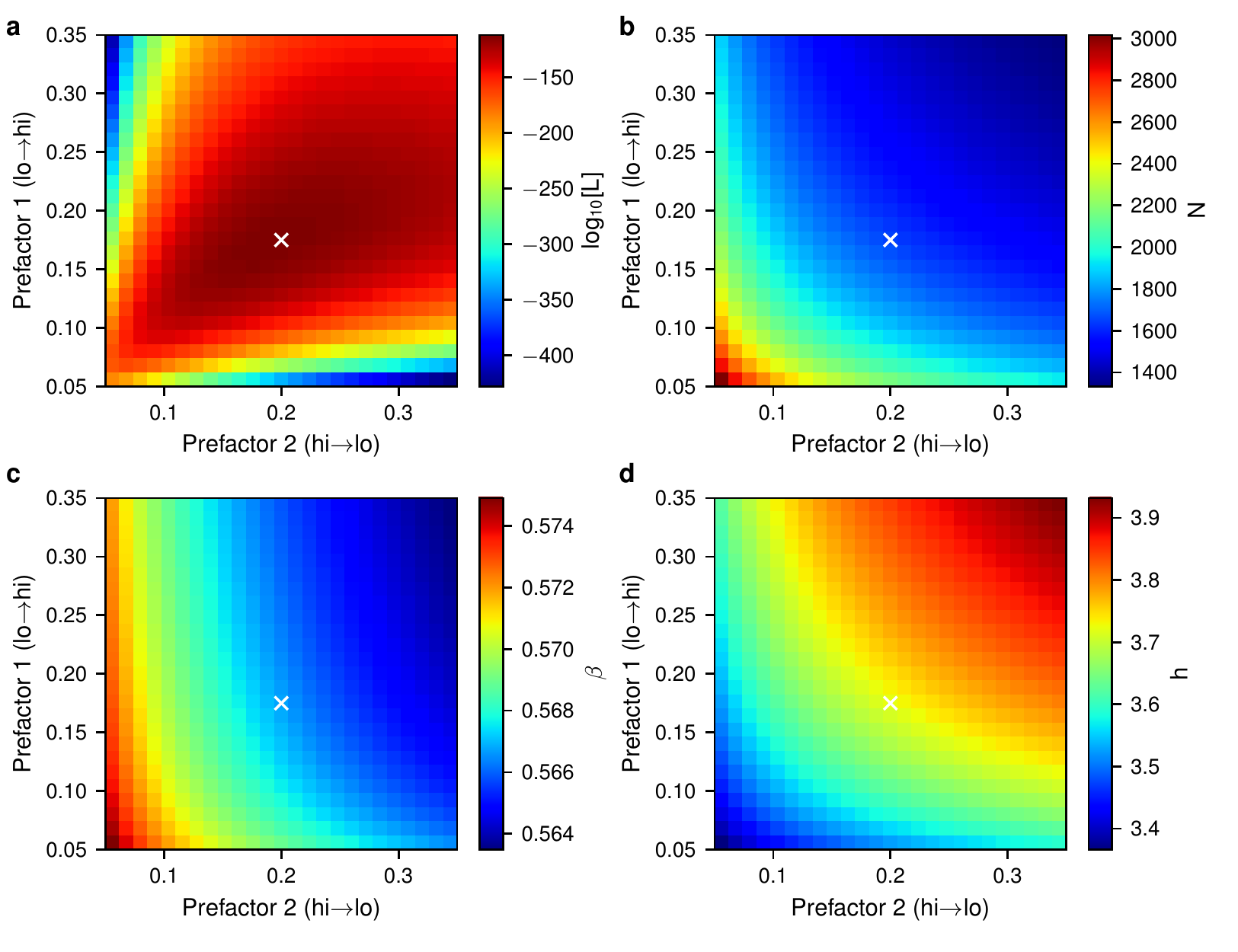}
    \end{center}
    \vspace{-0.2in}
    \caption{{\bf Prefactor dependence during maximum likelihood fitting of the SRG model with $h=3.75$.} (a) The maximum likelihood score obtained for each prefactor pair during fitting. Fitting was performed with $\alpha$ fixed to its true value, as described in the main text, using simulation data obtained from the SRG model with parameters $N=1500$, $\alpha_0=0.2$, $\beta=0.566$, $h=3.75$. The prefactor pair with the highest likelihood score is marked with a white $\times$ in each panel. (b-d) The dependence of the maximum likelihood estimate for the parameters $N$, $\beta$, and $h$, respectively, on the prefactors.}
    \label{fig:SRGPrefactorOptimization2D-h375}
\end{figure}
\newpage

\begin{figure}[h!]
    \begin{center}
    \includegraphics{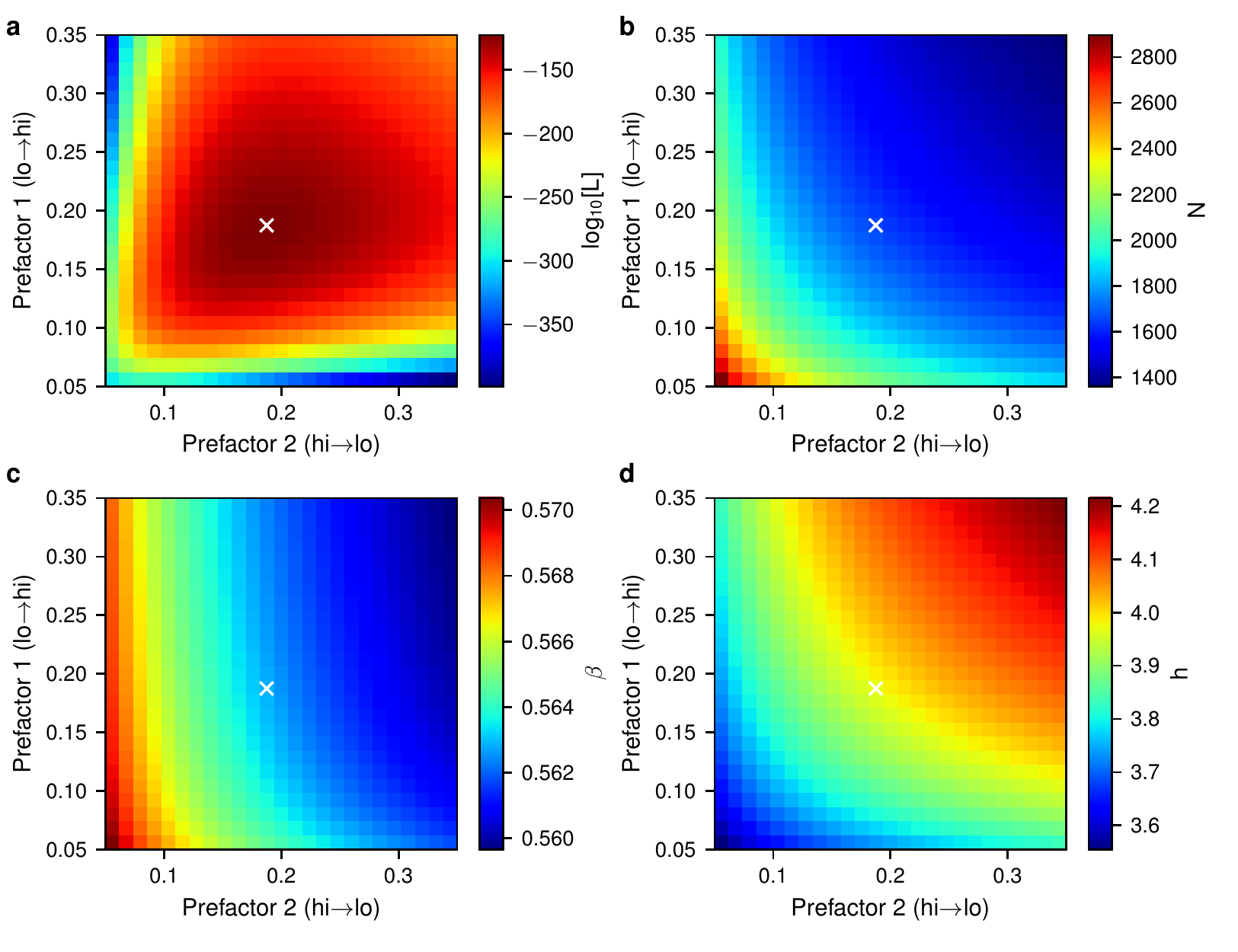}
    \end{center}
    \vspace{-0.2in}
    \caption{{\bf Prefactor dependence during maximum likelihood fitting of the SRG model with $h=4.0$.} (a) The maximum likelihood score obtained for each prefactor pair during fitting. Fitting was performed with $\alpha$ fixed to its true value, as described in the main text, using simulation data obtained from the SRG model with parameters $N=1500$, $\alpha_0=0.2$, $\beta=0.562$, $h=4.0$. The prefactor pair with the highest likelihood score is marked with a white $\times$ in each panel. (b-d) The dependence of the maximum likelihood estimate for the parameters $N$, $\beta$, and $h$, respectively, on the prefactors.}
    \label{fig:SRGPrefactorOptimization2D-h40}
\end{figure}
\newpage

\begin{figure}[h!]
    \begin{center}
    \includegraphics{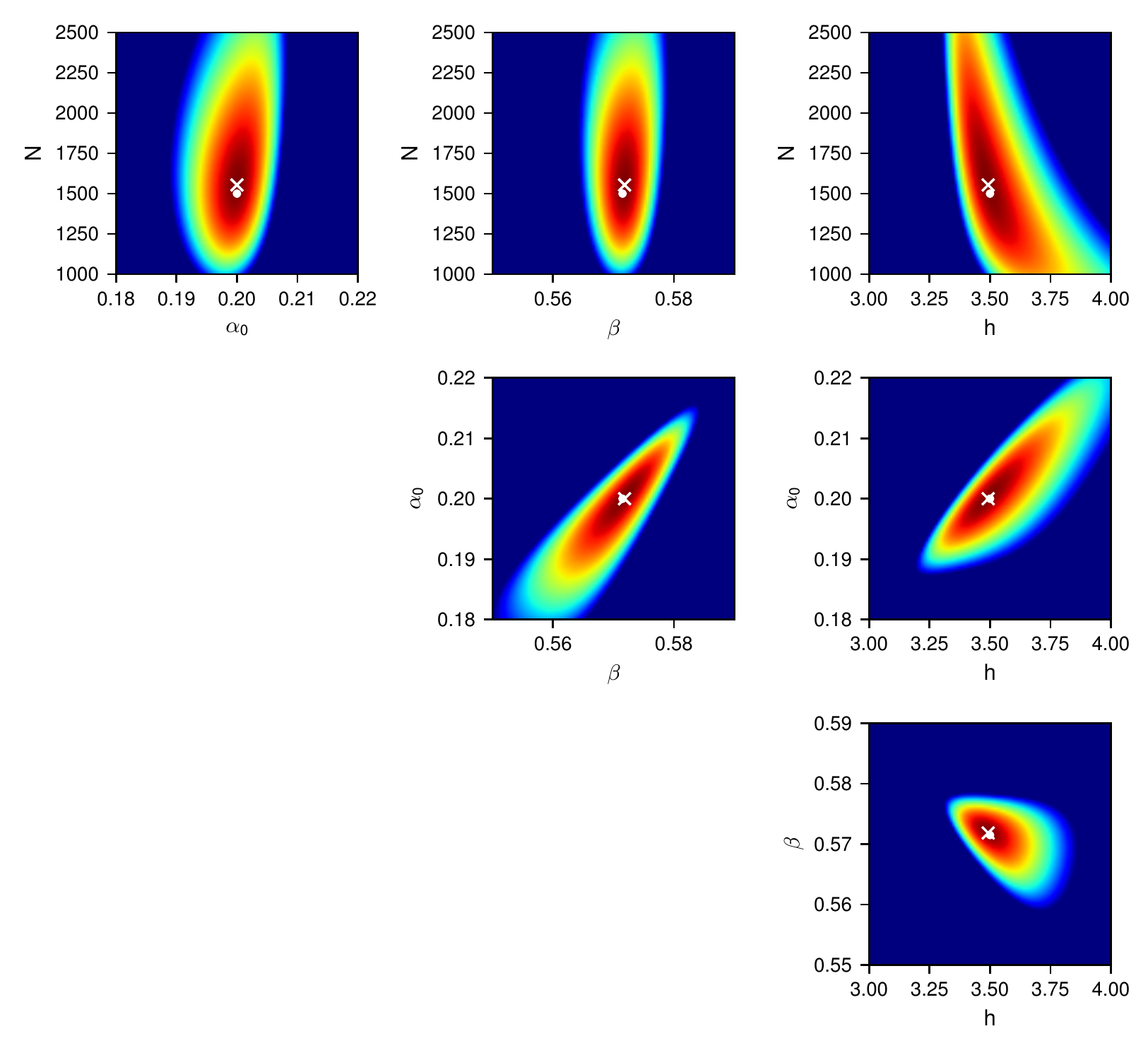}
    \end{center}
    \vspace{-0.2in}
    \caption{{\bf Likelihood distribution for the SRG model with $h=3.5$.} Likelihood distribution for inference of all pairs of parameters for the SRG model, using the optimized prefactors. For each plot, all other parameters are fixed to their MLE. The MLE is marked with a white $\times$ and the true parameter values are marked with a white $\bullet$. Colors show $\mathrm{log}_{10}[\mathrm{L}]$ and range from $-1\times 10^5$  (blue) to $0$ (red).}
    \label{fig:SRGLikelihood-h35}
\end{figure}
\newpage

\begin{figure}[h!]
    \begin{center}
    \includegraphics{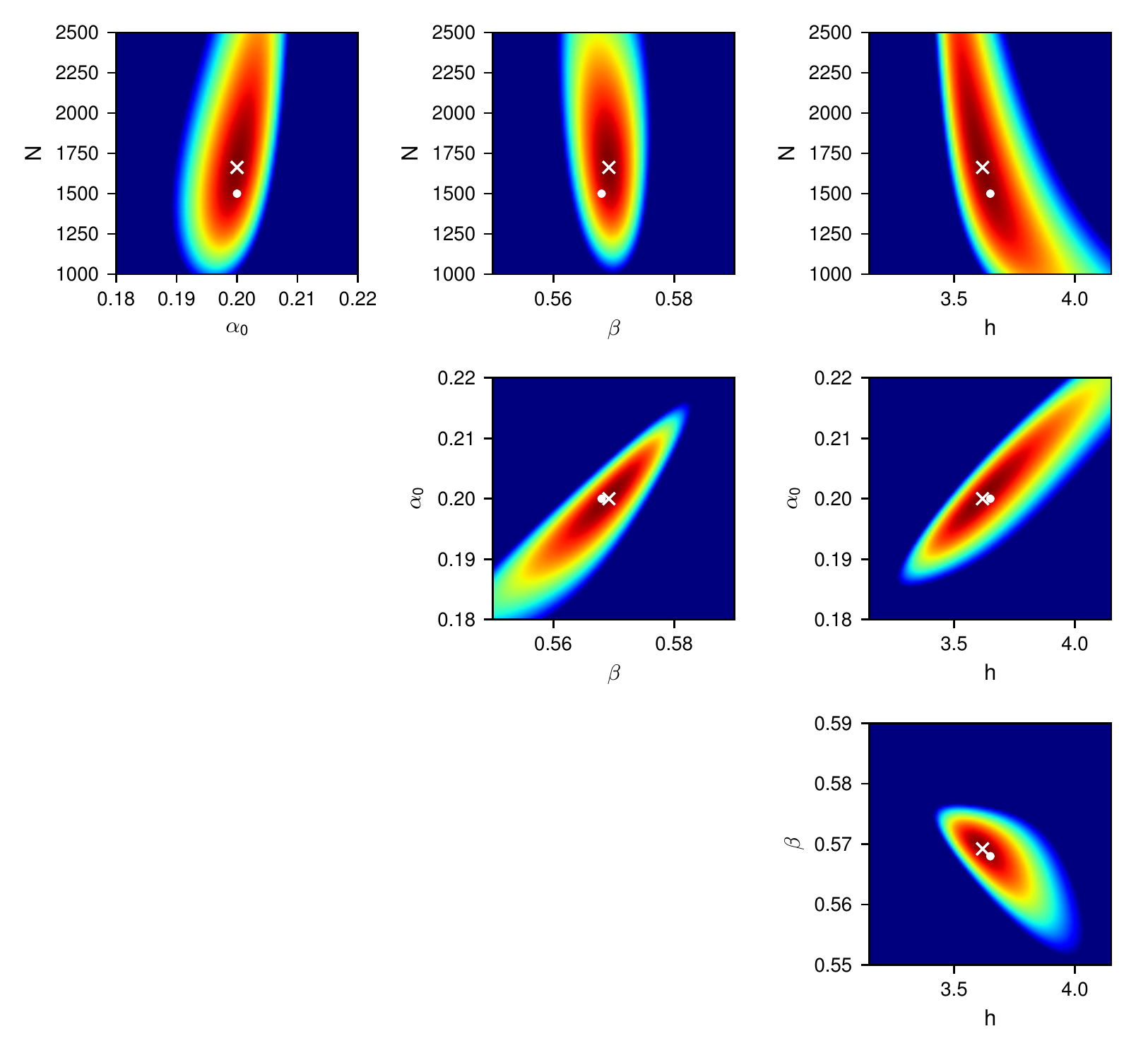}
    \end{center}
    \vspace{-0.2in}
    \caption{{\bf Likelihood distribution for the SRG model with $h=3.65$.} Likelihood distribution for inference of all pairs of parameters for the SRG model, using the optimized prefactors. For each plot, all other parameters are fixed to their MLE. The MLE is marked with a white $\times$ and the true parameter values are marked with a white $\bullet$. Colors show $\mathrm{log}_{10}[\mathrm{L}]$ and range from $-1\times 10^5$ (blue) to $0$ (red).}
    \label{fig:SRGLikelihood-h365}
\end{figure}
\newpage

\begin{figure}[h!]
    \begin{center}
    \includegraphics{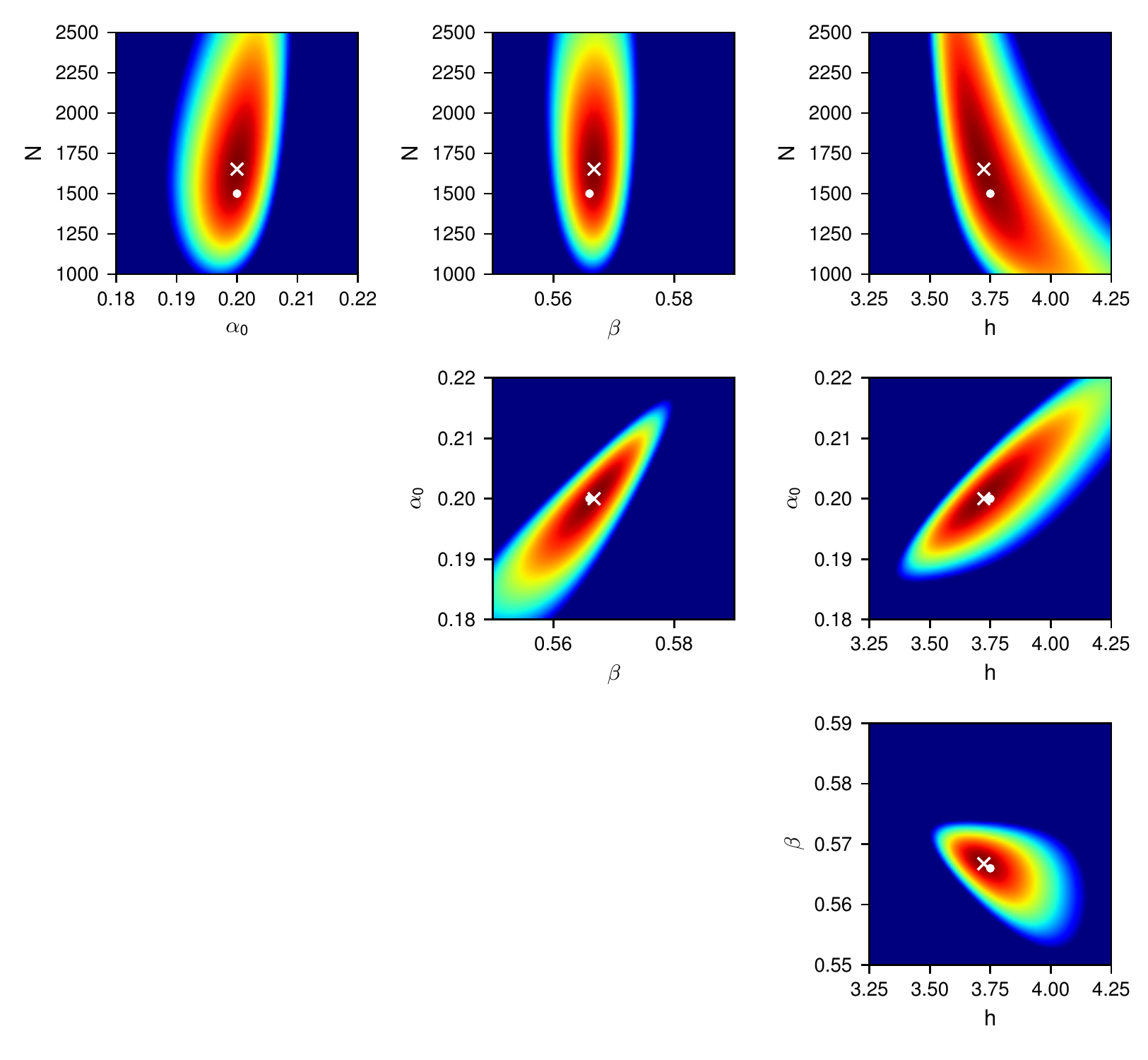}
    \end{center}
    \vspace{-0.2in}
    \caption{{\bf Likelihood distribution for the SRG model with $h=3.75$.} Likelihood distribution for inference of all pairs of parameters for the SRG model, using the optimized prefactors. For each plot, all other parameters are fixed to their MLE. The MLE is marked with a white $\times$ and the true parameter values are marked with a white $\bullet$. Colors show $\mathrm{log}_{10}[\mathrm{L}]$ and range from $-1\times 10^5$ (blue) to $0$ (red).}
    \label{fig:SRGLikelihood-h375}
\end{figure}
\newpage

\begin{figure}[h!]
    \begin{center}
    \includegraphics{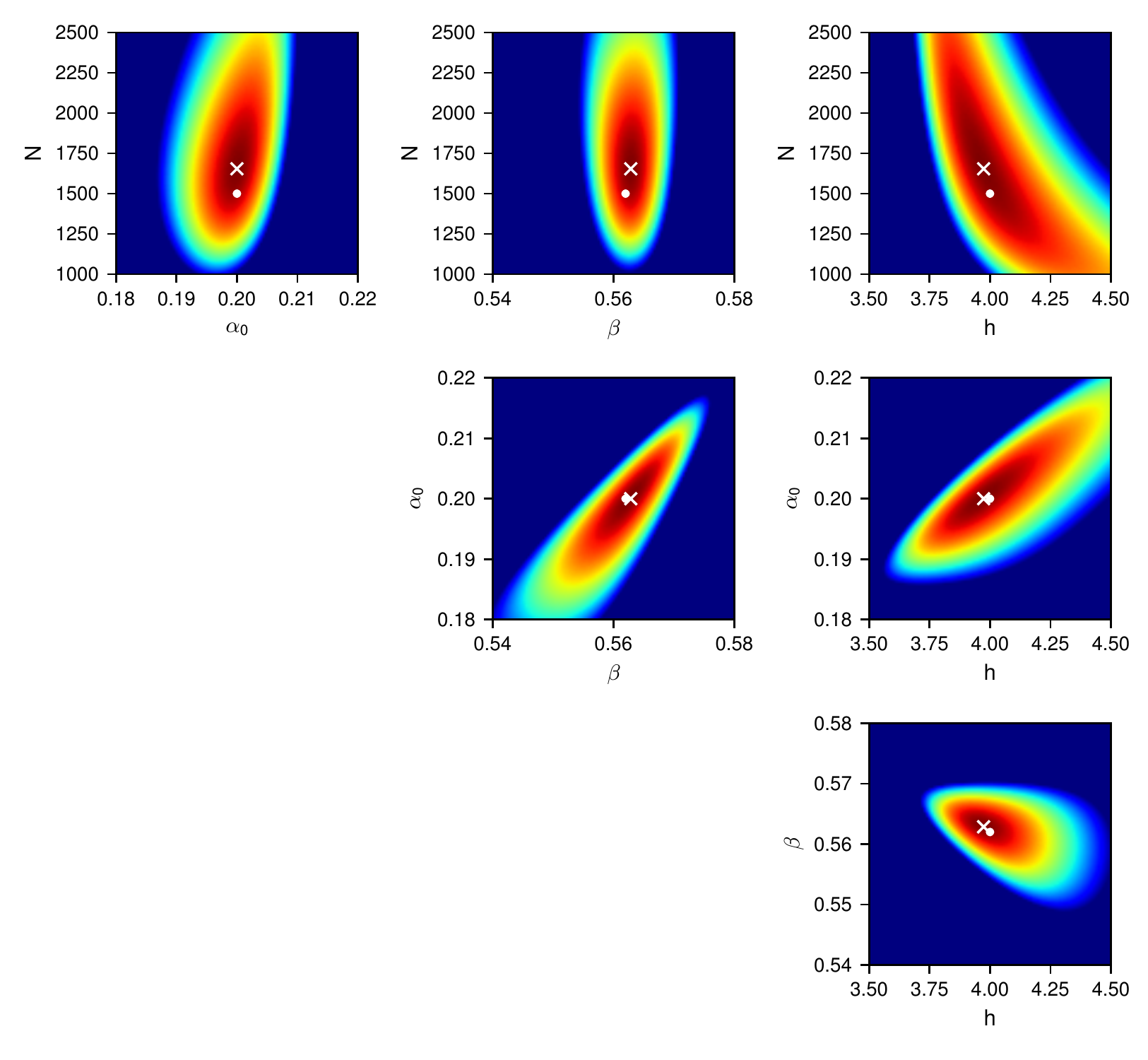}
    \end{center}
    \vspace{-0.2in}
    \caption{{\bf Likelihood distribution for the SRG model with $h=4.0$.} Likelihood distribution for inference of all pairs of parameters for the SRG model, using the optimized prefactors. For each plot, all other parameters are fixed to their MLE. The MLE is marked with a white $\times$ and the true parameter values are marked with a white $\bullet$.  Colors show $\mathrm{log}_{10}[\mathrm{L}]$ and range from $-1\times 10^5$ (blue) to $0$ (red).}
    \label{fig:SRGLikelihood-h40}
\end{figure}
\newpage

\begin{figure}[h!]
    \begin{center}
    \includegraphics{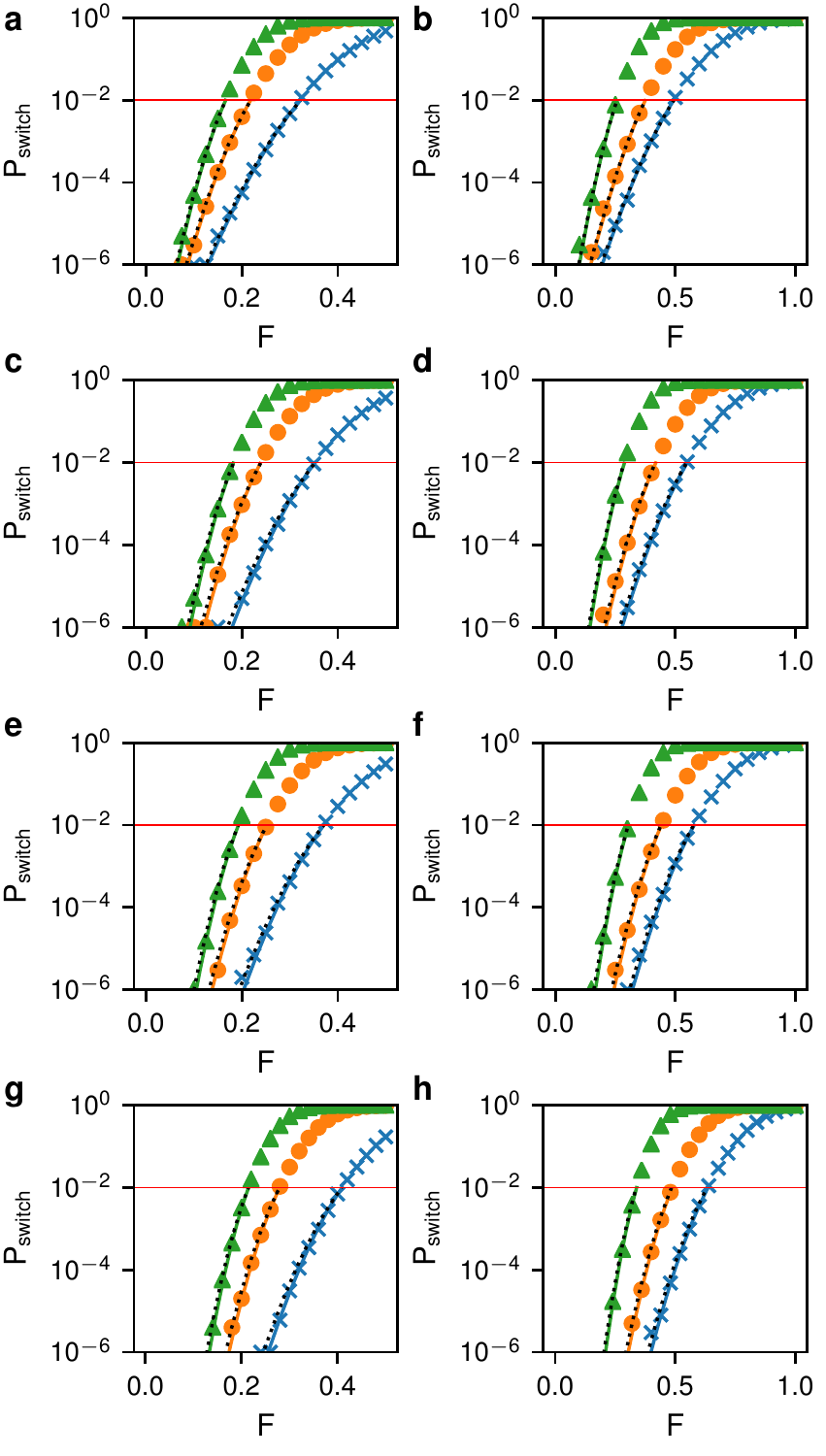}
    \end{center}
    \caption{{\bf Perturbation effect on the SRG model.} (a) Change in switching probability vs perturbation strength $F$ for the \textit{low} to \textit{high} switch  with $h=3.5$. Shown are the numerical solution (symbols), theory with MLE parameters (solid lines), and theory with true parameters and 0.15 prefactor (dotted lines). Colors give the perturbation time $T$: 0.5 (blue $\times$), 0.75 (orange $\circ$), 1.0 (green $\bigtriangleup$). (b)  Change in switching probability vs perturbation strength for the \textit{high} to \textit{low} switch with perturbation times 0.75 (blue $\times$), 1.0 (orange $\circ$), and 1.5 (green $\bigtriangleup$). Here the prefactor for the true parameter line was 0.2. (c+d) As in (a+b) except for $h=3.65$. (e+f) $h=3.75$. (g+h) $h=4.0$.}
    \label{fig:SRGSimTheoryCompSI}
\end{figure}
\newpage

\begin{figure}[h!]
    \begin{center}
    \includegraphics{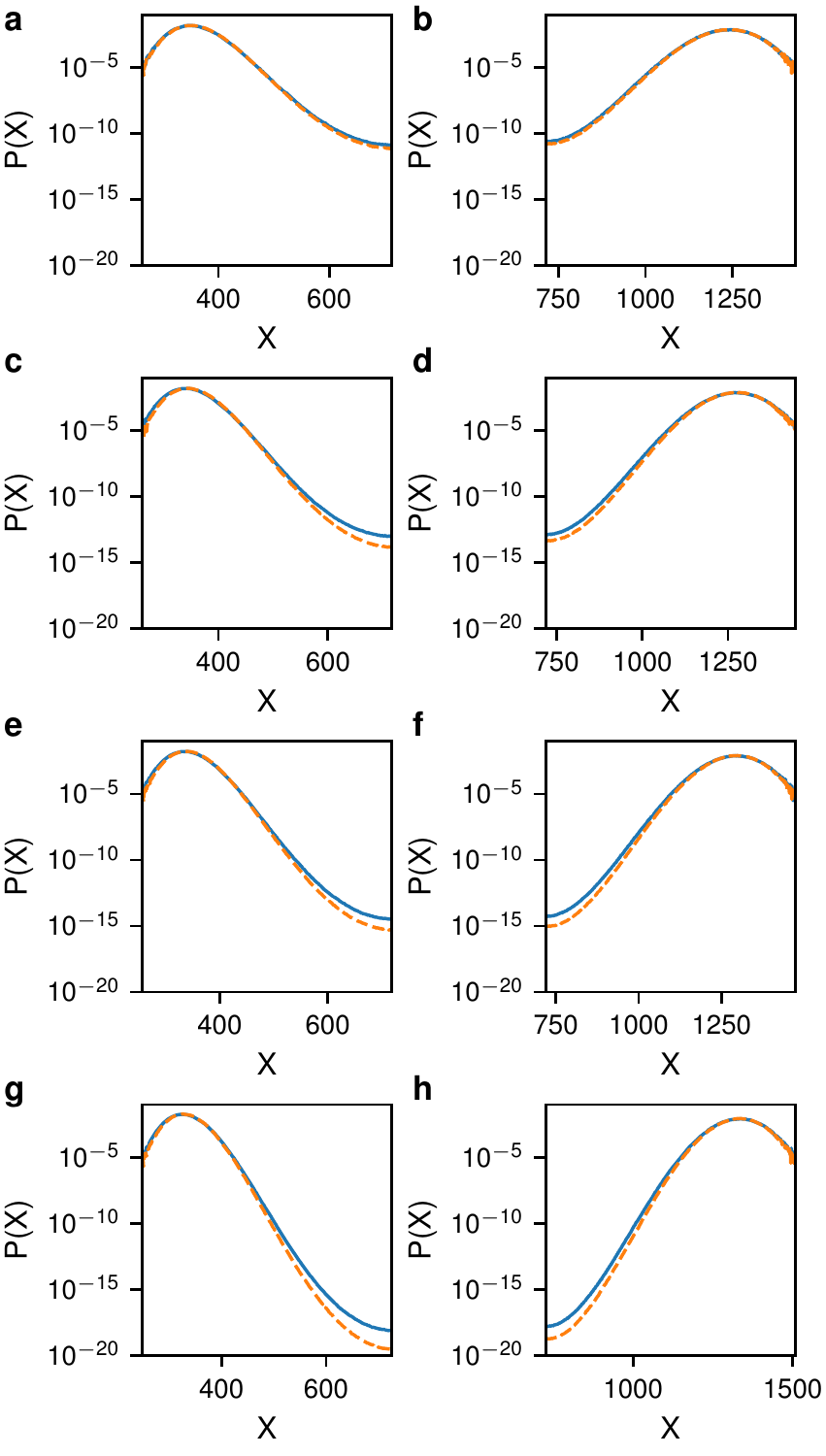}
    \end{center}
    \caption{{\bf Comparison of actual and inferred probability distributions for the self-regulating gene.} (a+b) The actual (solid blue) and inferred (dashed orange) PDFs for the \textit{low} state (a) and \textit{high} state (b) for $h=3.5$. (c+d) As in (a+b) except for $h=3.65$. (e+f) $h=3.75$. (g+h) $h=4.0$.}
    \label{fig:SRGFitPDFSI}
\end{figure}
\newpage


\begin{figure}[h!]
    \begin{center}
    \includegraphics{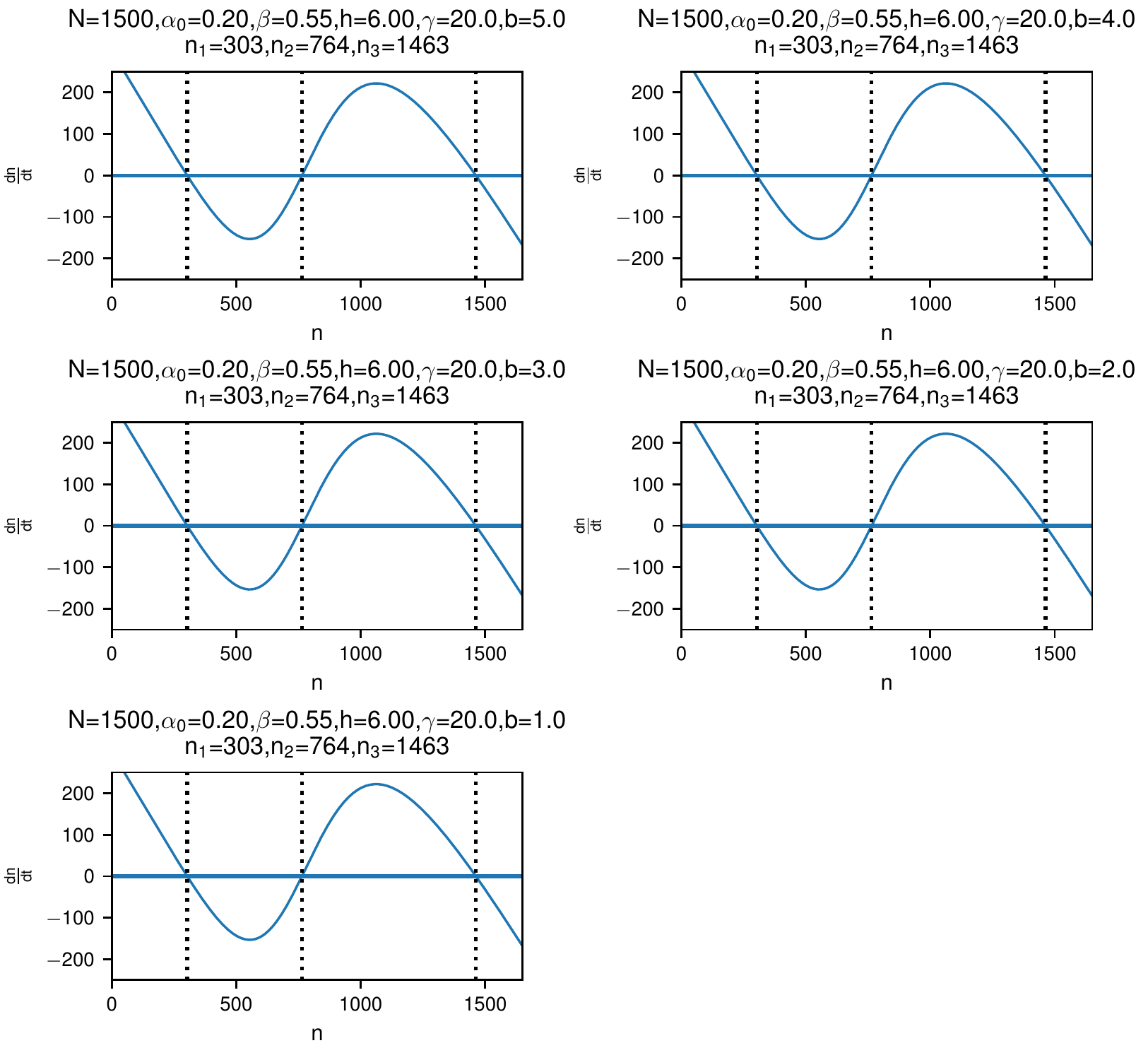}
    \end{center}
    \caption{{\bf Deterministic rate equations for the mrna-protein model.} Value of $\frac{dn}{dt}$ as a function of $n$ for the deterministic model of the mrna-protein switch. The positions of the three fixed points are given by the dotted lines. Parameters for each panel are as indicated.}
    \label{fig:MRNARateEquations}
\end{figure}
\newpage

\begin{figure}[h!]
    \begin{center}
    \includegraphics{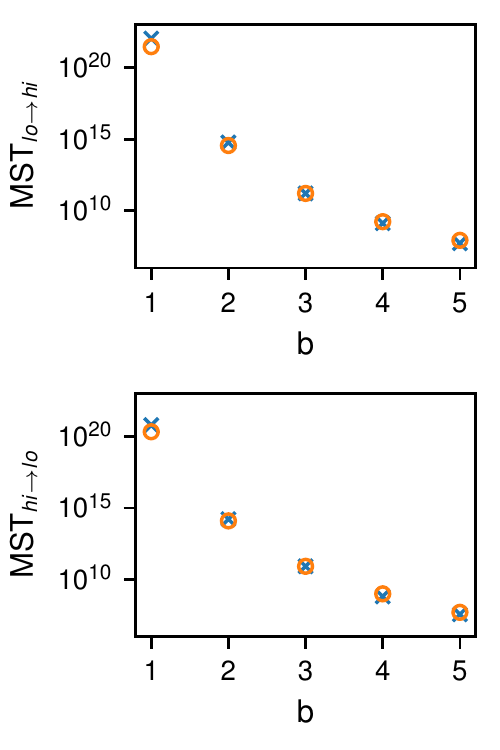}
    \end{center}
    \caption{{\bf Mean switching times for the mrna-protein model.} (top) The MST to go from the $low$ state to the $high$ state vs $b$ calculated from numerical simulations (blue $\times$) and WKB theory (orange $\circ$) as given by Eq.~(24) in the main text. The WKB points are multiplied by a constant preexponential factor of 37.33. (bottom) The same for the $high$ to $low$ state with a preexponent of 14.11. All other parameters are as in Figure \ref{fig:MRNARateEquations}.}
    \label{fig:MRNASwitchingTimes}
\end{figure}
\newpage

\begin{figure}[h!]
    \begin{center}
    \includegraphics{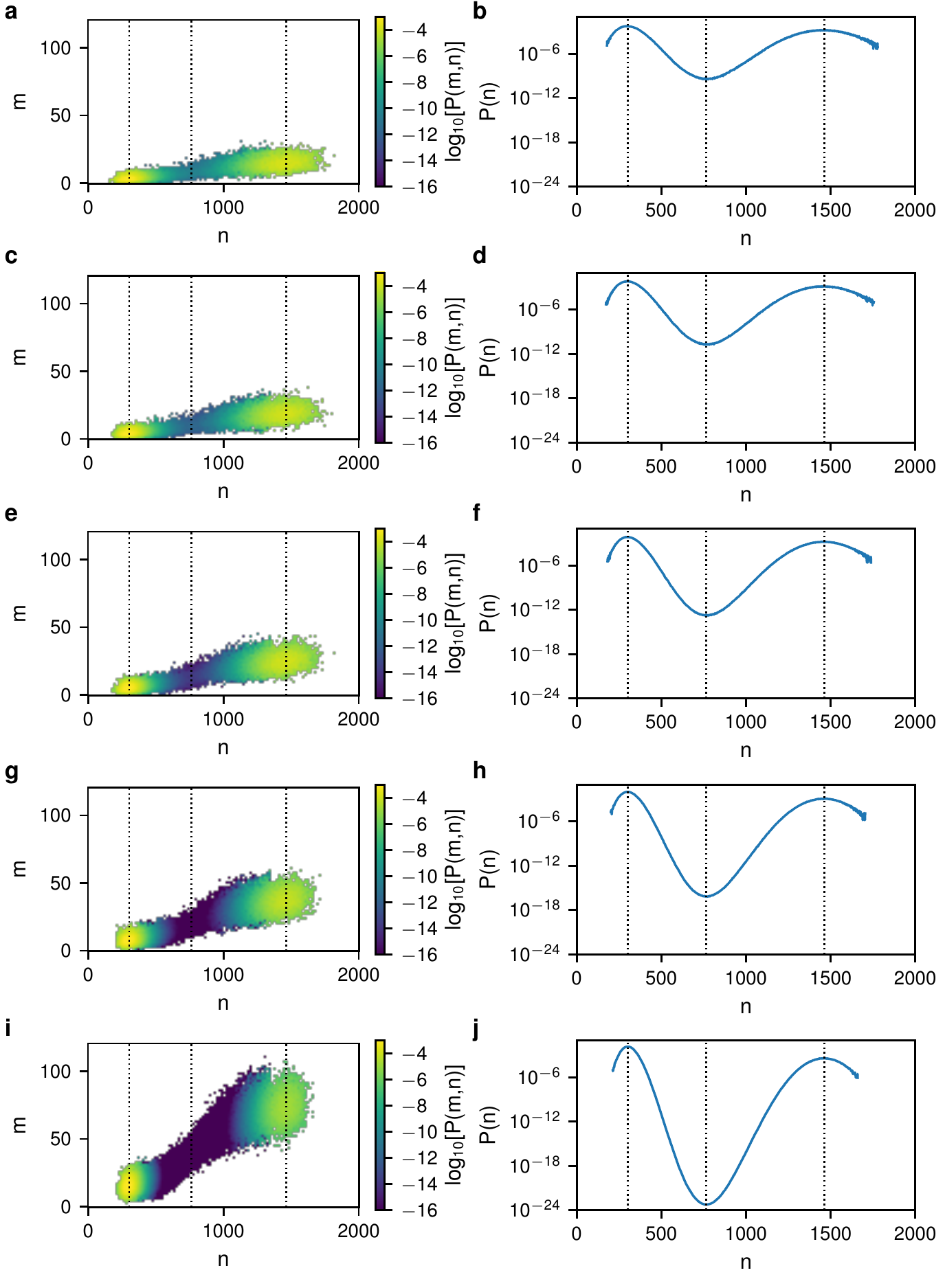}
    \end{center}
    \caption{{\bf Stationary probability distributions for the mrna-protein model.} (a) The joint probability density for a given number of mrna (m) and protein (n) molecules for the mrna-protein switch with $b=5.0$. (b) The marginal probability density for only the protein count (n) with $b=5$. (c+d) As in (a+b) except for $4=2$. (e+f) $b=3$. (g+h) $b=2$. (i+j) $b=1$.
    }
    \label{fig:MRNAPDFs}
\end{figure}
\newpage

\begin{figure}[h!]
    \begin{center}
    \includegraphics{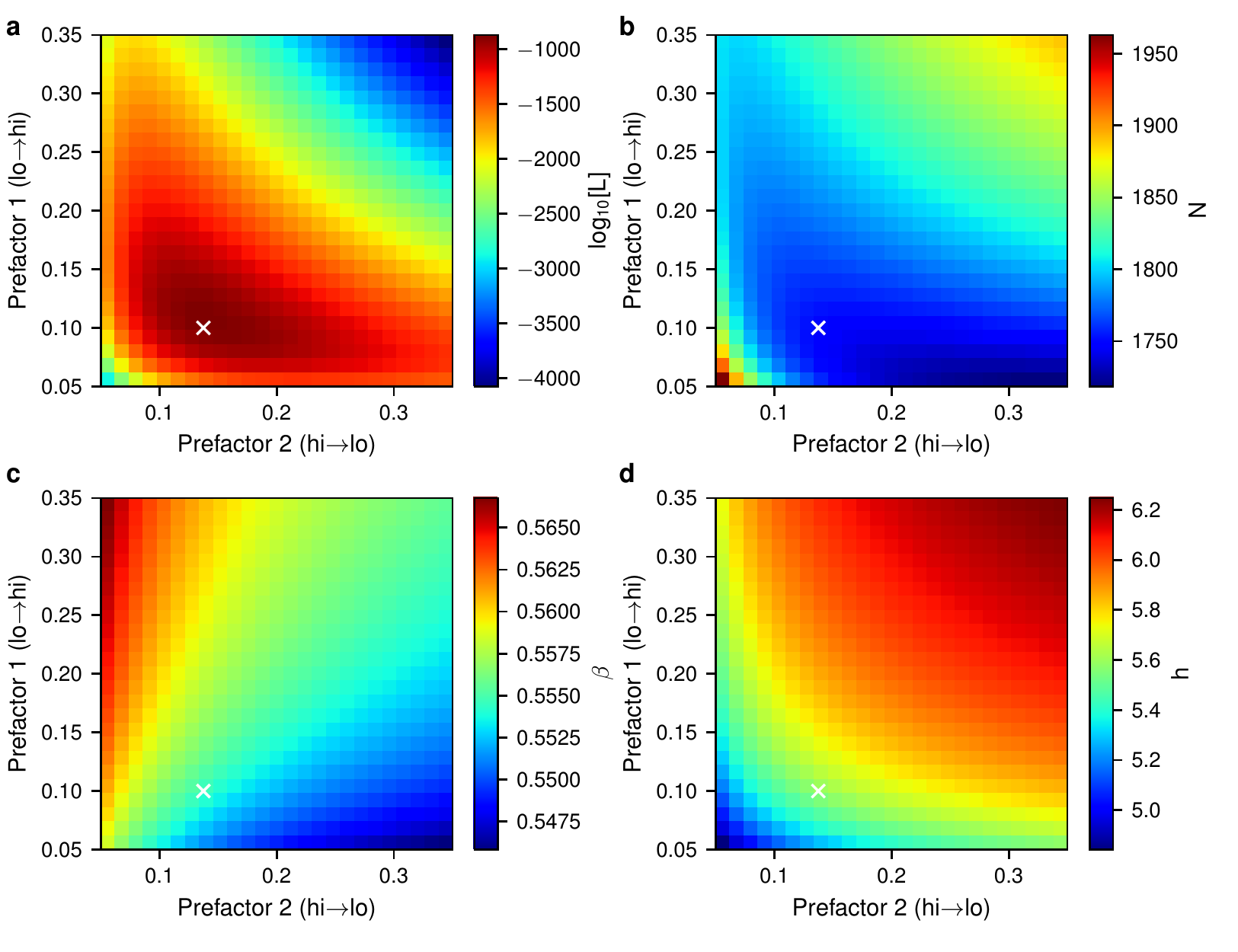}
    \end{center}
    \vspace{-0.2in}
    \caption{{\bf Prefactor dependence during maximum likelihood fitting of the mrna-protein model with $b=5$.} (a) The maximum likelihood score obtained for each prefactor pair during fitting. Fitting was performed with $\alpha$ and $b$ fixed to their true value, as described in the main text, using simulation data obtained from the mrna-protein model with parameters $N=1500$, $\alpha_0=0.2$, $\beta=0.55$, $h=6.0$, $\gamma=20$, $b=5$. The prefactor pair with the highest likelihood score is marked with a white $\times$ in each panel. (b-d) The dependence of the maximum likelihood estimate for the parameters $N$, $\beta$, and $h$, respectively, on the prefactors.}
    \label{fig:MRNAPrefactorOptimization2D-b5}
\end{figure}
\newpage

\begin{figure}[h!]
    \begin{center}
    \includegraphics{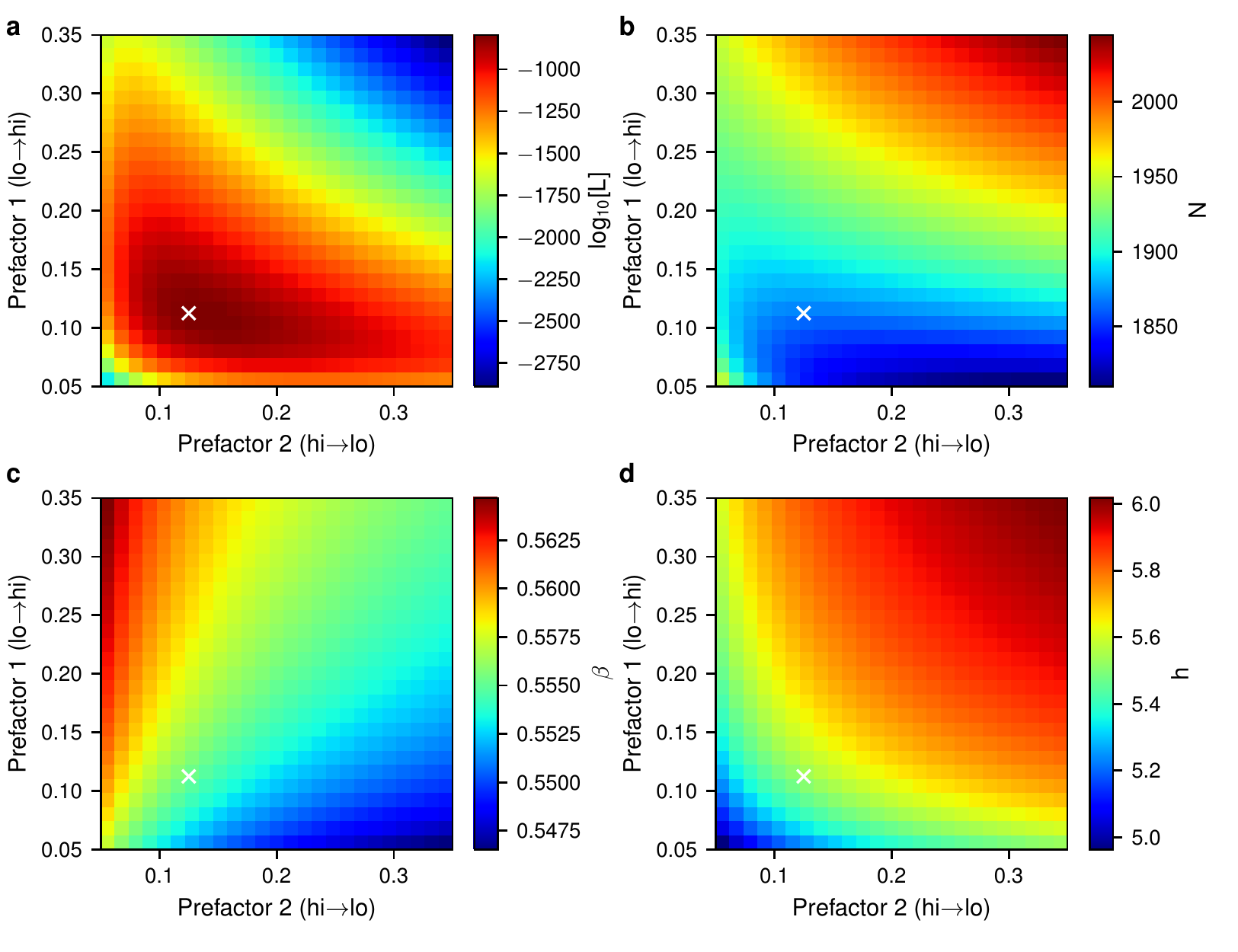}
    \end{center}
    \vspace{-0.2in}
    \caption{{\bf Prefactor dependence during maximum likelihood fitting of the mrna-protein model with $b=4$.} (a) The maximum likelihood score obtained for each prefactor pair during fitting. Fitting was performed with $\alpha$ and $b$ fixed to their true value, as described in the main text, using simulation data obtained from the mrna-protein model with parameters $N=1500$, $\alpha_0=0.2$, $\beta=0.55$, $h=6.0$, $\gamma=20$, $b=4$. The prefactor pair with the highest likelihood score is marked with a white $\times$ in each panel. (b-d) The dependence of the maximum likelihood estimate for the parameters $N$, $\beta$, and $h$, respectively, on the prefactors.}
    \label{fig:MRNAPrefactorOptimization2D-b4}
\end{figure}
\newpage

\begin{figure}[h!]
    \begin{center}
    \includegraphics{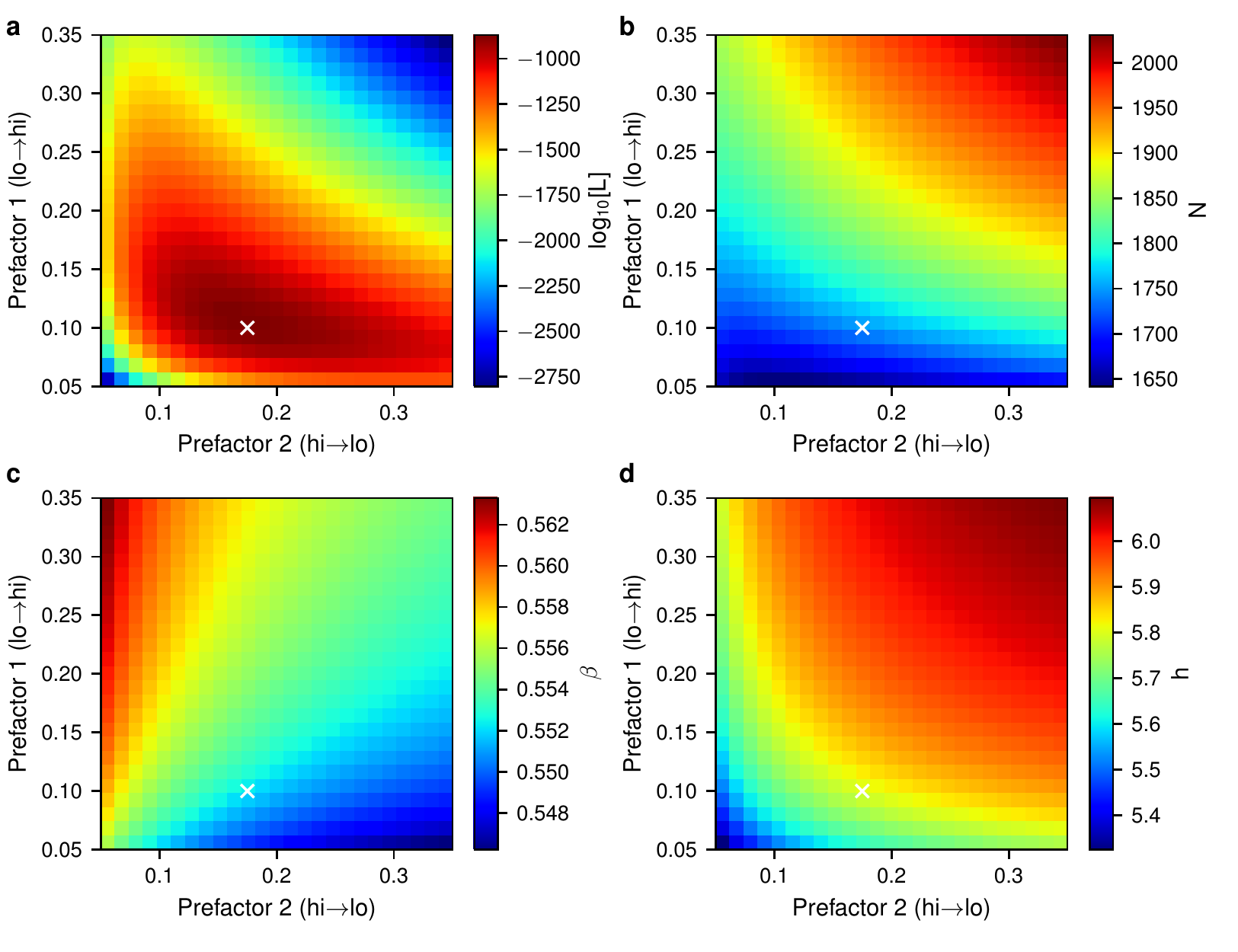}
    \end{center}
    \vspace{-0.2in}
    \caption{{\bf Prefactor dependence during maximum likelihood fitting of the mrna-protein model with $b=3$.} (a) The maximum likelihood score obtained for each prefactor pair during fitting. Fitting was performed with $\alpha$ and $b$ fixed to their true value, as described in the main text, using simulation data obtained from the mrna-protein model with parameters $N=1500$, $\alpha_0=0.2$, $\beta=0.55$, $h=6.0$, $\gamma=20$, $b=3$. The prefactor pair with the highest likelihood score is marked with a white $\times$ in each panel. (b-d) The dependence of the maximum likelihood estimate for the parameters $N$, $\beta$, and $h$, respectively, on the prefactors.}
    \label{fig:MRNAPrefactorOptimization2D-b3}
\end{figure}
\newpage

\begin{figure}[h!]
    \begin{center}
    \includegraphics{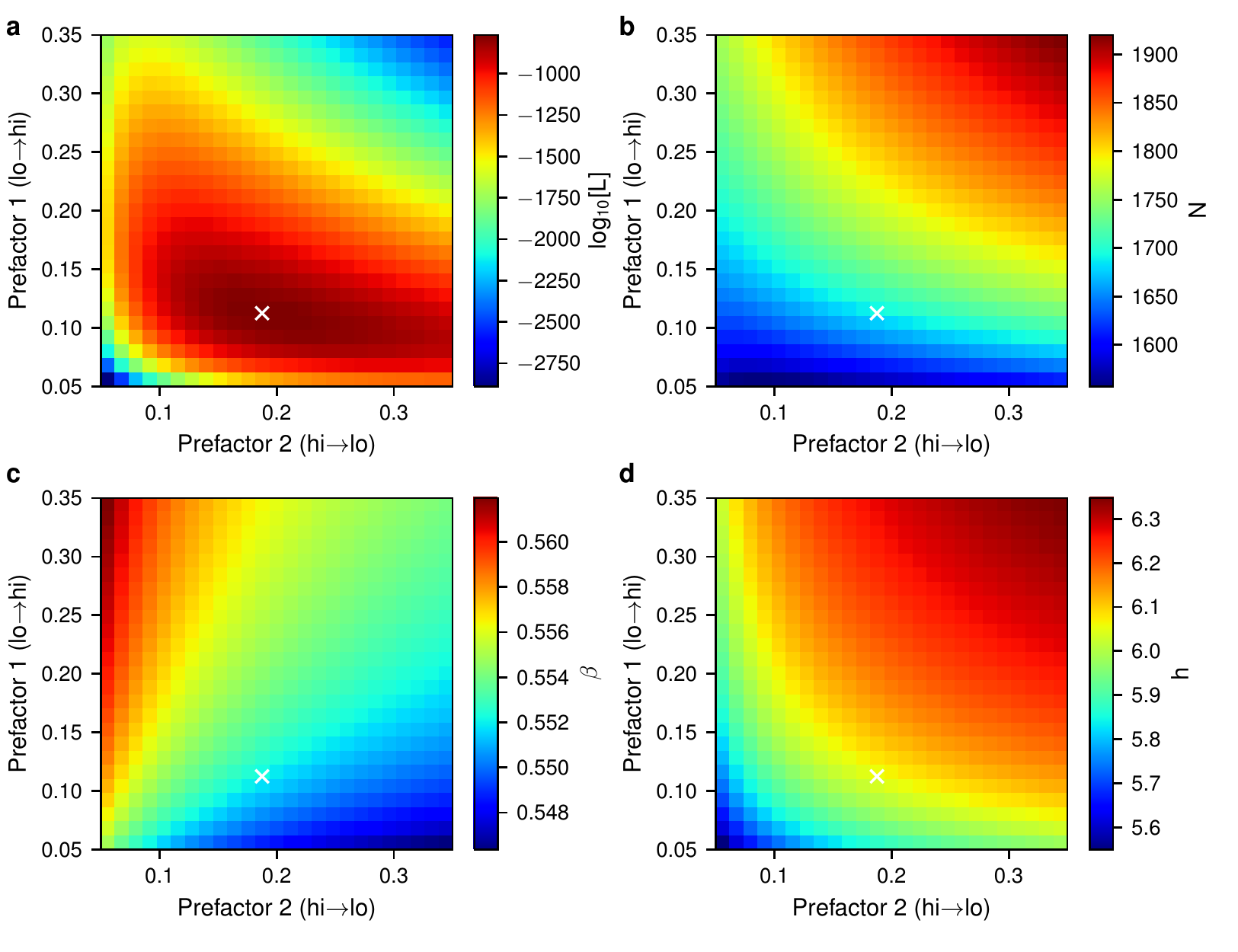}
    \end{center}
    \vspace{-0.2in}
    \caption{{\bf Prefactor dependence during maximum likelihood fitting of the mrna-protein model with $b=2$.} (a) The maximum likelihood score obtained for each prefactor pair during fitting. Fitting was performed with $\alpha$ and $b$ fixed to their true value, as described in the main text, using simulation data obtained from the mrna-protein model with parameters $N=1500$, $\alpha_0=0.2$, $\beta=0.55$, $h=6.0$, $\gamma=20$, $b=2$. The prefactor pair with the highest likelihood score is marked with a white $\times$ in each panel. (b-d) The dependence of the maximum likelihood estimate for the parameters $N$, $\beta$, and $h$, respectively, on the prefactors.}
    \label{fig:MRNAPrefactorOptimization2D-b2}
\end{figure}
\newpage

\begin{figure}[h!]
    \begin{center}
    \includegraphics{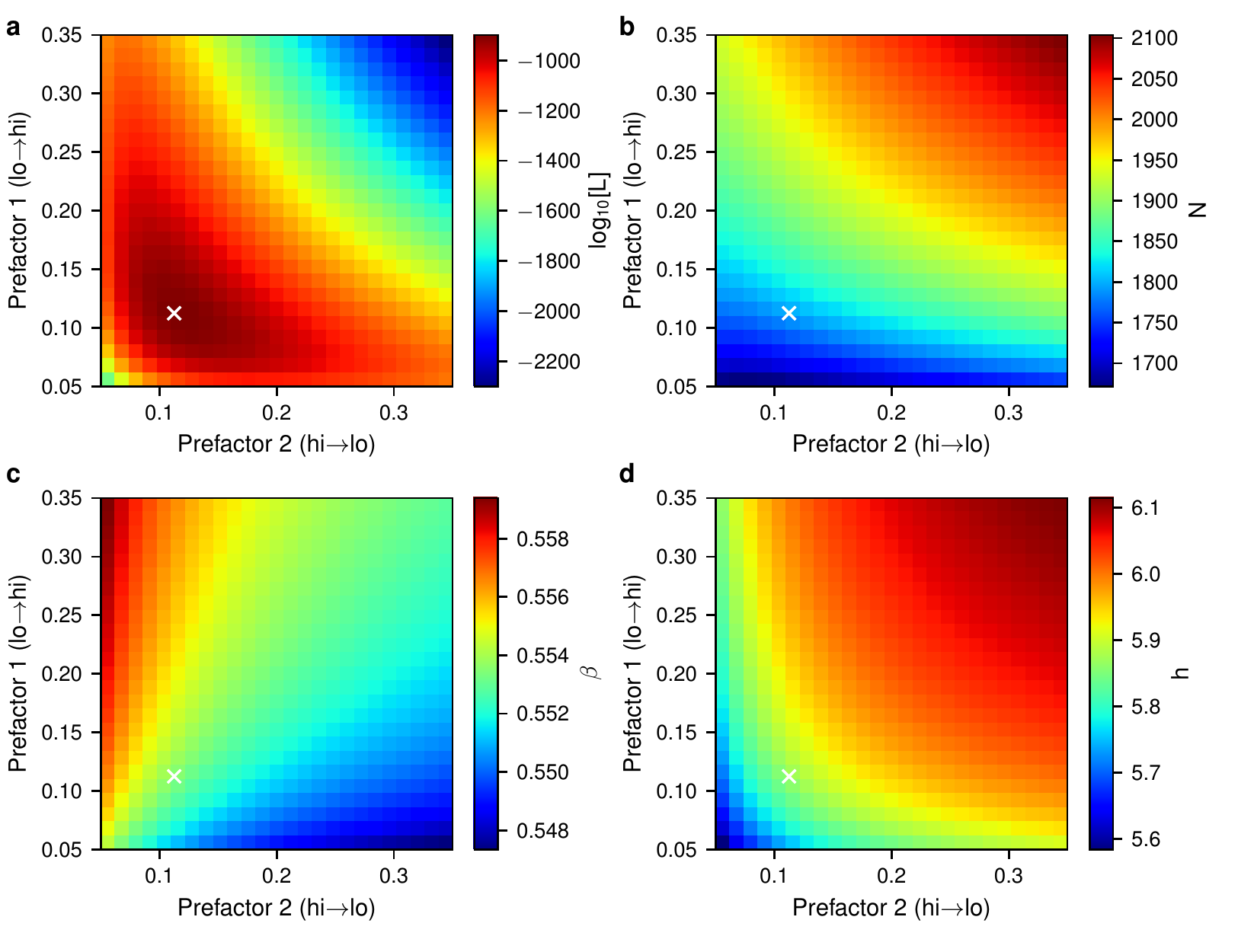}
    \end{center}
    \vspace{-0.2in}
    \caption{{\bf Prefactor dependence during maximum likelihood fitting of the mrna-protein model with $b=1$.} (a) The maximum likelihood score obtained for each prefactor pair during fitting. Fitting was performed with $\alpha$ and $b$ fixed to their true value, as described in the main text, using simulation data obtained from the mrna-protein model with parameters $N=1500$, $\alpha_0=0.2$, $\beta=0.55$, $h=6.0$, $\gamma=20$, $b=1$. The prefactor pair with the highest likelihood score is marked with a white $\times$ in each panel. (b-d) The dependence of the maximum likelihood estimate for the parameters $N$, $\beta$, and $h$, respectively, on the prefactors.}
    \label{fig:MRNAPrefactorOptimization2D-b1}
\end{figure}
\newpage

\begin{figure}[h!]
    \begin{center}
    \includegraphics{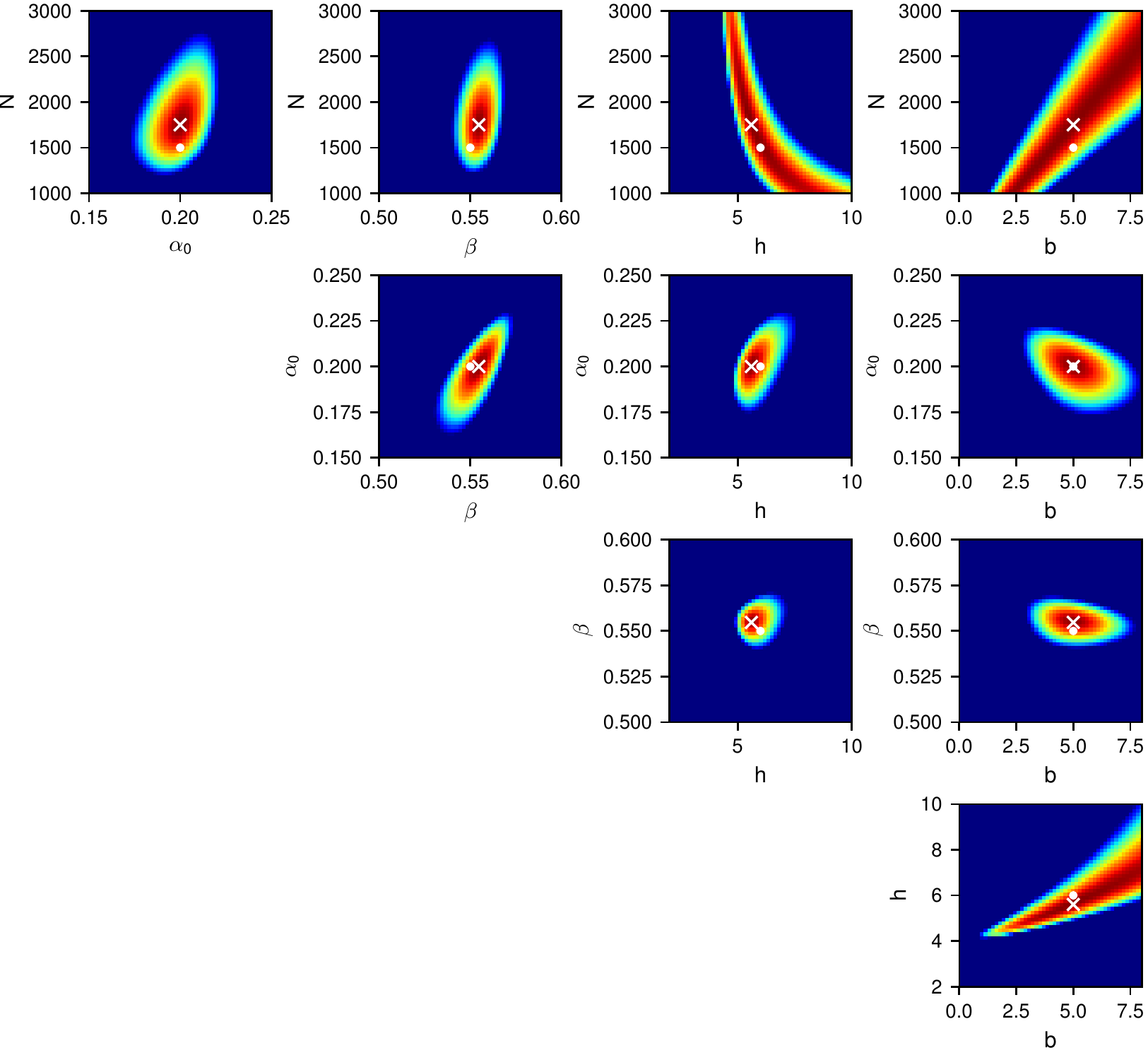}
    \end{center}
    \vspace{-0.2in}
    \caption{{\bf Likelihood distribution for the mrna-protein model with $b=5$.} Likelihood distribution for inference of all pairs of parameters for the mrna-protein model, using the optimized prefactors. For each plot, all other parameters are fixed to their MLE. The MLE is marked with a white $\times$ and the true parameter values are marked with a white $\bullet$. Colors show $\mathrm{log}_{10}[\mathrm{L}]$ and range from $-1\times 10^5$ (blue) to $0$ (red).}
    \label{fig:MRNALikelihood-b5}
\end{figure}
\newpage

\begin{figure}[h!]
    \begin{center}
    \includegraphics{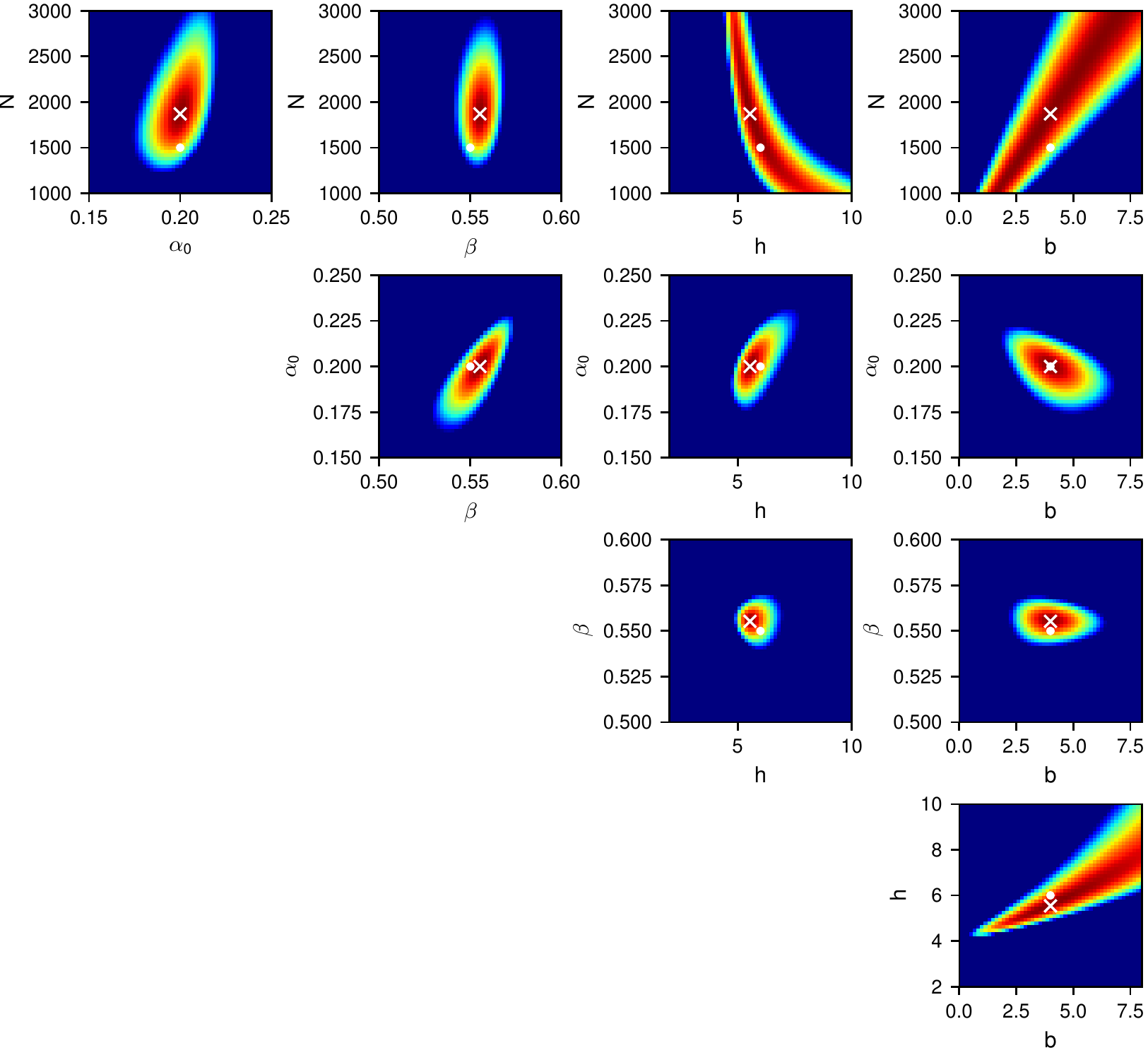}
    \end{center}
    \vspace{-0.2in}
    \caption{{\bf Likelihood distribution for the mrna-protein model with $b=4$.} Likelihood distribution for inference of all pairs of parameters for the mrna-protein model, using the optimized prefactors. For each plot, all other parameters are fixed to their MLE. The MLE is marked with a white $\times$ and the true parameter values are marked with a white $\bullet$. Colors show $\mathrm{log}_{10}[\mathrm{L}]$ and range from $-1\times 10^5$ (blue) to $0$ (red).}
    \label{fig:MRNALikelihood-b4}
\end{figure}
\newpage

\begin{figure}[h!]
    \begin{center}
    \includegraphics{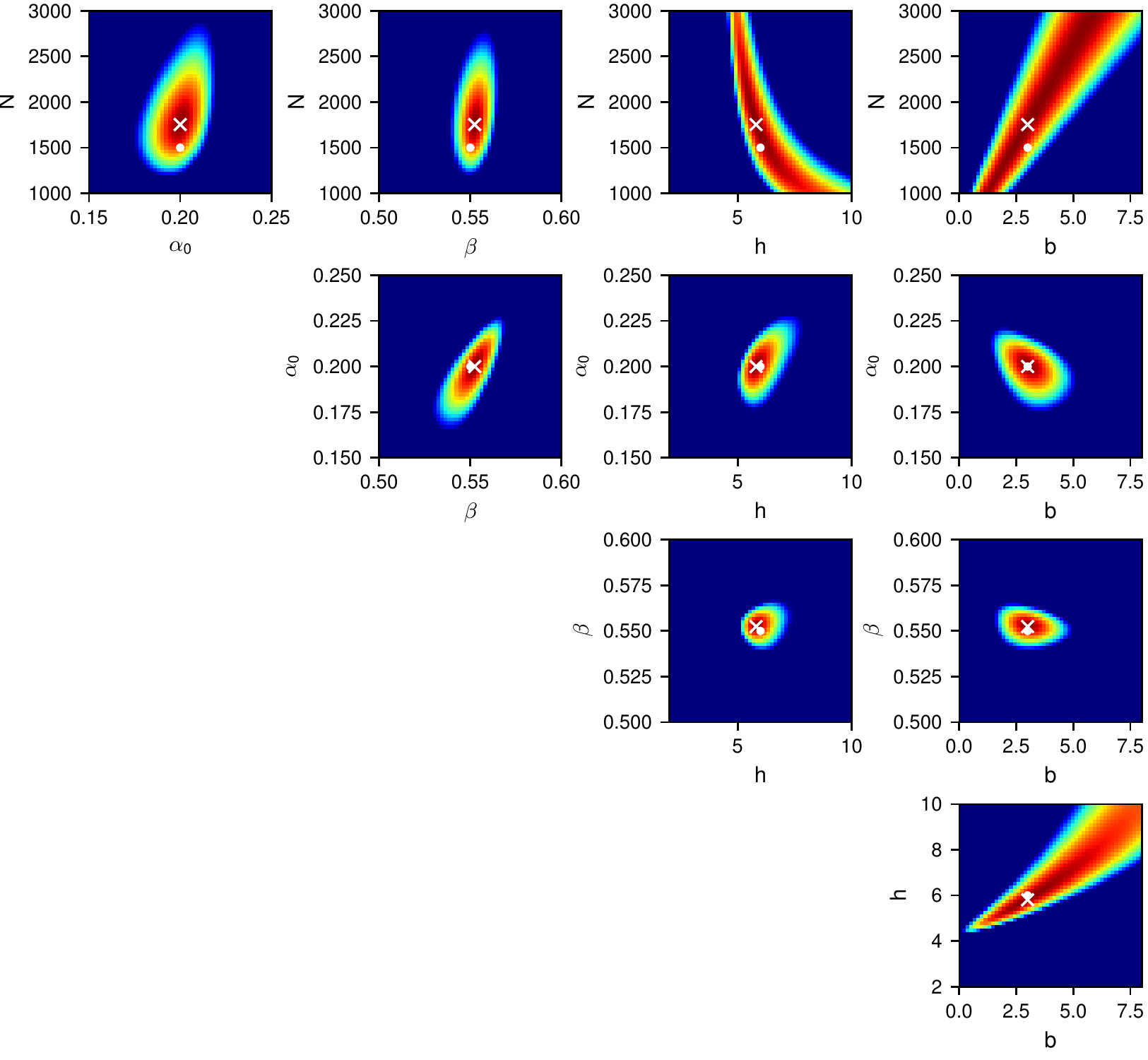}
    \end{center}
    \vspace{-0.2in}
    \caption{{\bf Likelihood distribution for the mrna-protein model with $b=3$.} Likelihood distribution for inference of all pairs of parameters for the mrna-protein model, using the optimized prefactors. For each plot, all other parameters are fixed to their MLE. The MLE is marked with a white $\times$ and the true parameter values are marked with a white $\bullet$. Colors show $\mathrm{log}_{10}[\mathrm{L}]$ and range from $-1\times 10^5$ (blue) to $0$ (red).}
    \label{fig:MRNALikelihood-b3}
\end{figure}
\newpage

\begin{figure}[h!]
    \begin{center}
    \includegraphics{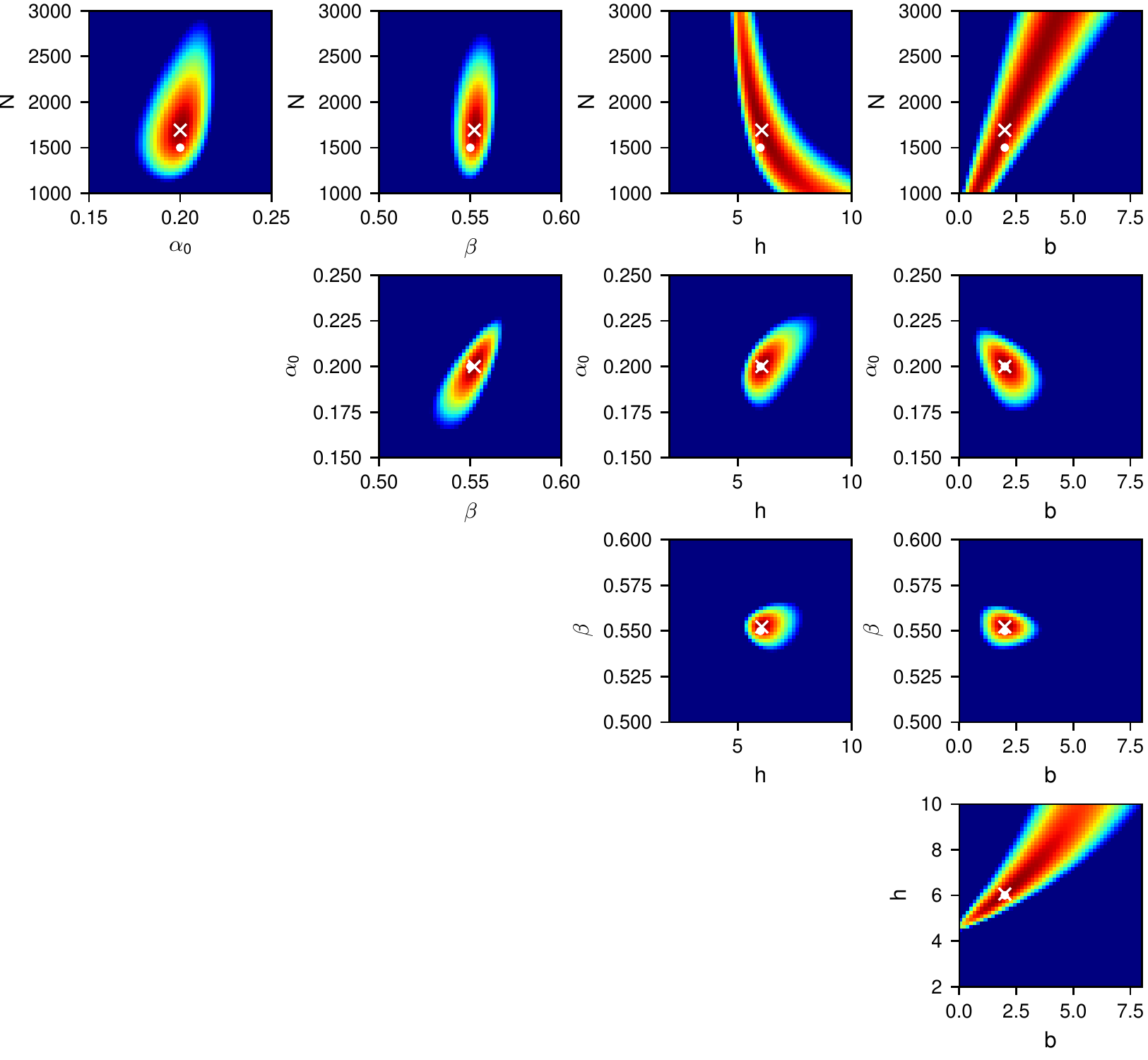}
    \end{center}
    \vspace{-0.2in}
    \caption{{\bf Likelihood distribution for the mrna-protein model with $b=2$.} Likelihood distribution for inference of all pairs of parameters for the mrna-protein model, using the optimized prefactors. For each plot, all other parameters are fixed to their MLE. The MLE is marked with a white $\times$ and the true parameter values are marked with a white $\bullet$. Colors show $\mathrm{log}_{10}[\mathrm{L}]$ and range from $-1\times 10^5$ (blue) to $0$ (red).}
    \label{fig:MRNALikelihood-b2}
\end{figure}
\newpage

\begin{figure}[h!]
    \begin{center}
    \includegraphics{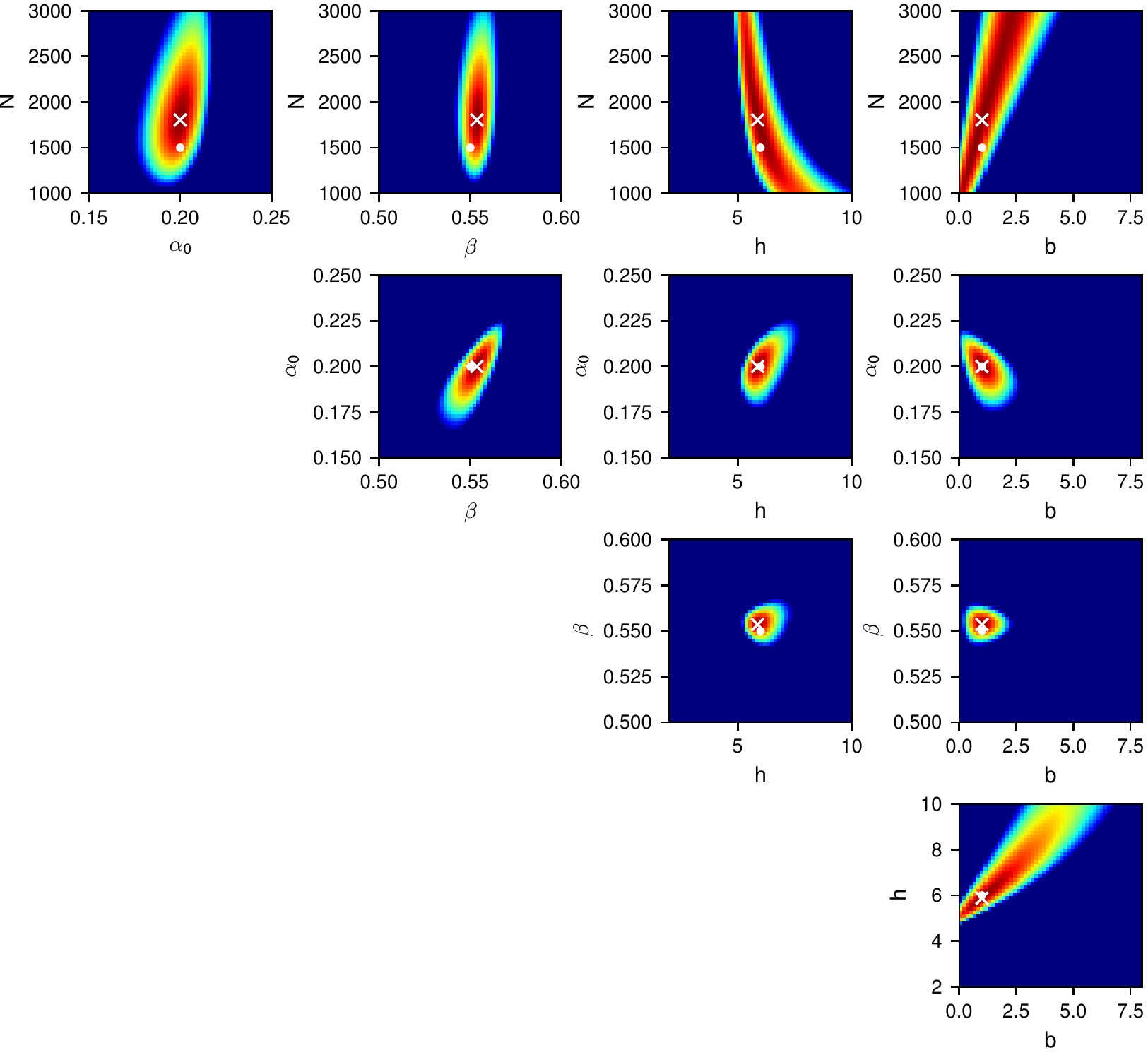}
    \end{center}
    \vspace{-0.2in}
    \caption{{\bf Likelihood distribution for the mrna-protein model with $b=1$.} Likelihood distribution for inference of all pairs of parameters for the mrna-protein model, using the optimized prefactors. For each plot, all other parameters are fixed to their MLE. The MLE is marked with a white $\times$ and the true parameter values are marked with a white $\bullet$. Colors show $\mathrm{log}_{10}[\mathrm{L}]$ and range from $-1\times 10^5$ (blue) to $0$ (red).}
    \label{fig:MRNALikelihood-b1}
\end{figure}
\newpage

\begin{figure}[h!]
    \begin{center}
    \includegraphics{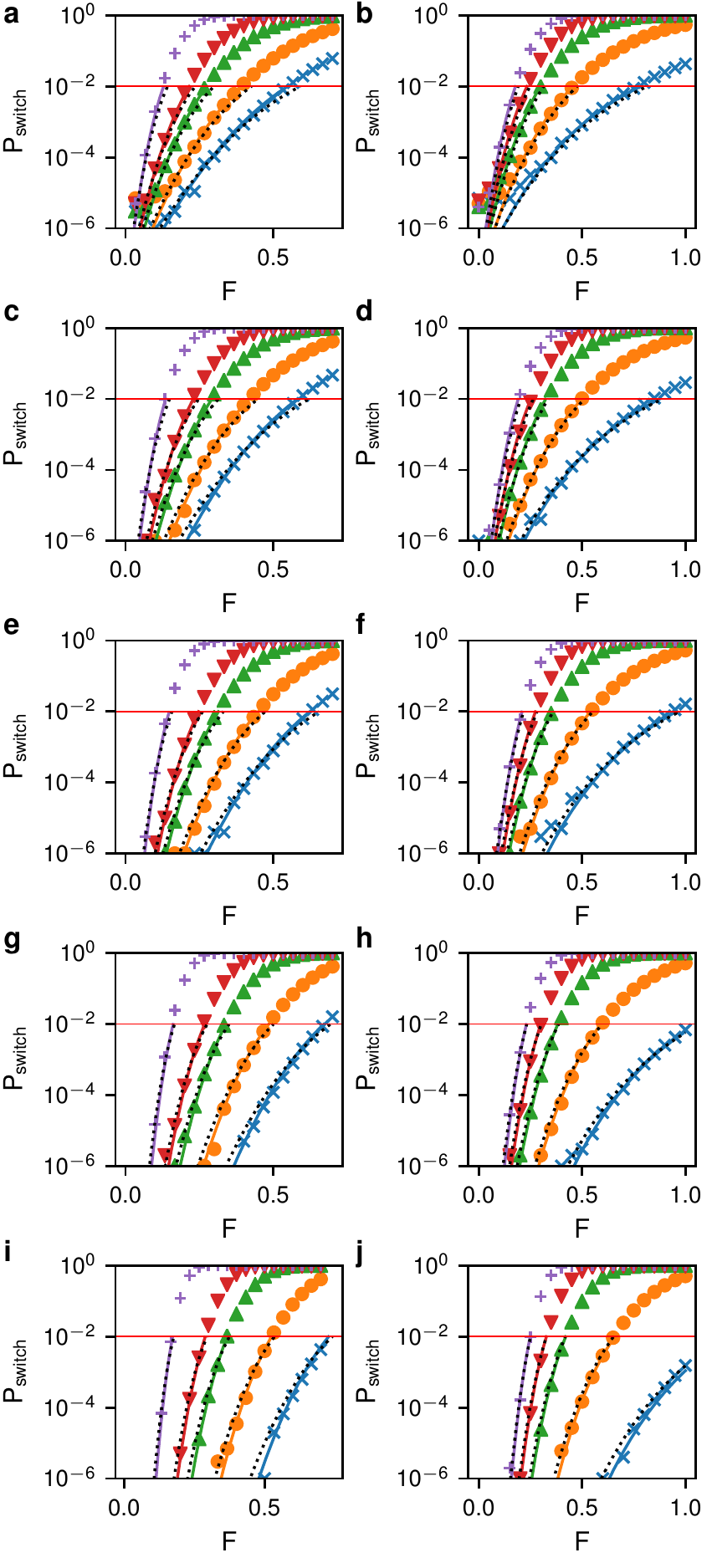}
    \end{center}
    \caption{{\bf Perturbation effect on the mrna-protein model.} (a) Change in switching probability vs perturbation strength $F$ for the \textit{low} to \textit{high} switch  with $b=5$. Shown are the numerical solution (symbols), theory with MLE parameters (solid lines), and theory with true parameters and 0.05 prefactor (dotted lines). Colors give the perturbation time $T$: 0.35 (blue $\times$), 0.5 (orange $\circ$), 0.75 (green $\bigtriangleup$), 1.0 (red $\bigtriangledown$), 2.0 (purple $+$). (b)  Change in switching probability vs perturbation strength for the \textit{high} to \textit{low} switch with perturbation times 1.0 (blue $\times$), 1.5 (orange $\circ$), 2.25 (green $\bigtriangleup$), 3.0 (red $\bigtriangledown$), 4.5 (purple $+$). Here the prefactor for the true parameter line was 0.15. (c+d) As in (a+b) except for $b=4$. (e+f) $b=3$. (g+h) $b=2$ (i+j) $b=1$.}
    \label{fig:MRNASimTheoryCompSI}
\end{figure}
\newpage

\begin{figure}[h!]
    \begin{center}
    \includegraphics{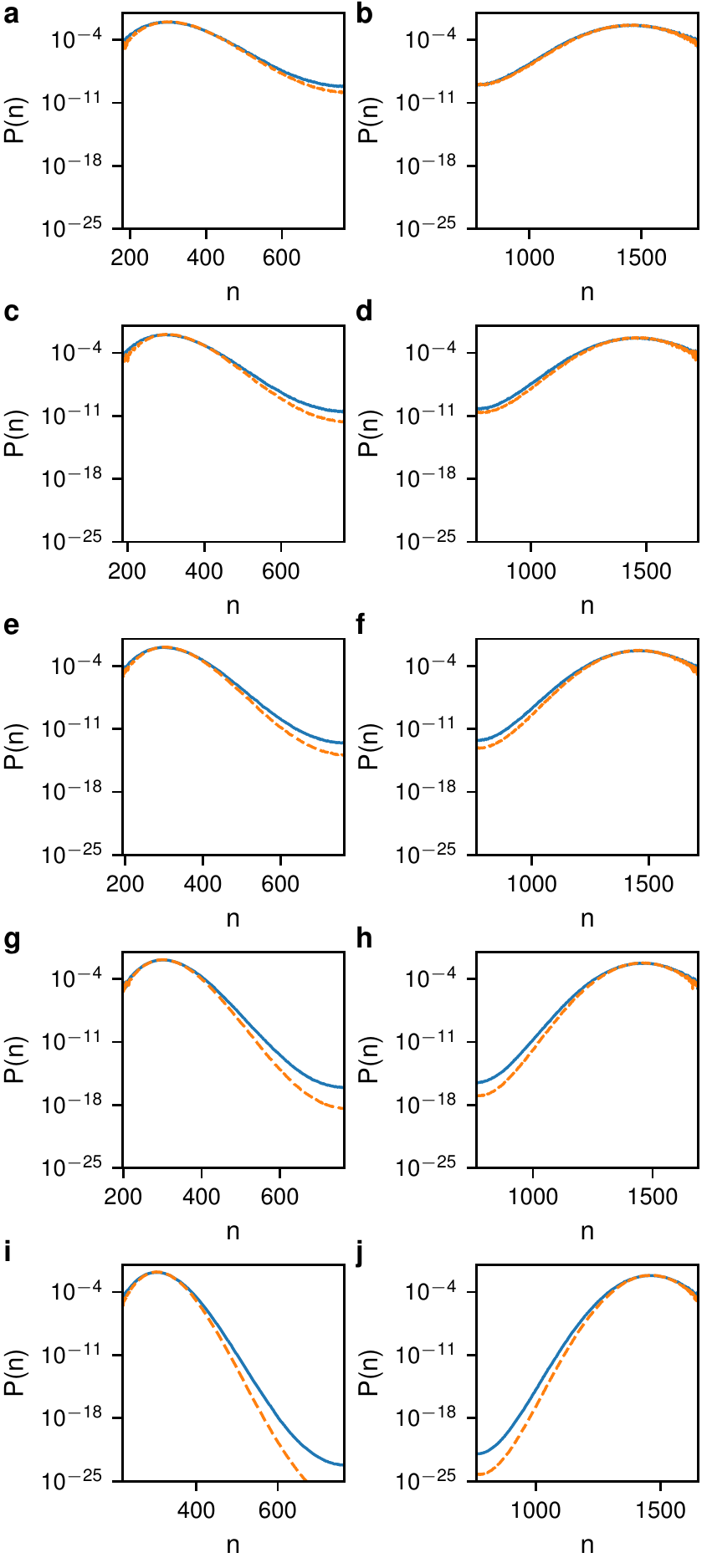}
    \end{center}
    \caption{{\bf Comparison of actual and inferred probability distributions for the mrna-protein model.} (a+b) The actual (solid blue) and inferred (dashed orange) PDFs for the \textit{low} state (a) and \textit{high} state (b) for $b=5$. (c+d) As in (a+b) except for $b=4$. (e+f) $b=3$. (g+h) $b=2$ (i+j) $b=1$.}
    \label{fig:MRNAFitPDFSI}
\end{figure}
\newpage

\begin{figure}[h!]
    \begin{center}
    \includegraphics{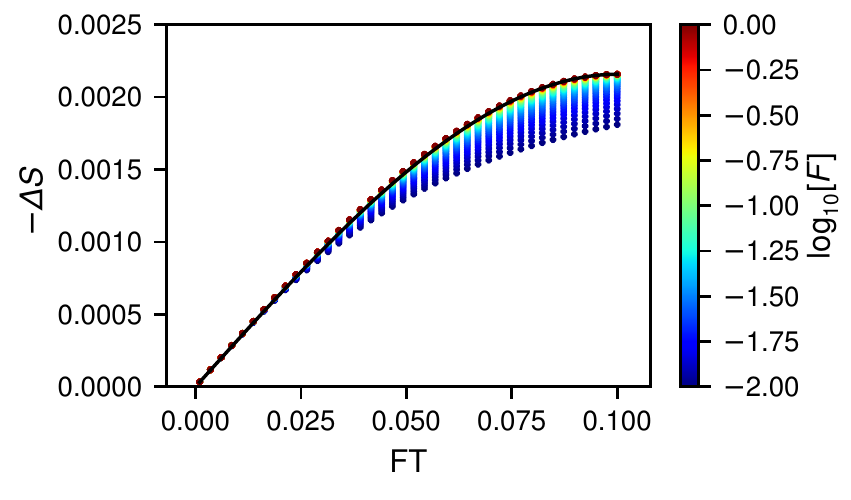}
    \end{center}
    \caption{{\bf Dependence of change in the switching barrier on impulse near bifurcation.} Distribution of the perturbed switching barrier $\Delta{}S = {\cal S}^{lh}-{\cal S}^{lh}_0$ versus the total applied impulse $F\,T$ for the SRG model. Each symbol represents a perturbation with a different $F$, which is given by the color. The solid line shows the bifurcation theory given in Eq.~(44) in the main text. The parameters were $\alpha_0=0.2$, $\beta=0.501$ and $h=4$}
    \label{fig:SRGBifurcation}
\end{figure}
\clearpage
\clearpage

\end{document}